\documentclass[aps,prl,twocolumn,superscriptaddress,showpacs,footnotebib]{revtex4-2}
\usepackage{amsmath}
\usepackage{bm}
\usepackage{color}
\usepackage{graphicx}
\usepackage{epstopdf}
\usepackage{epsfig}
\usepackage{amsfonts}
\usepackage[naturalnames]{hyperref}
\usepackage{hypcap}
\usepackage{verbatim}
\usepackage{tabularx}
\usepackage{bbm}
\usepackage{esvect}
\usepackage{subfigure}
\usepackage{slashed}

\usepackage{slashed}

\usepackage[T1]{fontenc} % if needed

\def\bea{\begin{eqnarray}}
\def\eea{\end{eqnarray}}

\def\ba{\begin{array}}
\def\ea{\end{array}}

\hypersetup{colorlinks=true, citecolor=blue, urlcolor=blue, linkcolor=blue}

\usepackage[dvipsnames]{xcolor}
\usepackage[normalem]{ulem} % for `\sout{}`

\begin{document}

\title{Magnetorotons in moir\'e fractional Chern insulators}

\author{Xiaoyang Shen}
\thanks{These authors contributed equally to the work.}

\affiliation{State Key Laboratory of Low Dimensional Quantum Physics and
Department of Physics, Tsinghua University, Beijing, 100084, China}

\author{Chonghao Wang}
\thanks{These authors contributed equally to the work.}
\affiliation{State Key Laboratory of Low Dimensional Quantum Physics and
Department of Physics, Tsinghua University, Beijing, 100084, China} 
\author{Xiaodong Hu}
\thanks{These authors contributed equally to the work.}
\affiliation{Department of Material Science and Engineering, University of Washington, Seattle, WA 98195, USA}
\author{Ruiping Guo}         
\affiliation{State Key Laboratory of Low Dimensional Quantum Physics and
Department of Physics, Tsinghua University, Beijing, 100084, China}
\affiliation{Institute for Advanced Study, Tsinghua University, Beijing 100084, China}
\author{\\ Hong Yao}
\affiliation{Institute for Advanced Study, Tsinghua University, Beijing 100084, China}
\author{Chong Wang}
\email{chongwang@mail.tsinghua.edu.cn}
\affiliation{State Key Laboratory of Low Dimensional Quantum Physics and
Department of Physics, Tsinghua University, Beijing, 100084, China}
\author{Wenhui Duan}
\email{duanw@tsinghua.edu.cn}
\affiliation{State Key Laboratory of Low Dimensional Quantum Physics and
Department of Physics, Tsinghua University, Beijing, 100084, China}
\affiliation{Institute for Advanced Study, Tsinghua University, Beijing 100084, China}
\affiliation{Frontier Science Center for Quantum Information, Beijing, China}
\affiliation{Beijing Academy of Quantum Information Sciences, Beijing 100193, China}
\author{Yong Xu}
\email{yongxu@mail.tsinghua.edu.cn}
\thanks{}
\affiliation{State Key Laboratory of Low Dimensional Quantum Physics and
Department of Physics, Tsinghua University, Beijing, 100084, China}
\affiliation{Frontier Science Center for Quantum Information, Beijing, China}
\affiliation{RIKEN Center for Emergent Matter Science (CEMS), Wako, Saitama 351-0198, Japan}

\date{\today} 

\begin{abstract}
The discovery of fractional Chern insulators (FCIs) unlocks exciting opportunities to explore emergent physical excitations arising from topological and geometric effects in novel phases of quantum matter. Here we investigate the intraband neutral excitations, namely magnetorotons, in moir\'e FCIs within twisted $\rm{MoTe}_2$ by applying the Girvin, MacDonald, and Platzman (GMP) ansatz together with the method of dynamical geometric response. We reveal the universal existence of the finite-momentum magnetorotons in moir\'e FCIs and predict their characteristic scales. Furthermore, we explore the geometric nature of magnetorotons in the long-wavelength limit, identifying their gapped chiral nature with angular momentum-2, which originates from the momentum-space incompressibility of FCIs. Utilizing the excellent tunability of moir\'e systems, we extend our analysis to other incompressible phases and uncover the dynamical properties of geometric excitations influenced by quantum phase transitions. Finally, we provide experimental proposals for detecting and advancing the study of intraband neutral excitations in moir\'e FCIs. 
\end{abstract}

\maketitle
{\it Introduction.---}
Recent experimental discovery of fractional Chern insulators (FCIs) in twisted transition metal dichalocogenide (TMD) homobilayers has sparked intensive interest in exploring novel quantum states in moir\'e systems~\cite{Cai2023,Park2023,LiTingxin2023PRX,JuLong2024,xie2021fractional,kang2024evidence,zeng2023thermodynamic,xie2405even}. FCIs can be regarded as the zero-field, lattice analogy of fractional quantum Hall (FQH) states, henceforth inheriting the intrinsic topological nature of the FQH phase and the potential for quantum computation~\cite{nayak2008non,kitaev2003fault}. Numerous theoretical studies~\cite{2021li,2021Devakul,abouelkomasan2020,2020Repellin,2021wilhelm,PhysRevLett.131.136501,PhysRevB.108.205144,wang2024fractional} have predicted the existence of FCIs in moiré materials, which are likely to give rise to emergent physics beyond the conventional lowest Landau level (LLL) framework~\cite{2014Roy,regnault2011fractional,2011Qi,Jackson_2015,Parameswaran_2013,2021wang,wang2024fractionalcherninsulatorsmoire,PhysRevLett.133.166503,PhysRevB.110.L161109}. 
While most research focused on studying the ground state properties of FCIs, the low-lying collective excitations that could encode the significant information related to the topological and geometric properties of FCIs remain largely unexplored~\cite{Lu_2024,Repellin_2014,PhysRevB.109.245125}.

A celebrated hallmark of FQH states is the existence of intraband neutral excitations, known as \textit{magnetorotons}, proposed by Girvin, MacDonald, and Platzman (GMP) \cite{PhysRevB.33.2481}. Such kind of neutral excitations can be effectively described using a density wave ansatz within the framework of the single mode approximation (SMA), which reveals a ``roton-like'' local minimum at finite momentum. 
The GMP theory further demonstrates that in the long-wavelength limit, the energy dispersion of magnetorotons is gapped and can be interpreted as bi-excitons with quadrupole structure due to the $(q\ell_B)^4$ behavior of cross section ~\cite{PhysRev.123.1242,PhysRevLett.114.156802,wolf2024intrabandcollectiveexcitationsfractional,cavicchi2024opticalpropertiesplasmonsorbital}. Recent studies~\cite{Gromov:2017qeb,Haldane_2021,PhysRevX.5.031027,haldane2011selfdualitylongwavelengthbehaviorlandaulevel,PhysRevB.90.014435,PhysRevB.88.125137} have revealed that magnetorotons of FQH in the long-wavelength limit can be interpreted as excitations of the internal quantum metric. These excitations exhibit a chiral spin-two nature, earning the name of ``massive chiral gravitons''~\cite{CGFQHLiou,MMSSNguyen,yuzhu2023geometric,geometric_yang,PhysRevB.98.155140,Du_2022}. The remarkable features of geometric excitations have been experimentally observed a few months ago \cite{PhysRevLett.70.3983,liang2024evidence,PhysRevLett.86.2637}. The study of magnetorotons in FQH systems naturally motivates the exploration of similar excitations in FCIs hosted in moir\'e flatbands. The exceptional tunability of moir\'e systems facilitates the study of magnetorotons not only in ideal FCIs but also in less ideal FCIs undergoing transitions to competing phases, offering opportunities to uncover intriguing physics beyond the conventional LLL framework.

%The study of magnetorotons in FQH naturally raises interest in the exotic behaviors of magnetorotons in FCI hosted in moir\'e flatband. The prominent tunability of moir\'e system enables one to study the magnetorotons not only in ideal FCIs, but also in less ideal FCIs and transition into other competing phases, where the intriguing physics beyond the LLL might emerge.

% All these naturally raise the question: does there exist analogous neutral excitation in moir\'e FCIs? What are the similarities and differences of the intraband neutral excitations between FQHs and FCIs, and the other strongly-correlated phases?
\begin{figure*}[t]
    \centering
    \includegraphics[width=0.88\linewidth]{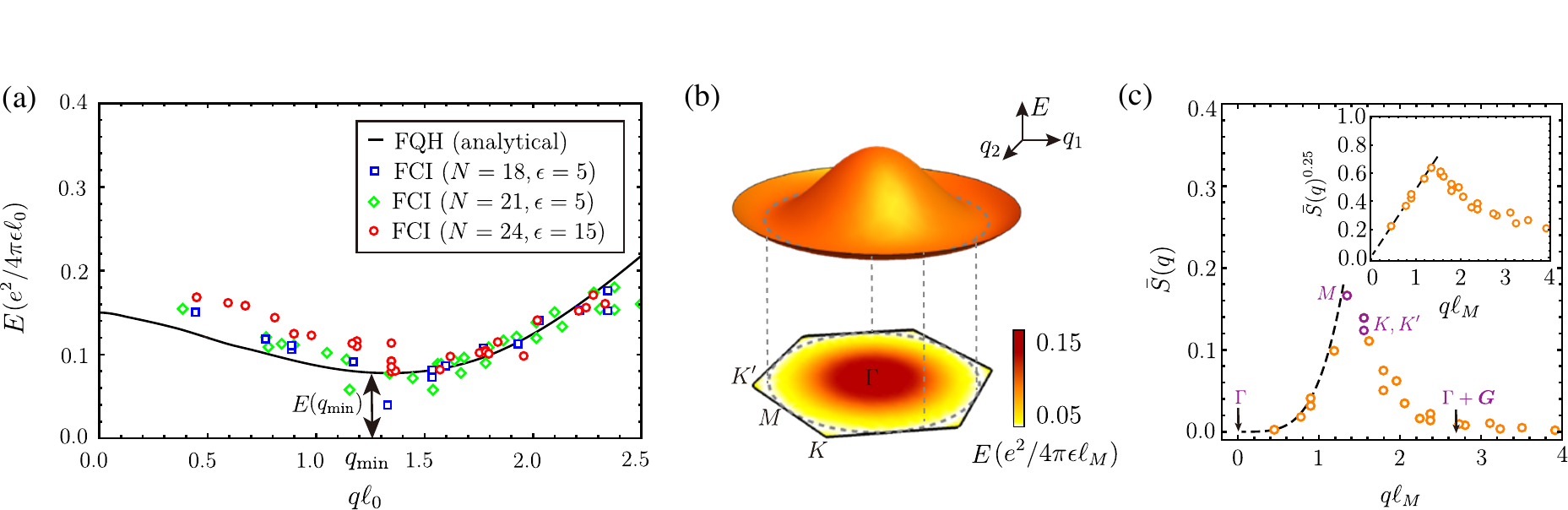}
    \caption{(a) Energy dispersions of magnetorotons in FCIs obtained for different numbers of k-points ($N$) used in ED and varying dielectric constant $\epsilon$. The black curve denotes the analytical results deduced from the GMP algebra for disk FQH. $\ell_0$ represents $\ell_B$ in FQH and $\ell_M$ in FCIs. (b) Three-dimensional diagram illustrating the magnetoroton dispersion. The local minimum resides close to the $M$ points. (c) The projected structure factor $\bar{S}(\boldsymbol{q})$ at $N = 18, \epsilon = 5$. $\bar{S}(\boldsymbol{q})$ scales as $(q\ell_M)^4$ in the long-wavelength limit. High symmetry points are labeled in the figure. The inset is the linear fitting of $\bar{S}^{0.25}(q)$ versus $q$.}
    \label{fig:magentorotons_idealFCI}
\end{figure*}

In this Letter, we provide a comprehensive study of the magnetorotons in FCIs hosted by the twisted $\mathrm{MoTe}_2$. We evaluate the magnetoroton dispersion by generalizing the GMP ansatz to FCIs, and find the universal existence of magnetorotons at finite momentum. We then analyze the dynamical geometric response of magnetorotons in the long-wavelength limit.
In ideal FCIs, the geometric excitations process definite chirality near the zero-momentum magnetoroton energy. Those geometric excitations originate from the momentum-space incompressibility of FCI fluid and can be interpreted as multipolar quantum fluctuations on the uniform charge density background in momentum space. The geometric excitations in FQH and FCI fluid can be unified into the incompressibility of the space-momentum phase space. We find new indicators of FCI stability from the perspective of excitations.
% When the FCI transitions into a charge density wave (CDW) phase, the magnetoroton softens at specific commensurate momenta corresponding to the density wave. 
We extend the analysis to the geometric excitations in other incompressible phases and find the Berry curvature and quantum metric play decisive roles in the diverse behaviors of excitations. Finally, we give proposals for detecting these intraband neutral excitations, providing a pathway for experimental exploration.
% \begin{widetext}
% in revex 4.2, two-column figure is achieved with figure*-environment with [t] parameter

{\it Model.---}
Recent experiments~\cite{Cai2023,Park2023,LiTingxin2023PRX,JuLong2024,xie2021fractional,kang2024evidence,zeng2023thermodynamic,xie2405even} have observed prominent signatures of FCI in twisted homobilayer MoTe$_2$ around a twisted angle $\theta\sim 4^\circ$ and a fractional filling $\nu = -\frac{2}{3}$, with an estimated dielectric constant $\epsilon \sim 5-15$. We start with the continuum model of twisted bilayer $\mathrm{MoTe}_2$ \cite{wu2019topological} using the aforementioned parameters: $\mathcal{H} = \mathcal{H}_0+V$. Here $\mathcal{H}_0$ is the single particle moir\'e Hamiltonian as described in Sec. I of the Supplemental Material (SM). $V$ is the Coulomb interaction $V = \frac{1}{2 A} \sum_{\boldsymbol{q}} V(\boldsymbol{q}) :\hat{\bar{\rho}}(\boldsymbol{q}) \hat{\bar{\rho}}(-\boldsymbol{q}):$. $A$ denotes the area of the two-dimensional system, $\boldsymbol{q}$ is the momentum, $\quad V(\boldsymbol{q})  =\frac{2 \pi \tanh (q d)}{\epsilon\epsilon_0 q}$ is the dual-gate screened Coulomb interaction, 
$d$ is the gate-to-sample distance, and ``$::$'' means normal ordering. $\hat{\bar{\rho}}(\boldsymbol{q})=\sum_{\boldsymbol{k}} 
 \Lambda_{\boldsymbol{k},\boldsymbol{q}} \hat{c}_{\boldsymbol{k}+\boldsymbol{q}}^{\dagger} \hat{c}_{\boldsymbol{k}}$ is the density operator projected into the highest valence band that hosts the FCI state, where $\Lambda_{\boldsymbol{k},\boldsymbol{q}}= \left\langle u_{\boldsymbol{k}+\boldsymbol{q}} |u_{\boldsymbol{k}}\right\rangle $ is the form factor and $|u_{\boldsymbol{k}}\rangle$ is the periodic part of Bloch eigenstates of $\mathcal{H}_0$, and $\hat{c}_{\boldsymbol{k}}^\dagger$ is the corresponding creation operator. Starting from this continuum model, previous studies \cite{2021li,2021Devakul,abouelkomasan2020,2020Repellin,2021wilhelm,PhysRevLett.131.136501,PhysRevB.108.205144,wang2024fractional} have revealed that the topmost valence band is an ideally flat band with non-trivial Chern number $\mathcal{C} = 1$, demonstrating the emergence of FCI phases. We provide the detailed band structure and the exact diagonalization (ED) spectrum based on the continuum model, and further verify the existence of an FCI as its ground state (see Sec. I of the SM).

{\it Magnetorotons in ideal moir\'e FCIs.---}
To investigate the magnetorotons in FCI, the GMP ansatz in FQH with the SMA~\cite{PhysRevB.33.2481} is used, 
and a density-wave-like excitation of FCIs is assumed as $|\psi_{\boldsymbol{q}}\rangle = \hat{\bar{\rho}}(\boldsymbol{q})|\psi_0\rangle$, $|\psi_0\rangle$ is the many-body ground state from the ED, and $\hat{\bar{\rho}}(\boldsymbol{q})$ is the projected density operator. The GMP ansatz is able to properly describe the low-lying intraband neutral excitations in FCI on the following aspects: (i) $|\psi_{\boldsymbol{q}}\rangle$ is orthogonal to ground state $|\psi_0\rangle$ when $\boldsymbol{q}\neq 0$ and thus is truly an excited state. (ii) The density operator projected to the topmost valence band characterizes the low energy intraband excitons. (iii) The ansatz is build from the ground state of FCI, inheriting the topological and geometric nature of the FCI. The variational expectation energy of $|\psi_{\boldsymbol{q}}\rangle$ is $E_{\boldsymbol{q}} = \frac{\langle\psi_{\boldsymbol{q}}|H|\psi_{\boldsymbol{q}}\rangle}{\langle\psi_{\boldsymbol{q}}|\psi_{\boldsymbol{q}}\rangle}-E_0$, where the ground state energy $E_0$ is chosen as the energy reference.

%We focus on the ground state with the lowest energy in the $\Gamma$ point to erase ambiguity among three-fold degeneracy in the ground states. (See SM Sec. II and Sec. IV for numerical details and relations with other trial wavefunction.) 

% \blue{why it works? zero-overlap with the ground state, intraband operators, neutral excitations}

The energy dispersions of magnetorotons in FCIs calculated by using different numbers $N$ of k points and varying dielectric constant $\epsilon$ are shown in Fig.~\ref{fig:magentorotons_idealFCI}(b). After rescaling, we observe that the dispersions are full-range gapped and there exists a universal local minimum of the dispersion at a mediate value of momentum $q_{\text{min}}\sim 1.3/\ell_M$, $\ell_M\equiv \sqrt{\sqrt{3}/4\pi}a_M\sim 2\text{nm}$ ($a_M \approx a_0/\theta$ is the length of a moir\'e unit cell) is the characteristic length analogous to the magnetic length $\ell_B(\ell_B = \sqrt{{\hbar}/{e B}} \sim 20 \mathrm{~nm})$ in FQH. 
This matches with the unclosed GMP-like algebra at the long-wavelength limit in FCI, $[\hat{\bar{\rho}}(\boldsymbol{q}), \hat{\bar{\rho}}(\boldsymbol{q}^\prime)] \approx i\Omega \boldsymbol{q}\times \boldsymbol{q}^\prime\hat{\bar{\rho}}(\boldsymbol{q}+\boldsymbol{q}^\prime)+\mathcal{O}(q^3), \Omega \approx 2\pi/S_{\text{BZ}} = \ell_M^2$ is the flatband Berry curvature and $S_{\text{BZ}}$ is the area of 1st Brillouin zone (BZ). The minimum resides between the $M$ point and the $K$ point in the first BZ, showing a magnetoroton gap $E(q = q_{\text{min}}) \sim 0.08e^2/(4\pi \epsilon\ell_M)\approx \Delta_{\text{mb}}$, where $\Delta_{\text{mb}}$ is the many-body gap or the energy of the first excited state in ED spectrum. At $q = q_{\text{min}}$ the overlap between the GMP wavefunction and the exact first excitation state reaches $98.22\%$ at $N = 18$, confirming the reliability of the GMP ansatz in studying FCIs. 
\begin{figure}[htp]
    \centering
\includegraphics[width = 0.49\textwidth]{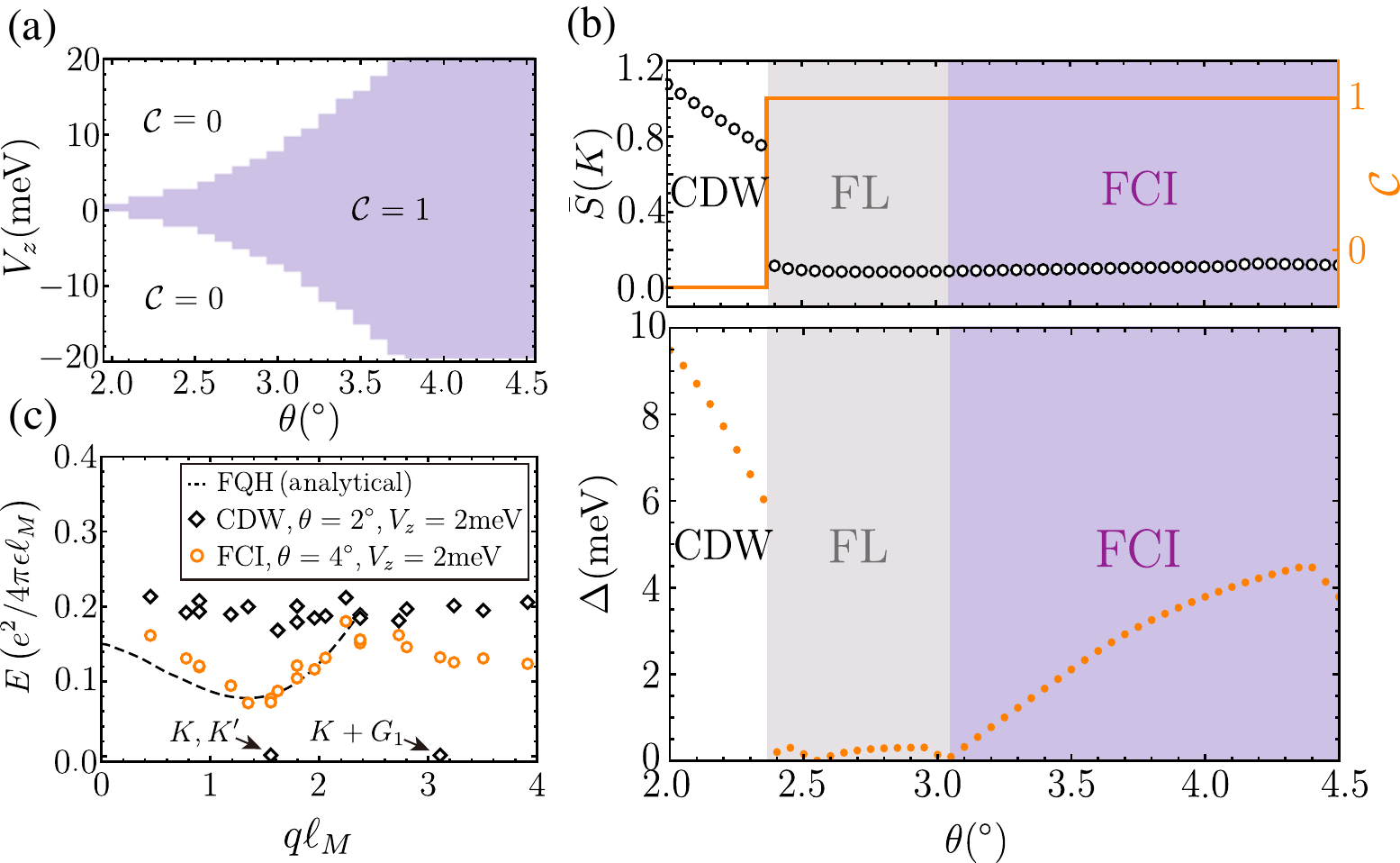}
\caption{(a) Topology of the topmost valence band  in twisted $\mathrm{MoTe}_2$ as a function of vertical displacement field $V_z$ and twisted angle $\theta$. (b) Phase structure indicators at small displacement field $V_z  = 2\text{meV}$ as a function of twisted angle $\theta$. In the upper panel of (b) the dots are the projected static structure factor $\bar{S}(q)$ at $K$ points that exhibit a peak at CDW phases and are featureless at gapless and FCI phases. The orange line is the Chern number of the topmost valence band. In the bottom panel of (b) the solid orange dots are the many-body gap $\Delta$ computed by ED. (c) Variational energy from the SMA for $V_z = 2\text{meV}$ with $\theta = 2^\circ$ (FCI) and $\theta = 4^\circ$ (CDW). At the CDW phase, the magnetoroton is softened at commensurate momenta. }
    \label{fig:fci2cdw}
\end{figure} 
The extrapolation to zero-momentum gives a gap $E(q = 0)\sim 0.17e^2/(4\pi \epsilon\ell_M) \approx 2.125\Delta_{\text{mb}}$. The zero momentum gap remains robust for various parameters, interaction strengths, and finite-size effects. At small $q$, we also extract the projected static structure factor
$\bar{S}(q) \equiv \langle\psi_{\boldsymbol{q}}|\psi_{\boldsymbol{q}}\rangle$. At the long wavelength limit, the projected form factor $\bar{S}(q)$ vanishes as $(q\ell_M)^4$ at $q\to 0$ limit, indicating the intraband excitations at long-wavelength limit have the quadruple structure.

\textit{Magnetorotons in FCI-CDW transition}.--- Charge density wave (CDW) is one of FCI's major competitors in moir\'e systems~\cite{xu2024maximally,sharma2024topological,reddy2023fractional,shen2024stabilizingfractionalcherninsulators,lu2024vestigialgaplessbosondensity}. In the twisted $\mathrm{MoTe}_2$ system, a topological transition from $\mathcal{C} = 1$ to $\mathcal{C} = 0$ can be induced by either tuning the twisted angle $\theta$ or the vertical displacement field $V_z$, inducing the possibility of FCI-CDW phase transition \cite{yu2024fractional,Lu_2024,lu2024interactiondrivenrotoncondensationc,waters2024interplay,PhysRevB.15.1959,tan2024parentberrycurvatureideal,dong2024anomaloushallcrystalsrhombohedral,PhysRevLett.133.206502,lu2024fractionalquantumanomaloushall,PhysRevB.110.205130}. It is intriguing to investigate the GMP ansatz after FCI-CDW phase transition. 

%At smaller twist angle, the gapped phase emerges with $\mathcal{C} = 0$, with the projected structure factor $\bar{S}(q)$ strongly peaks at $K, K^\prime$, signaling a topological trivial charge density wave. As the twisted angle increases, a topological transition emerges, endowing the Chern number $\mathcal{C} = 1$ to the valence band. In the vicinity of topological transition, both the band gap and the many-body gap are small and the system is in gapless or metal phases, where the strong band-mixing effect occurs. At the large twisted angle, an FCI phase emerges in a topological flatband (See SM Sec.I for the many-body spectrum).
A typical phase transition induced by changing the twisted angle from $4^\circ$ to $2^\circ$ at a fixed displacement field $V_z = 2~\text{meV}$ is shown in Fig.~\ref{fig:fci2cdw}(d). At $\theta = 4^\circ$, the system is in the FCI phase hosted by a $\mathcal{C} = 1$ band, and the dispersion exhibits a local minimum at finite $\bm{q}$. From $\theta = 4^\circ$ to $\theta = 2^\circ$, in the vicinity of gap closing, the spectrum is gapless and the ground state is Fermi liquid (FL). At $\theta = 2^\circ$, the system is in the $\mathcal{C} = 0$ CDW phase with a three-fold ground state $|0\rangle = |\psi_0\rangle,|\psi_{K}\rangle,|\psi_{K^\prime}\rangle$, and the dispersion of magnetorotons is nearly flat and features a sizable gap, except at the commensurate momenta $K$ and $K^\prime+n\boldsymbol{G}$ where the neutral excitation energy is nearly zero. This means that in the CDW phase, the excitations become soft at those momenta, as expected from the fact that the static form factor $\bar{S}(q)$ exhibits peaks at those momenta and thus leads to small $E_{\boldsymbol{q}}$. The overlap between the GMP wavefunction $\hat{\bar{\rho}}(K)|\psi_0\rangle$ and the ED ground state $|\psi_K\rangle$ at $K$ reaches 98.98\%, indicating that in CDW phase, the GMP ansatz $\hat{\bar{\rho}}(K)|\psi_0\rangle = |\psi_K\rangle$ is the charge density wave with wavevector $K$.

{\it Geometric excitations in FCIs and their proximate phases.---} 
So far, the property of long-wavelength magnetorotons in FCI remains elusive. In the following, we demonstrate that the excitations come from the momentum-space incompressibility of the FCI fluid. 

We start with the single-particle 4D phase space $\text{x} = (x_1,x_2,k_1,k_2)$ of a 2D system. The canonical transformation $f$ preserving commutator $[x_i,k_j] = \delta_{ij}, i,j = 1,2$ in general is expressed to be $f_i: x_i^\prime\to x_i+\nabla_{k_i}h(\text{x}),k_i^\prime\to k_i-\nabla_{x_i}h(\text{x}),i = 1,2$, $h(\text{x})$ is the differentiable function of $\text{x}$. Mapping $f_i$ also preserves the volume element $\mathrm{d}x_i\wedge \mathrm{d}k_i$ according to Liouville's theorem.

% The incompressibility of the fluid naturally means in the space-momentum phase space $\textbf{x} = (\boldsymbol{x},\boldsymbol{k})$, the volume element $\text{Vol}_{\textbf{x}} = \mathrm{d}x\wedge \mathrm{d}k$ is invariant under mapping $f$, $\text{Vol}_{\textbf{x}} = \text{Vol}_{f(\textbf{x})} $. Such mapping, termed volume-preserving diffeomorphism (VPD), is described by  $f: {\textbf{x}^\prime}^{\mu}\to \textbf{x}^\mu+\varepsilon^{\mu\nu}\partial_{\nu}h(\textbf{x})$, where $h(\textbf{x})$ is a differentiable
% function of $\textbf{x}$. The canonical mapping $f$ preserves the canonical commutator $[\hat{\boldsymbol{k}}^\prime,\hat{\boldsymbol{x}}^\prime] = 1$ and forms a VPD Lie group in 4D phase space. 

In FQH, due to LLL band projection, the momentum is related to position via $k_z\to \bar{z}, k_{\bar{z}}\to z$. The 4D phase space is then projected to the 2D real space \footnote{One can also project the phase space into the momentum space in FQH and consider the LLL in the momentum space.}. $f$ thus keeps the volume element $\mathrm{d}x_1\wedge \mathrm{d}k_1\to\mathrm{d}x_1\wedge \mathrm{d}x_2$ invariant, rendering real-space incompressibility.  $f$ forms a Lie group consisting volume-preserving diffeomorphism (VPD) in 2D, and the Lie algebra is the classical $w_\infty$ algebra with generators $\hat{\mathcal{L}}_{m}^\ell \sim z^{ m+1} \bar{z}^{\ell+1},m,\ell\ge -1$,$[z,\bar{z}] =1$. The generators generate the geometric excitations $\hat{\mathcal{L}}_m^\ell|0\rangle$ with higher angular momentum $l$, for example $l=2$ mode generated by $\hat{\mathcal{L}}_{1}^{-1},\hat{\mathcal{L}}_{-1}^1$  \cite{haldane2011selfdualitylongwavelengthbehaviorlandaulevel}.

In ideal FCI with LLL-like flatband \footnote{In this work, the ideal FCI refers to the FCI hosted in a perfect flatband with topology and geometry identical with LLL.}, via band projection 4D phase space is reduced to the 2D momentum space and recover the momentum-space volume-perserving symmetry. The geometric excitations from the momentum-space incompressibility of FCI then behave like multipolar charge distribution fluctuation $\hat{n}_{\boldsymbol{k}}-\langle \hat{n}_{\boldsymbol{k}} \rangle$ on the uniform background in the momentum space. 

For the systems that are not residing on the LLL-like flatband, the band projection does not relate the momentum with the position, henceforth the phase space volume-preserving symmetry can not reduce to real (or momentum) space volume-preserving symmetry, rendering the breakdown of $w_\infty$ algebra (or GMP algebra). Henceforth geometric excitations fundamentally come from the volume-preserving symmetry of enlarged phase space and can only be reduced to 2D subspace in LLL physics (See SM Sec. IV for details).

To investigate the energy of the excitations, we study the change of energy (band-projected Hamiltonian) after distortion $f$,  $\hat{\mathcal{O}}:= \delta_f H\approx \delta_f V $. Generally in a Bloch band, the Berry connection and quantum metric encoded in the band-projected interaction $V(A,g)$ are different from LLL-like flatband and fluctuate in $k$ space, giving rise to distinct behaviors of geometric excitations in FCI and other incompressible phases.

To be concrete, we consider the distortion $f$ on an LLL-like flatband. In the monomial basis, the parametrized function is expanded as $h(\text{x}) = \sum_{\ell,m}h_{\ell,m}k_1^{m+1}k_{2}^{\ell+1}$ to generate a superposition of higher angular momentum modes. We choose $h(\text{x})  = \epsilon (k_1\pm ik_2)^2$ to generate the angular momentum-2 mode with opposite chirality. The Berry connection and quantum metric then transform as $\delta_fA = 0,\delta_f g = \epsilon(\sigma_z\pm i\sigma_x)$. The resulting operator is $\hat{\mathcal{O}}^{}_{\pm} = \sum_{\boldsymbol{q}}\mathcal{D}_{\pm}(\boldsymbol{q})V(\boldsymbol{q}):\hat{\bar{\rho}}(-\boldsymbol{q})\hat{\bar{\rho}}(\boldsymbol{q}):,\mathcal{D}_{\pm}(\boldsymbol{q}) = \delta_f g_{\mu\nu} q^\mu q^\nu = (q_1\pm iq_2)^2$, which has been derived in \cite{geometric_yang} using anisotropy deformation.
In the twisted $\text{MoTe}_2$ system, we further incorporate the $C_{3v}$ lattice symmetry and the translation invariance \cite{wu2019topological}.
% The underlying form of angular momentum-2 operators with opposite chiralities is determined to be 
% $\hat{\mathcal{O}}^{}_{\pm} = \sum_{\boldsymbol{q}}\mathcal{D}_{\pm}(\boldsymbol{q})V(\boldsymbol{q}):\hat{\bar{\rho}}(-\boldsymbol{q})\hat{\bar{\rho}}(\boldsymbol{q}):$
% where 
The underlying form of $\mathcal{D}_\pm(\boldsymbol{q})$ are periodic functions in each BZ and behave as $(q_x\pm iq_y)^2$ in the long-wavelength limit in the $E$ representations of $C_{3v}$ \footnote{In practice, the truncation is set to be $3|\boldsymbol{G}|$ to ensure the convergence of $\boldsymbol{q}$ summation. In fact, due to the exponential decay of the form factor, the contribution outside the 1st BZ in summation of $\boldsymbol{q}$ is strongly suppressed, and hence one can directly apply the $(q_x\pm iq_y)^2$ and set the truncation of the summation up to the 1st BZ. }.
    \begin{figure}[t]
        \centering
        \includegraphics[width=0.9\linewidth]{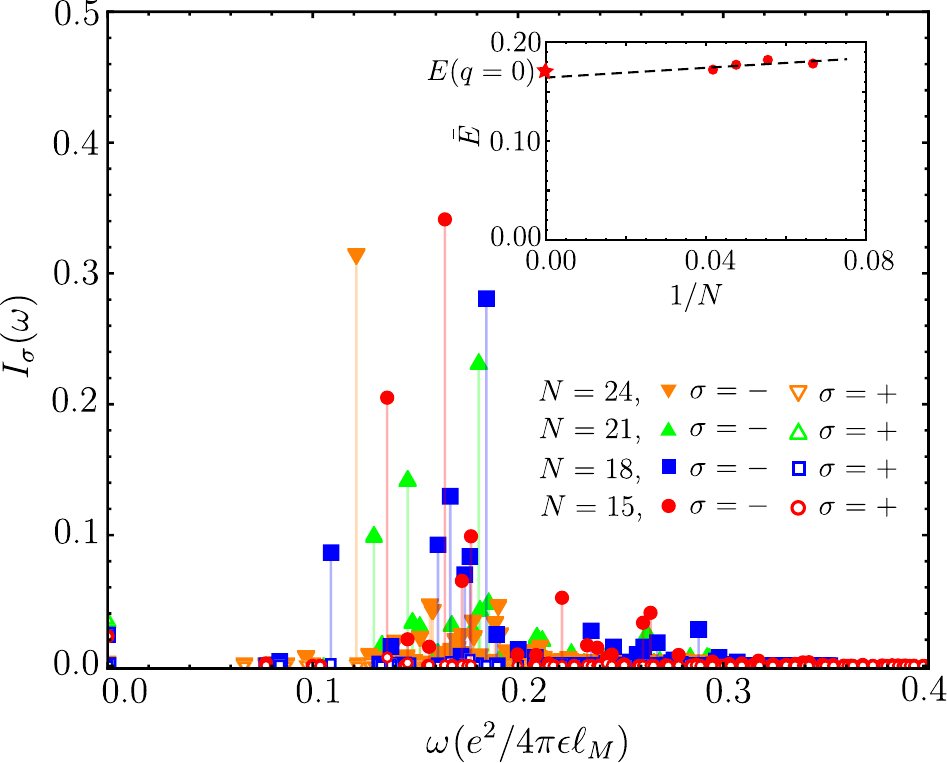}
        \caption{Spectral weight functions $I_{\sigma}(\omega)$ for the angular momentum-2 operators with opposite chiralities ($\sigma = +, -$) in FCI, computed by using different numbers of k-points ($N$), corresponding to different system sizes. The inset is the extrapolation of average energy $\bar{E} = \int \omega I(\omega)d\omega$ at the large system size limit. The red star $E(q = 0)$ is the extrapolation of zero momentum magnetoroton energy in the SMA.}
        \label{fig:chiralspin2}
    \end{figure}
We consider the spectral function $
    I^{}_{\pm}(E) = \sum_n|\langle\psi_n|\hat{\mathcal{O}}^{}_{\pm}|\psi_0\rangle|^2\delta(E_n-E) $ of operator $\hat{\mathcal{O}}_{\pm}^{}$,
$|\psi_0\rangle$ is the ground state and $|\psi_n\rangle$ is the eigenstates. To compare the spectral weight of different chiralities, $I_\pm(E)$ is normalized by $\langle\psi_{0}|\hat{\mathcal{O}}_-^\dagger\hat{\mathcal{O}}_-|\psi_{0}\rangle$. 

As shown in Fig.~\ref{fig:chiralspin2}, over $90\%$ weight of spectral function resides in the region of $0.10\sim 0.20e^2/(4\pi\epsilon\ell_M)$.  In the inset of Fig.~\ref{fig:chiralspin2}, the extrapolation to the large system size indicates an excitation energy $\bar{E}  = \int \mathrm{d}\omega \omega I(\omega)\approx 0.1642e^2/(4\pi\epsilon\ell_M)$, which is close to the long-wavelength magnetoroton energy $E(q = 0)\approx 0.165 e^2/(4\pi\epsilon\ell_M)$ derived from the GMP ansatz, suggesting the long-wavelength magnetoroton $\hat{\bar{\rho}}(\boldsymbol{q})|\psi_0\rangle,\boldsymbol{q}\to 0$ in Fig.~\ref{fig:magentorotons_idealFCI} can be interpreted as the geometric excitations in FCI. The results unequivocally show that the negative chiral excitations dominate while the positive chiral excitations are negligible, indicating that those excitations are featured with a chiral nature.

\begin{figure}[h]
    \centering
    \includegraphics[width=0.9\linewidth]{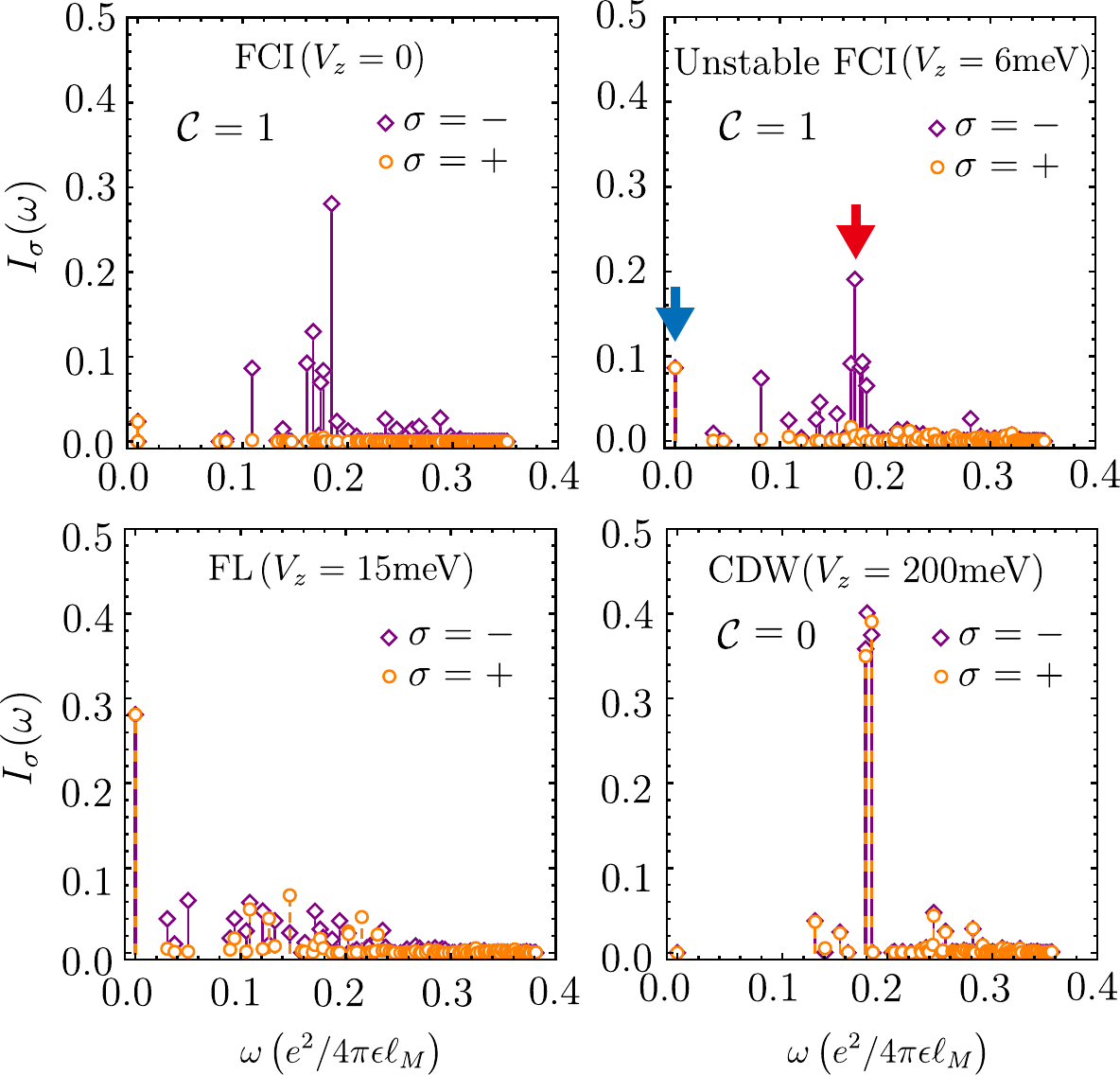}
    \caption{Spectral weight function of the angular momentum-2 operator for FCI ($V_z = 0$), unstable FCI ($V_z = 6 \text{meV}$), FL ($V_z = 15 \text{meV}$) and CDW ($ V_z = 200 \text{meV}$). $\sigma$ denotes the chirality of the excitations. For ideal FCI, the geometric excitations are chiral and gapped. The gapped chiral excitations (red arrow) are diminished for unstable FCI, and gapless signals with equal chirality (blue arrow) appear. The gapless modes dominate in FL. In CDW, the geometric excitations become gapped and nonchiral.}
    \label{fig:fci_cdw_graviton}
\end{figure}

By tuning the displacement field or twisted angle, when the low-energy band gets dispersive, FCI becomes less stable and the highly concentrated peaks in the spectral weight functions 
are broadened (Fig.~\ref{fig:fci_cdw_graviton}), signaling the reduced stability of geometric excitations. Meanwhile, a growing gapless nonchiral signal is observed. The competition of gapless and gapped modes can be considered the new indicators of FCI stability from the perspective of excitations. When the gap closes, the system enters into FL phases, which can still be considered as incompressible fluid in phase space according to Luttinger theorem \cite{PhysRev.119.1153,PhysRevResearch.4.033131}, the gapless signals dominate due to the vanishing renormalized interaction, signaling that the vanishing of geometric excitations in FL. 

At the flatband CDW phase, the spectrum exhibits a sharp peak near the many-body gap. Moreover, a prominent feature of the spectrum is that the opposite chiral branches become almost identical, suggesting that the long-wavelength excitations are nonchiral. A possible physical picture is that the electron residing at $k$ point locally experiences the Berry curvature $\Omega_{\boldsymbol{k}}$ as the effective magnetic field. Electrons that experience magnetic fields in opposite directions contribute to opposite chiralities. For a $\mathcal{C} = 0$ band, the Berry curvature can be somewhere positive and negative, and the numbers of electrons that experience the opposite magnetic field are nearly equal to each other when opposite chiral branches are equivalently involved in the geometric excitations. We verify it by calculating $\bar{B}= \sum_{\boldsymbol{k}}n_{\boldsymbol{k}}\Omega_{\boldsymbol{k}}/N$, where $n_{\boldsymbol{k}}$ is the charge density at $k$ point. We find that $\bar{B}\sim 10^{-3}$ in the CDW $(V_z = 200\text{meV})$. For comparison, $\bar{B}\sim 1$ in the FCI ($V_z = 0$) \footnote{We are thankful for Steven Simon for pointing out this.}. 

{\it Experiment realization.---}
To probe the signal of neutral excitations, it is feasible to apply the resonant inelastic light scattering (RILS) in the twisted $\mathrm{MoTe}_2$ or multilayer graphenes. RILS is able to detect the low-lying neutral excitations in FQH liquids \cite{liang2024evidence,Hirjibehedin_2005} and observe the interband excitons in the twisted $\mathrm{WSe}_2$~\cite{saigal2024collective}. It is feasible to identify magnetoroton modes via the recognition of peaks of the density of states at the predicted characteristic scale. It is also likely to find other co-existing excitations, such as magnons \cite{wang2024diversemagneticordersquantum}  in the spin-valley polarized states, or competing excitations which tend to destroy the magnetorotons such as interband excitations due to the prominent band mixing effects \cite{yu2023fractionalcherninsulatorsvs}. We will leave the study of these possible excitations to future work. To capture the features of excitations in the long-wavelength limit, the circularly polarized resonant inelastic light scattering (CP-RILS) \cite{liang2024evidence} is suggested. CP-RILS can effectively probe the angular momenta of the excitations and thus has the potential to detect the magnetorotons in the long-wavelength limit. 

% \hxd{Any justification for this (what is the meaning of ``spin'' here)? Note: any the physical excitations would manifest as the irreps of the crystalline symmetry --- this is true for both intra-Chern band excitation and the magnons. Therefore, even for graviton mode that is spin-2, it can still be hard to tell the difference from the rotation eigenvalue, which is directly related to the optical selection rules.}

{\it Discussion.---} In this work, we study the magnetorotons in the moir\'e FCIs, using the twisted $\mathrm{MoTe}_2$ as an example. We identify the universal existence of the magnetoroton and predict their characteristic energy and wavelength scale, which is instructive for future experiments. We elucidate the geometric nature of the long-wavelength magnetorotons in FCI and find it can be associated with incompressibility in momentum space. Utilizing the excellent tunability of the moir\'e system, we extend the discussion to the other incompressible phases, where we find such geometric excitations naturally come from the volume-preserving symmetry in phase space and reduce to position (momentum) subspace in FQH or ideal FCI. The behaviors of geometric excitations relate closely to the band geometry and topology. In a nutshell, the study provides a comprehensive study of the magnetorotons in the moir\'e FCI, making essential progress in understanding moir\'e FCI from the perspective of excitations, and evoking further interest in searching for the geometric excitations in other fractionally filled phases. 
% We therefore extend our analysis the the proximate incompressible phase, for instance, the chiral-mixing effect becomes significant in the angular momentum-2 long-wavelength excitations when the FCIs approach phase transitions. We further find evidence of nonchiral angular momentum-2 excitations within a topologically trivial band hosting CDW, showing that the geometric excitations can exist even without nontrivial topology.
% In conclusion, 
% utilizing the excellent tunability, the properties of moir\'e FCIs are expected to extend beyond the established physics of FQH. This tunability could be instrumental in developing a unifying theoretical framework for fractionally filled phases, thereby inspiring further experimental research in moir\'e systems.
% We conclude with some future directions of study. First of all, it would be interesting to study the neutral excitations in systems with bands with higher Chern numbers, generalized Jain's states, anomalous composite Fermi liquid, and non-abelian fractional quantum anomalous Hall systems. Second, the prominent band mixing rises as the realistic effects in FCIs and is likely to lead to the instability of intraband excitations and deserves further investigation. 

{\it Acknowledgements.---}
We thank the stimulating discussion with Lingjie Du, Yiyang Jiang, Steven Simon, Dam Thanh Son, Yuzhu Wang, Zhengzhi Wu, Bo Yang, Kun Yang and Zijian Zhou. This work was supported by the National Key Basic Research and Development Program of China (grant no. 2024YFA1409100), the Basic Science Center Project of NSFC (grant no. 52388201), the National Natural Science Foundation of China (grants no. 12334003, no. 12421004, and no. 12361141826), the National Key Basic Research and Development Program of China (grant no.2023YFA1406400), and the National Science Fund for Distinguished Young Scholars (grant no. 12025405). The calculations were performed at National Supercomputer Center in Tianjin using the Tianhe new generation supercomputer. 

\bibliography{reference.bib}

%apsrev4-2.bst 2019-01-14 (MD) hand-edited version of apsrev4-1.bst
%Control: key (0)
%Control: author (8) initials jnrlst
%Control: editor formatted (1) identically to author
%Control: production of article title (0) allowed
%Control: page (0) single
%Control: year (1) truncated
%Control: production of eprint (0) enabled
\begin{thebibliography}{84}%
\makeatletter
\providecommand \@ifxundefined [1]{%
 \@ifx{#1\undefined}
}%
\providecommand \@ifnum [1]{%
 \ifnum #1\expandafter \@firstoftwo
 \else \expandafter \@secondoftwo
 \fi
}%
\providecommand \@ifx [1]{%
 \ifx #1\expandafter \@firstoftwo
 \else \expandafter \@secondoftwo
 \fi
}%
\providecommand \natexlab [1]{#1}%
\providecommand \enquote  [1]{``#1''}%
\providecommand \bibnamefont  [1]{#1}%
\providecommand \bibfnamefont [1]{#1}%
\providecommand \citenamefont [1]{#1}%
\providecommand \href@noop [0]{\@secondoftwo}%
\providecommand \href [0]{\begingroup \@sanitize@url \@href}%
\providecommand \@href[1]{\@@startlink{#1}\@@href}%
\providecommand \@@href[1]{\endgroup#1\@@endlink}%
\providecommand \@sanitize@url [0]{\catcode `\\12\catcode `\$12\catcode
  `\&12\catcode `\#12\catcode `\^12\catcode `\_12\catcode `\%12\relax}%
\providecommand \@@startlink[1]{}%
\providecommand \@@endlink[0]{}%
\providecommand \url  [0]{\begingroup\@sanitize@url \@url }%
\providecommand \@url [1]{\endgroup\@href {#1}{\urlprefix }}%
\providecommand \urlprefix  [0]{URL }%
\providecommand \Eprint [0]{\href }%
\providecommand \doibase [0]{https://doi.org/}%
\providecommand \selectlanguage [0]{\@gobble}%
\providecommand \bibinfo  [0]{\@secondoftwo}%
\providecommand \bibfield  [0]{\@secondoftwo}%
\providecommand \translation [1]{[#1]}%
\providecommand \BibitemOpen [0]{}%
\providecommand \bibitemStop [0]{}%
\providecommand \bibitemNoStop [0]{.\EOS\space}%
\providecommand \EOS [0]{\spacefactor3000\relax}%
\providecommand \BibitemShut  [1]{\csname bibitem#1\endcsname}%
\let\auto@bib@innerbib\@empty
%</preamble>
\bibitem [{\citenamefont {Cai}\ \emph {et~al.}(2023)\citenamefont {Cai},
  \citenamefont {Anderson}, \citenamefont {Wang}, \citenamefont {Zhang},
  \citenamefont {Liu}, \citenamefont {Holtzmann}, \citenamefont {Zhang},
  \citenamefont {Fan}, \citenamefont {Taniguchi}, \citenamefont {Watanabe},
  \citenamefont {Ran}, \citenamefont {Cao}, \citenamefont {Fu}, \citenamefont
  {Xiao}, \citenamefont {Yao},\ and\ \citenamefont {Xu}}]{Cai2023}%
  \BibitemOpen
  \bibfield  {author} {\bibinfo {author} {\bibfnamefont {J.}~\bibnamefont
  {Cai}}, \bibinfo {author} {\bibfnamefont {E.}~\bibnamefont {Anderson}},
  \bibinfo {author} {\bibfnamefont {C.}~\bibnamefont {Wang}}, \bibinfo {author}
  {\bibfnamefont {X.}~\bibnamefont {Zhang}}, \bibinfo {author} {\bibfnamefont
  {X.}~\bibnamefont {Liu}}, \bibinfo {author} {\bibfnamefont {W.}~\bibnamefont
  {Holtzmann}}, \bibinfo {author} {\bibfnamefont {Y.}~\bibnamefont {Zhang}},
  \bibinfo {author} {\bibfnamefont {F.}~\bibnamefont {Fan}}, \bibinfo {author}
  {\bibfnamefont {T.}~\bibnamefont {Taniguchi}}, \bibinfo {author}
  {\bibfnamefont {K.}~\bibnamefont {Watanabe}}, \bibinfo {author}
  {\bibfnamefont {Y.}~\bibnamefont {Ran}}, \bibinfo {author} {\bibfnamefont
  {T.}~\bibnamefont {Cao}}, \bibinfo {author} {\bibfnamefont {L.}~\bibnamefont
  {Fu}}, \bibinfo {author} {\bibfnamefont {D.}~\bibnamefont {Xiao}}, \bibinfo
  {author} {\bibfnamefont {W.}~\bibnamefont {Yao}},\ and\ \bibinfo {author}
  {\bibfnamefont {X.}~\bibnamefont {Xu}},\ }\bibfield  {title} {\bibinfo
  {title} {Signatures of fractional quantum anomalous hall states in twisted
  mote2},\ }\href {https://doi.org/10.1038/s41586-023-06289-w} {\bibfield
  {journal} {\bibinfo  {journal} {Nature}\ }\textbf {\bibinfo {volume} {622}},\
  \bibinfo {pages} {63} (\bibinfo {year} {2023})}\BibitemShut {NoStop}%
\bibitem [{\citenamefont {Park}\ \emph {et~al.}(2023)\citenamefont {Park},
  \citenamefont {Cai}, \citenamefont {Anderson}, \citenamefont {Zhang},
  \citenamefont {Zhu}, \citenamefont {Liu}, \citenamefont {Wang}, \citenamefont
  {Holtzmann}, \citenamefont {Hu}, \citenamefont {Liu}, \citenamefont
  {Taniguchi}, \citenamefont {Watanabe}, \citenamefont {Chu}, \citenamefont
  {Cao}, \citenamefont {Fu}, \citenamefont {Yao}, \citenamefont {Chang},
  \citenamefont {Cobden}, \citenamefont {Xiao},\ and\ \citenamefont
  {Xu}}]{Park2023}%
  \BibitemOpen
  \bibfield  {author} {\bibinfo {author} {\bibfnamefont {H.}~\bibnamefont
  {Park}}, \bibinfo {author} {\bibfnamefont {J.}~\bibnamefont {Cai}}, \bibinfo
  {author} {\bibfnamefont {E.}~\bibnamefont {Anderson}}, \bibinfo {author}
  {\bibfnamefont {Y.}~\bibnamefont {Zhang}}, \bibinfo {author} {\bibfnamefont
  {J.}~\bibnamefont {Zhu}}, \bibinfo {author} {\bibfnamefont {X.}~\bibnamefont
  {Liu}}, \bibinfo {author} {\bibfnamefont {C.}~\bibnamefont {Wang}}, \bibinfo
  {author} {\bibfnamefont {W.}~\bibnamefont {Holtzmann}}, \bibinfo {author}
  {\bibfnamefont {C.}~\bibnamefont {Hu}}, \bibinfo {author} {\bibfnamefont
  {Z.}~\bibnamefont {Liu}}, \bibinfo {author} {\bibfnamefont {T.}~\bibnamefont
  {Taniguchi}}, \bibinfo {author} {\bibfnamefont {K.}~\bibnamefont {Watanabe}},
  \bibinfo {author} {\bibfnamefont {J.-H.}\ \bibnamefont {Chu}}, \bibinfo
  {author} {\bibfnamefont {T.}~\bibnamefont {Cao}}, \bibinfo {author}
  {\bibfnamefont {L.}~\bibnamefont {Fu}}, \bibinfo {author} {\bibfnamefont
  {W.}~\bibnamefont {Yao}}, \bibinfo {author} {\bibfnamefont {C.-Z.}\
  \bibnamefont {Chang}}, \bibinfo {author} {\bibfnamefont {D.}~\bibnamefont
  {Cobden}}, \bibinfo {author} {\bibfnamefont {D.}~\bibnamefont {Xiao}},\ and\
  \bibinfo {author} {\bibfnamefont {X.}~\bibnamefont {Xu}},\ }\bibfield
  {title} {\bibinfo {title} {Observation of fractionally quantized anomalous
  hall effect},\ }\href {https://doi.org/10.1038/s41586-023-06536-0} {\bibfield
   {journal} {\bibinfo  {journal} {Nature}\ }\textbf {\bibinfo {volume}
  {622}},\ \bibinfo {pages} {74} (\bibinfo {year} {2023})}\BibitemShut
  {NoStop}%
\bibitem [{\citenamefont {Xu}\ \emph {et~al.}(2023)\citenamefont {Xu},
  \citenamefont {Sun}, \citenamefont {Jia}, \citenamefont {Liu}, \citenamefont
  {Xu}, \citenamefont {Li}, \citenamefont {Gu}, \citenamefont {Watanabe},
  \citenamefont {Taniguchi}, \citenamefont {Tong}, \citenamefont {Jia},
  \citenamefont {Shi}, \citenamefont {Jiang}, \citenamefont {Zhang},
  \citenamefont {Liu},\ and\ \citenamefont {Li}}]{LiTingxin2023PRX}%
  \BibitemOpen
  \bibfield  {author} {\bibinfo {author} {\bibfnamefont {F.}~\bibnamefont
  {Xu}}, \bibinfo {author} {\bibfnamefont {Z.}~\bibnamefont {Sun}}, \bibinfo
  {author} {\bibfnamefont {T.}~\bibnamefont {Jia}}, \bibinfo {author}
  {\bibfnamefont {C.}~\bibnamefont {Liu}}, \bibinfo {author} {\bibfnamefont
  {C.}~\bibnamefont {Xu}}, \bibinfo {author} {\bibfnamefont {C.}~\bibnamefont
  {Li}}, \bibinfo {author} {\bibfnamefont {Y.}~\bibnamefont {Gu}}, \bibinfo
  {author} {\bibfnamefont {K.}~\bibnamefont {Watanabe}}, \bibinfo {author}
  {\bibfnamefont {T.}~\bibnamefont {Taniguchi}}, \bibinfo {author}
  {\bibfnamefont {B.}~\bibnamefont {Tong}}, \bibinfo {author} {\bibfnamefont
  {J.}~\bibnamefont {Jia}}, \bibinfo {author} {\bibfnamefont {Z.}~\bibnamefont
  {Shi}}, \bibinfo {author} {\bibfnamefont {S.}~\bibnamefont {Jiang}}, \bibinfo
  {author} {\bibfnamefont {Y.}~\bibnamefont {Zhang}}, \bibinfo {author}
  {\bibfnamefont {X.}~\bibnamefont {Liu}},\ and\ \bibinfo {author}
  {\bibfnamefont {T.}~\bibnamefont {Li}},\ }\bibfield  {title} {\bibinfo
  {title} {Observation of integer and fractional quantum anomalous hall effects
  in twisted bilayer ${\mathrm{mote}}_{2}$},\ }\href
  {https://doi.org/10.1103/PhysRevX.13.031037} {\bibfield  {journal} {\bibinfo
  {journal} {Phys. Rev. X}\ }\textbf {\bibinfo {volume} {13}},\ \bibinfo
  {pages} {031037} (\bibinfo {year} {2023})}\BibitemShut {NoStop}%
\bibitem [{\citenamefont {Lu}\ \emph {et~al.}(2024{\natexlab{a}})\citenamefont
  {Lu}, \citenamefont {Han}, \citenamefont {Yao}, \citenamefont {Reddy},
  \citenamefont {Yang}, \citenamefont {Seo}, \citenamefont {Watanabe},
  \citenamefont {Taniguchi}, \citenamefont {Fu},\ and\ \citenamefont
  {Ju}}]{JuLong2024}%
  \BibitemOpen
  \bibfield  {author} {\bibinfo {author} {\bibfnamefont {Z.}~\bibnamefont
  {Lu}}, \bibinfo {author} {\bibfnamefont {T.}~\bibnamefont {Han}}, \bibinfo
  {author} {\bibfnamefont {Y.}~\bibnamefont {Yao}}, \bibinfo {author}
  {\bibfnamefont {A.~P.}\ \bibnamefont {Reddy}}, \bibinfo {author}
  {\bibfnamefont {J.}~\bibnamefont {Yang}}, \bibinfo {author} {\bibfnamefont
  {J.}~\bibnamefont {Seo}}, \bibinfo {author} {\bibfnamefont {K.}~\bibnamefont
  {Watanabe}}, \bibinfo {author} {\bibfnamefont {T.}~\bibnamefont {Taniguchi}},
  \bibinfo {author} {\bibfnamefont {L.}~\bibnamefont {Fu}},\ and\ \bibinfo
  {author} {\bibfnamefont {L.}~\bibnamefont {Ju}},\ }\bibfield  {title}
  {\bibinfo {title} {Fractional quantum anomalous hall effect in multilayer
  graphene},\ }\href {https://doi.org/10.1038/s41586-023-07010-7} {\bibfield
  {journal} {\bibinfo  {journal} {Nature}\ }\textbf {\bibinfo {volume} {626}},\
  \bibinfo {pages} {759} (\bibinfo {year} {2024}{\natexlab{a}})}\BibitemShut
  {NoStop}%
\bibitem [{\citenamefont {Xie}\ \emph {et~al.}(2021)\citenamefont {Xie},
  \citenamefont {Pierce}, \citenamefont {Park}, \citenamefont {Parker},
  \citenamefont {Khalaf}, \citenamefont {Ledwith}, \citenamefont {Cao},
  \citenamefont {Lee}, \citenamefont {Chen}, \citenamefont {Forrester} \emph
  {et~al.}}]{xie2021fractional}%
  \BibitemOpen
  \bibfield  {author} {\bibinfo {author} {\bibfnamefont {Y.}~\bibnamefont
  {Xie}}, \bibinfo {author} {\bibfnamefont {A.~T.}\ \bibnamefont {Pierce}},
  \bibinfo {author} {\bibfnamefont {J.~M.}\ \bibnamefont {Park}}, \bibinfo
  {author} {\bibfnamefont {D.~E.}\ \bibnamefont {Parker}}, \bibinfo {author}
  {\bibfnamefont {E.}~\bibnamefont {Khalaf}}, \bibinfo {author} {\bibfnamefont
  {P.}~\bibnamefont {Ledwith}}, \bibinfo {author} {\bibfnamefont
  {Y.}~\bibnamefont {Cao}}, \bibinfo {author} {\bibfnamefont {S.~H.}\
  \bibnamefont {Lee}}, \bibinfo {author} {\bibfnamefont {S.}~\bibnamefont
  {Chen}}, \bibinfo {author} {\bibfnamefont {P.~R.}\ \bibnamefont {Forrester}},
  \emph {et~al.},\ }\bibfield  {title} {\bibinfo {title} {Fractional chern
  insulators in magic-angle twisted bilayer graphene},\ }\href
  {https://doi.org/10.1038/s41586-021-04002-3} {\bibfield  {journal} {\bibinfo
  {journal} {Nature}\ }\textbf {\bibinfo {volume} {600}},\ \bibinfo {pages}
  {439} (\bibinfo {year} {2021})}\BibitemShut {NoStop}%
\bibitem [{\citenamefont {Kang}\ \emph {et~al.}(2024)\citenamefont {Kang},
  \citenamefont {Shen}, \citenamefont {Qiu}, \citenamefont {Zeng},
  \citenamefont {Xia}, \citenamefont {Watanabe}, \citenamefont {Taniguchi},
  \citenamefont {Shan},\ and\ \citenamefont {Mak}}]{kang2024evidence}%
  \BibitemOpen
  \bibfield  {author} {\bibinfo {author} {\bibfnamefont {K.}~\bibnamefont
  {Kang}}, \bibinfo {author} {\bibfnamefont {B.}~\bibnamefont {Shen}}, \bibinfo
  {author} {\bibfnamefont {Y.}~\bibnamefont {Qiu}}, \bibinfo {author}
  {\bibfnamefont {Y.}~\bibnamefont {Zeng}}, \bibinfo {author} {\bibfnamefont
  {Z.}~\bibnamefont {Xia}}, \bibinfo {author} {\bibfnamefont {K.}~\bibnamefont
  {Watanabe}}, \bibinfo {author} {\bibfnamefont {T.}~\bibnamefont {Taniguchi}},
  \bibinfo {author} {\bibfnamefont {J.}~\bibnamefont {Shan}},\ and\ \bibinfo
  {author} {\bibfnamefont {K.~F.}\ \bibnamefont {Mak}},\ }\bibfield  {title}
  {\bibinfo {title} {Evidence of the fractional quantum spin hall effect in
  moir{\'e} mote2},\ }\href {https://doi.org/10.1038/s41586-024-07214-5}
  {\bibfield  {journal} {\bibinfo  {journal} {Nature}\ }\textbf {\bibinfo
  {volume} {628}},\ \bibinfo {pages} {522} (\bibinfo {year}
  {2024})}\BibitemShut {NoStop}%
\bibitem [{\citenamefont {Zeng}\ \emph {et~al.}(2023)\citenamefont {Zeng},
  \citenamefont {Xia}, \citenamefont {Kang}, \citenamefont {Zhu}, \citenamefont
  {Kn{\"u}ppel}, \citenamefont {Vaswani}, \citenamefont {Watanabe},
  \citenamefont {Taniguchi}, \citenamefont {Mak},\ and\ \citenamefont
  {Shan}}]{zeng2023thermodynamic}%
  \BibitemOpen
  \bibfield  {author} {\bibinfo {author} {\bibfnamefont {Y.}~\bibnamefont
  {Zeng}}, \bibinfo {author} {\bibfnamefont {Z.}~\bibnamefont {Xia}}, \bibinfo
  {author} {\bibfnamefont {K.}~\bibnamefont {Kang}}, \bibinfo {author}
  {\bibfnamefont {J.}~\bibnamefont {Zhu}}, \bibinfo {author} {\bibfnamefont
  {P.}~\bibnamefont {Kn{\"u}ppel}}, \bibinfo {author} {\bibfnamefont
  {C.}~\bibnamefont {Vaswani}}, \bibinfo {author} {\bibfnamefont
  {K.}~\bibnamefont {Watanabe}}, \bibinfo {author} {\bibfnamefont
  {T.}~\bibnamefont {Taniguchi}}, \bibinfo {author} {\bibfnamefont {K.~F.}\
  \bibnamefont {Mak}},\ and\ \bibinfo {author} {\bibfnamefont {J.}~\bibnamefont
  {Shan}},\ }\bibfield  {title} {\bibinfo {title} {Thermodynamic evidence of
  fractional chern insulator in moir{\'e} mote2},\ }\href
  {https://doi.org/10.1038/s41586-023-06452-3} {\bibfield  {journal} {\bibinfo
  {journal} {Nature}\ }\textbf {\bibinfo {volume} {622}},\ \bibinfo {pages}
  {69} (\bibinfo {year} {2023})}\BibitemShut {NoStop}%
\bibitem [{\citenamefont {Xie}\ \emph {et~al.}(2024)\citenamefont {Xie},
  \citenamefont {Huo}, \citenamefont {Lu}, \citenamefont {Feng}, \citenamefont
  {Zhang}, \citenamefont {Wang}, \citenamefont {Yang}, \citenamefont
  {Watanabe}, \citenamefont {Taniguchi}, \citenamefont {Liu}, \citenamefont
  {Song}, \citenamefont {Xie}, \citenamefont {Liu},\ and\ \citenamefont
  {Lu}}]{xie2405even}%
  \BibitemOpen
  \bibfield  {author} {\bibinfo {author} {\bibfnamefont {J.}~\bibnamefont
  {Xie}}, \bibinfo {author} {\bibfnamefont {Z.}~\bibnamefont {Huo}}, \bibinfo
  {author} {\bibfnamefont {X.}~\bibnamefont {Lu}}, \bibinfo {author}
  {\bibfnamefont {Z.}~\bibnamefont {Feng}}, \bibinfo {author} {\bibfnamefont
  {Z.}~\bibnamefont {Zhang}}, \bibinfo {author} {\bibfnamefont
  {W.}~\bibnamefont {Wang}}, \bibinfo {author} {\bibfnamefont {Q.}~\bibnamefont
  {Yang}}, \bibinfo {author} {\bibfnamefont {K.}~\bibnamefont {Watanabe}},
  \bibinfo {author} {\bibfnamefont {T.}~\bibnamefont {Taniguchi}}, \bibinfo
  {author} {\bibfnamefont {K.}~\bibnamefont {Liu}}, \bibinfo {author}
  {\bibfnamefont {Z.}~\bibnamefont {Song}}, \bibinfo {author} {\bibfnamefont
  {X.~C.}\ \bibnamefont {Xie}}, \bibinfo {author} {\bibfnamefont
  {J.}~\bibnamefont {Liu}},\ and\ \bibinfo {author} {\bibfnamefont
  {X.}~\bibnamefont {Lu}},\ }\href@noop {} {\bibinfo {title} {Even- and
  odd-denominator fractional quantum anomalous hall effect in graphene moire
  superlattices}} (\bibinfo {year} {2024}),\ \Eprint
  {https://arxiv.org/abs/2405.16944} {arXiv:2405.16944} \BibitemShut {NoStop}%
\bibitem [{\citenamefont {Nayak}\ \emph {et~al.}(2008)\citenamefont {Nayak},
  \citenamefont {Simon}, \citenamefont {Stern}, \citenamefont {Freedman},\ and\
  \citenamefont {Das~Sarma}}]{nayak2008non}%
  \BibitemOpen
  \bibfield  {author} {\bibinfo {author} {\bibfnamefont {C.}~\bibnamefont
  {Nayak}}, \bibinfo {author} {\bibfnamefont {S.~H.}\ \bibnamefont {Simon}},
  \bibinfo {author} {\bibfnamefont {A.}~\bibnamefont {Stern}}, \bibinfo
  {author} {\bibfnamefont {M.}~\bibnamefont {Freedman}},\ and\ \bibinfo
  {author} {\bibfnamefont {S.}~\bibnamefont {Das~Sarma}},\ }\bibfield  {title}
  {\bibinfo {title} {Non-abelian anyons and topological quantum computation},\
  }\href {https://doi.org/10.1103/RevModPhys.80.1083} {\bibfield  {journal}
  {\bibinfo  {journal} {Rev. Mod. Phys.}\ }\textbf {\bibinfo {volume} {80}},\
  \bibinfo {pages} {1083} (\bibinfo {year} {2008})}\BibitemShut {NoStop}%
\bibitem [{\citenamefont {Kitaev}(2003)}]{kitaev2003fault}%
  \BibitemOpen
  \bibfield  {author} {\bibinfo {author} {\bibfnamefont {A.~Y.}\ \bibnamefont
  {Kitaev}},\ }\bibfield  {title} {\bibinfo {title} {Fault-tolerant quantum
  computation by anyons},\ }\href@noop {} {\bibfield  {journal} {\bibinfo
  {journal} {Annals of physics}\ }\textbf {\bibinfo {volume} {303}},\ \bibinfo
  {pages} {2} (\bibinfo {year} {2003})}\BibitemShut {NoStop}%
\bibitem [{\citenamefont {{Li}}\ \emph {et~al.}(2021)\citenamefont {{Li}},
  \citenamefont {{Kumar}}, \citenamefont {{Sun}},\ and\ \citenamefont
  {{Lin}}}]{2021li}%
  \BibitemOpen
  \bibfield  {author} {\bibinfo {author} {\bibfnamefont {H.}~\bibnamefont
  {{Li}}}, \bibinfo {author} {\bibfnamefont {U.}~\bibnamefont {{Kumar}}},
  \bibinfo {author} {\bibfnamefont {K.}~\bibnamefont {{Sun}}},\ and\ \bibinfo
  {author} {\bibfnamefont {S.-Z.}\ \bibnamefont {{Lin}}},\ }\bibfield  {title}
  {\bibinfo {title} {{Spontaneous fractional Chern insulators in transition
  metal dichalcogenide moir{\'e} superlattices}},\ }\href
  {https://doi.org/10.1103/PhysRevResearch.3.L032070} {\bibfield  {journal}
  {\bibinfo  {journal} {Phys. Rev. Res.}\ }\textbf {\bibinfo {volume} {3}},\
  \bibinfo {eid} {L032070} (\bibinfo {year} {2021})}\BibitemShut {NoStop}%
\bibitem [{\citenamefont {{Devakul}}\ \emph {et~al.}(2021)\citenamefont
  {{Devakul}}, \citenamefont {{Cr{\'e}pel}}, \citenamefont {{Zhang}},\ and\
  \citenamefont {{Fu}}}]{2021Devakul}%
  \BibitemOpen
  \bibfield  {author} {\bibinfo {author} {\bibfnamefont {T.}~\bibnamefont
  {{Devakul}}}, \bibinfo {author} {\bibfnamefont {V.}~\bibnamefont
  {{Cr{\'e}pel}}}, \bibinfo {author} {\bibfnamefont {Y.}~\bibnamefont
  {{Zhang}}},\ and\ \bibinfo {author} {\bibfnamefont {L.}~\bibnamefont
  {{Fu}}},\ }\bibfield  {title} {\bibinfo {title} {{Magic in twisted transition
  metal dichalcogenide bilayers}},\ }\href
  {https://doi.org/10.1038/s41467-021-27042-9} {\bibfield  {journal} {\bibinfo
  {journal} {Nat. Commun.}\ }\textbf {\bibinfo {volume} {12}},\ \bibinfo {eid}
  {6730} (\bibinfo {year} {2021})}\BibitemShut {NoStop}%
\bibitem [{\citenamefont {Abouelkomsan}\ \emph {et~al.}(2020)\citenamefont
  {Abouelkomsan}, \citenamefont {Liu},\ and\ \citenamefont
  {Bergholtz}}]{abouelkomasan2020}%
  \BibitemOpen
  \bibfield  {author} {\bibinfo {author} {\bibfnamefont {A.}~\bibnamefont
  {Abouelkomsan}}, \bibinfo {author} {\bibfnamefont {Z.}~\bibnamefont {Liu}},\
  and\ \bibinfo {author} {\bibfnamefont {E.~J.}\ \bibnamefont {Bergholtz}},\
  }\bibfield  {title} {\bibinfo {title} {Particle-hole duality, emergent fermi
  liquids, and fractional chern insulators in moir\'e flatbands},\ }\href
  {https://doi.org/10.1103/PhysRevLett.124.106803} {\bibfield  {journal}
  {\bibinfo  {journal} {Phys. Rev. Lett.}\ }\textbf {\bibinfo {volume} {124}},\
  \bibinfo {pages} {106803} (\bibinfo {year} {2020})}\BibitemShut {NoStop}%
\bibitem [{\citenamefont {Repellin}\ and\ \citenamefont
  {Senthil}(2020)}]{2020Repellin}%
  \BibitemOpen
  \bibfield  {author} {\bibinfo {author} {\bibfnamefont {C.}~\bibnamefont
  {Repellin}}\ and\ \bibinfo {author} {\bibfnamefont {T.}~\bibnamefont
  {Senthil}},\ }\bibfield  {title} {\bibinfo {title} {Chern bands of twisted
  bilayer graphene: Fractional chern insulators and spin phase transition},\
  }\href {https://doi.org/10.1103/PhysRevResearch.2.023238} {\bibfield
  {journal} {\bibinfo  {journal} {Phys. Rev. Res.}\ }\textbf {\bibinfo {volume}
  {2}},\ \bibinfo {pages} {023238} (\bibinfo {year} {2020})}\BibitemShut
  {NoStop}%
\bibitem [{\citenamefont {Wilhelm}\ \emph {et~al.}(2021)\citenamefont
  {Wilhelm}, \citenamefont {Lang},\ and\ \citenamefont
  {L\"auchli}}]{2021wilhelm}%
  \BibitemOpen
  \bibfield  {author} {\bibinfo {author} {\bibfnamefont {P.}~\bibnamefont
  {Wilhelm}}, \bibinfo {author} {\bibfnamefont {T.~C.}\ \bibnamefont {Lang}},\
  and\ \bibinfo {author} {\bibfnamefont {A.~M.}\ \bibnamefont {L\"auchli}},\
  }\bibfield  {title} {\bibinfo {title} {Interplay of fractional chern
  insulator and charge density wave phases in twisted bilayer graphene},\
  }\href {https://doi.org/10.1103/PhysRevB.103.125406} {\bibfield  {journal}
  {\bibinfo  {journal} {Phys. Rev. B}\ }\textbf {\bibinfo {volume} {103}},\
  \bibinfo {pages} {125406} (\bibinfo {year} {2021})}\BibitemShut {NoStop}%
\bibitem [{\citenamefont {Goldman}\ \emph {et~al.}(2023)\citenamefont
  {Goldman}, \citenamefont {Reddy}, \citenamefont {Paul},\ and\ \citenamefont
  {Fu}}]{PhysRevLett.131.136501}%
  \BibitemOpen
  \bibfield  {author} {\bibinfo {author} {\bibfnamefont {H.}~\bibnamefont
  {Goldman}}, \bibinfo {author} {\bibfnamefont {A.~P.}\ \bibnamefont {Reddy}},
  \bibinfo {author} {\bibfnamefont {N.}~\bibnamefont {Paul}},\ and\ \bibinfo
  {author} {\bibfnamefont {L.}~\bibnamefont {Fu}},\ }\bibfield  {title}
  {\bibinfo {title} {Zero-field composite fermi liquid in twisted semiconductor
  bilayers},\ }\href {https://doi.org/10.1103/PhysRevLett.131.136501}
  {\bibfield  {journal} {\bibinfo  {journal} {Phys. Rev. Lett.}\ }\textbf
  {\bibinfo {volume} {131}},\ \bibinfo {pages} {136501} (\bibinfo {year}
  {2023})}\BibitemShut {NoStop}%
\bibitem [{\citenamefont {Ledwith}\ \emph {et~al.}(2023)\citenamefont
  {Ledwith}, \citenamefont {Vishwanath},\ and\ \citenamefont
  {Parker}}]{PhysRevB.108.205144}%
  \BibitemOpen
  \bibfield  {author} {\bibinfo {author} {\bibfnamefont {P.~J.}\ \bibnamefont
  {Ledwith}}, \bibinfo {author} {\bibfnamefont {A.}~\bibnamefont
  {Vishwanath}},\ and\ \bibinfo {author} {\bibfnamefont {D.~E.}\ \bibnamefont
  {Parker}},\ }\bibfield  {title} {\bibinfo {title} {Vortexability: A unifying
  criterion for ideal fractional chern insulators},\ }\href
  {https://doi.org/10.1103/PhysRevB.108.205144} {\bibfield  {journal} {\bibinfo
   {journal} {Phys. Rev. B}\ }\textbf {\bibinfo {volume} {108}},\ \bibinfo
  {pages} {205144} (\bibinfo {year} {2023})}\BibitemShut {NoStop}%
\bibitem [{\citenamefont {Wang}\ \emph
  {et~al.}(2024{\natexlab{a}})\citenamefont {Wang}, \citenamefont {Zhang},
  \citenamefont {Liu}, \citenamefont {He}, \citenamefont {Xu}, \citenamefont
  {Ran}, \citenamefont {Cao},\ and\ \citenamefont {Xiao}}]{wang2024fractional}%
  \BibitemOpen
  \bibfield  {author} {\bibinfo {author} {\bibfnamefont {C.}~\bibnamefont
  {Wang}}, \bibinfo {author} {\bibfnamefont {X.-W.}\ \bibnamefont {Zhang}},
  \bibinfo {author} {\bibfnamefont {X.}~\bibnamefont {Liu}}, \bibinfo {author}
  {\bibfnamefont {Y.}~\bibnamefont {He}}, \bibinfo {author} {\bibfnamefont
  {X.}~\bibnamefont {Xu}}, \bibinfo {author} {\bibfnamefont {Y.}~\bibnamefont
  {Ran}}, \bibinfo {author} {\bibfnamefont {T.}~\bibnamefont {Cao}},\ and\
  \bibinfo {author} {\bibfnamefont {D.}~\bibnamefont {Xiao}},\ }\bibfield
  {title} {\bibinfo {title} {Fractional chern insulator in twisted bilayer
  ${\mathrm{mote}}_{2}$},\ }\href
  {https://doi.org/10.1103/PhysRevLett.132.036501} {\bibfield  {journal}
  {\bibinfo  {journal} {Phys. Rev. Lett.}\ }\textbf {\bibinfo {volume} {132}},\
  \bibinfo {pages} {036501} (\bibinfo {year} {2024}{\natexlab{a}})}\BibitemShut
  {NoStop}%
\bibitem [{\citenamefont {Roy}(2014)}]{2014Roy}%
  \BibitemOpen
  \bibfield  {author} {\bibinfo {author} {\bibfnamefont {R.}~\bibnamefont
  {Roy}},\ }\bibfield  {title} {\bibinfo {title} {Band geometry of fractional
  topological insulators},\ }\href {https://doi.org/10.1103/PhysRevB.90.165139}
  {\bibfield  {journal} {\bibinfo  {journal} {Phys. Rev. B}\ }\textbf {\bibinfo
  {volume} {90}},\ \bibinfo {pages} {165139} (\bibinfo {year}
  {2014})}\BibitemShut {NoStop}%
\bibitem [{\citenamefont {Regnault}\ and\ \citenamefont
  {Bernevig}(2011)}]{regnault2011fractional}%
  \BibitemOpen
  \bibfield  {author} {\bibinfo {author} {\bibfnamefont {N.}~\bibnamefont
  {Regnault}}\ and\ \bibinfo {author} {\bibfnamefont {B.~A.}\ \bibnamefont
  {Bernevig}},\ }\bibfield  {title} {\bibinfo {title} {Fractional chern
  insulator},\ }\href {https://doi.org/10.1103/PhysRevX.1.021014} {\bibfield
  {journal} {\bibinfo  {journal} {Phys. Rev. X}\ }\textbf {\bibinfo {volume}
  {1}},\ \bibinfo {pages} {021014} (\bibinfo {year} {2011})}\BibitemShut
  {NoStop}%
\bibitem [{\citenamefont {Qi}(2011)}]{2011Qi}%
  \BibitemOpen
  \bibfield  {author} {\bibinfo {author} {\bibfnamefont {X.-L.}\ \bibnamefont
  {Qi}},\ }\bibfield  {title} {\bibinfo {title} {Generic wave-function
  description of fractional quantum anomalous hall states and fractional
  topological insulators},\ }\href
  {https://doi.org/10.1103/PhysRevLett.107.126803} {\bibfield  {journal}
  {\bibinfo  {journal} {Phys. Rev. Lett.}\ }\textbf {\bibinfo {volume} {107}},\
  \bibinfo {pages} {126803} (\bibinfo {year} {2011})}\BibitemShut {NoStop}%
\bibitem [{\citenamefont {Jackson}\ \emph {et~al.}(2015)\citenamefont
  {Jackson}, \citenamefont {Möller},\ and\ \citenamefont
  {Roy}}]{Jackson_2015}%
  \BibitemOpen
  \bibfield  {author} {\bibinfo {author} {\bibfnamefont {T.~S.}\ \bibnamefont
  {Jackson}}, \bibinfo {author} {\bibfnamefont {G.}~\bibnamefont {Möller}},\
  and\ \bibinfo {author} {\bibfnamefont {R.}~\bibnamefont {Roy}},\ }\bibfield
  {title} {\bibinfo {title} {Geometric stability of topological lattice
  phases},\ }\href {http://dx.doi.org/10.1038/ncomms9629} {\bibfield  {journal}
  {\bibinfo  {journal} {Nat. Commun.}\ }\textbf {\bibinfo {volume} {6}}
  (\bibinfo {year} {2015})}\BibitemShut {NoStop}%
\bibitem [{\citenamefont {Parameswaran}\ \emph {et~al.}(2013)\citenamefont
  {Parameswaran}, \citenamefont {Roy},\ and\ \citenamefont
  {Sondhi}}]{Parameswaran_2013}%
  \BibitemOpen
  \bibfield  {author} {\bibinfo {author} {\bibfnamefont {S.~A.}\ \bibnamefont
  {Parameswaran}}, \bibinfo {author} {\bibfnamefont {R.}~\bibnamefont {Roy}},\
  and\ \bibinfo {author} {\bibfnamefont {S.~L.}\ \bibnamefont {Sondhi}},\
  }\bibfield  {title} {\bibinfo {title} {Fractional quantum hall physics in
  topological flat bands},\ }\href {https://doi.org/10.1016/j.crhy.2013.04.003}
  {\bibfield  {journal} {\bibinfo  {journal} {Comptes Rendus. Physique}\
  }\textbf {\bibinfo {volume} {14}},\ \bibinfo {pages} {816–839} (\bibinfo
  {year} {2013})}\BibitemShut {NoStop}%
\bibitem [{\citenamefont {Wang}\ \emph {et~al.}(2021)\citenamefont {Wang},
  \citenamefont {Cano}, \citenamefont {Millis}, \citenamefont {Liu},\ and\
  \citenamefont {Yang}}]{2021wang}%
  \BibitemOpen
  \bibfield  {author} {\bibinfo {author} {\bibfnamefont {J.}~\bibnamefont
  {Wang}}, \bibinfo {author} {\bibfnamefont {J.}~\bibnamefont {Cano}}, \bibinfo
  {author} {\bibfnamefont {A.~J.}\ \bibnamefont {Millis}}, \bibinfo {author}
  {\bibfnamefont {Z.}~\bibnamefont {Liu}},\ and\ \bibinfo {author}
  {\bibfnamefont {B.}~\bibnamefont {Yang}},\ }\bibfield  {title} {\bibinfo
  {title} {Exact landau level description of geometry and interaction in a
  flatband},\ }\href {https://doi.org/10.1103/PhysRevLett.127.246403}
  {\bibfield  {journal} {\bibinfo  {journal} {Phys. Rev. Lett.}\ }\textbf
  {\bibinfo {volume} {127}},\ \bibinfo {pages} {246403} (\bibinfo {year}
  {2021})}\BibitemShut {NoStop}%
\bibitem [{\citenamefont {Wang}\ \emph
  {et~al.}(2024{\natexlab{b}})\citenamefont {Wang}, \citenamefont {Shen},
  \citenamefont {Guo}, \citenamefont {Wang}, \citenamefont {Duan},\ and\
  \citenamefont {Xu}}]{wang2024fractionalcherninsulatorsmoire}%
  \BibitemOpen
  \bibfield  {author} {\bibinfo {author} {\bibfnamefont {C.}~\bibnamefont
  {Wang}}, \bibinfo {author} {\bibfnamefont {X.}~\bibnamefont {Shen}}, \bibinfo
  {author} {\bibfnamefont {R.}~\bibnamefont {Guo}}, \bibinfo {author}
  {\bibfnamefont {C.}~\bibnamefont {Wang}}, \bibinfo {author} {\bibfnamefont
  {W.}~\bibnamefont {Duan}},\ and\ \bibinfo {author} {\bibfnamefont
  {Y.}~\bibnamefont {Xu}},\ }\href@noop {} {\bibinfo {title} {Fractional chern
  insulators in moir\'e flat bands with high chern numbers}} (\bibinfo {year}
  {2024}{\natexlab{b}}),\ \Eprint {https://arxiv.org/abs/2408.03305}
  {arXiv:2408.03305} \BibitemShut {NoStop}%
\bibitem [{\citenamefont {Reddy}\ \emph {et~al.}(2024)\citenamefont {Reddy},
  \citenamefont {Paul}, \citenamefont {Abouelkomsan},\ and\ \citenamefont
  {Fu}}]{PhysRevLett.133.166503}%
  \BibitemOpen
  \bibfield  {author} {\bibinfo {author} {\bibfnamefont {A.~P.}\ \bibnamefont
  {Reddy}}, \bibinfo {author} {\bibfnamefont {N.}~\bibnamefont {Paul}},
  \bibinfo {author} {\bibfnamefont {A.}~\bibnamefont {Abouelkomsan}},\ and\
  \bibinfo {author} {\bibfnamefont {L.}~\bibnamefont {Fu}},\ }\bibfield
  {title} {\bibinfo {title} {Non-abelian fractionalization in topological
  minibands},\ }\href {https://doi.org/10.1103/PhysRevLett.133.166503}
  {\bibfield  {journal} {\bibinfo  {journal} {Phys. Rev. Lett.}\ }\textbf
  {\bibinfo {volume} {133}},\ \bibinfo {pages} {166503} (\bibinfo {year}
  {2024})}\BibitemShut {NoStop}%
\bibitem [{\citenamefont {Ahn}\ \emph {et~al.}(2024)\citenamefont {Ahn},
  \citenamefont {Lee}, \citenamefont {Yananose}, \citenamefont {Kim},\ and\
  \citenamefont {Cho}}]{PhysRevB.110.L161109}%
  \BibitemOpen
  \bibfield  {author} {\bibinfo {author} {\bibfnamefont {C.-E.}\ \bibnamefont
  {Ahn}}, \bibinfo {author} {\bibfnamefont {W.}~\bibnamefont {Lee}}, \bibinfo
  {author} {\bibfnamefont {K.}~\bibnamefont {Yananose}}, \bibinfo {author}
  {\bibfnamefont {Y.}~\bibnamefont {Kim}},\ and\ \bibinfo {author}
  {\bibfnamefont {G.~Y.}\ \bibnamefont {Cho}},\ }\bibfield  {title} {\bibinfo
  {title} {Non-abelian fractional quantum anomalous hall states and first
  landau level physics of the second moir\'e band of twisted bilayer
  ${\mathrm{mote}}_{2}$},\ }\href
  {https://doi.org/10.1103/PhysRevB.110.L161109} {\bibfield  {journal}
  {\bibinfo  {journal} {Phys. Rev. B}\ }\textbf {\bibinfo {volume} {110}},\
  \bibinfo {pages} {L161109} (\bibinfo {year} {2024})}\BibitemShut {NoStop}%
\bibitem [{\citenamefont {Lu}\ \emph {et~al.}(2024{\natexlab{b}})\citenamefont
  {Lu}, \citenamefont {Chen}, \citenamefont {Wu}, \citenamefont {Sun},\ and\
  \citenamefont {Meng}}]{Lu_2024}%
  \BibitemOpen
  \bibfield  {author} {\bibinfo {author} {\bibfnamefont {H.}~\bibnamefont
  {Lu}}, \bibinfo {author} {\bibfnamefont {B.-B.}\ \bibnamefont {Chen}},
  \bibinfo {author} {\bibfnamefont {H.-Q.}\ \bibnamefont {Wu}}, \bibinfo
  {author} {\bibfnamefont {K.}~\bibnamefont {Sun}},\ and\ \bibinfo {author}
  {\bibfnamefont {Z.~Y.}\ \bibnamefont {Meng}},\ }\bibfield  {title} {\bibinfo
  {title} {Thermodynamic response and neutral excitations in integer and
  fractional quantum anomalous hall states emerging from correlated flat
  bands},\ }\href {http://dx.doi.org/10.1103/PhysRevLett.132.236502} {\bibfield
   {journal} {\bibinfo  {journal} {Phys. Rev. Lett.}\ }\textbf {\bibinfo
  {volume} {132}} (\bibinfo {year} {2024}{\natexlab{b}})}\BibitemShut {NoStop}%
\bibitem [{\citenamefont {Repellin}\ \emph {et~al.}(2014)\citenamefont
  {Repellin}, \citenamefont {Neupert}, \citenamefont {Papić},\ and\
  \citenamefont {Regnault}}]{Repellin_2014}%
  \BibitemOpen
  \bibfield  {author} {\bibinfo {author} {\bibfnamefont {C.}~\bibnamefont
  {Repellin}}, \bibinfo {author} {\bibfnamefont {T.}~\bibnamefont {Neupert}},
  \bibinfo {author} {\bibfnamefont {Z.}~\bibnamefont {Papić}},\ and\ \bibinfo
  {author} {\bibfnamefont {N.}~\bibnamefont {Regnault}},\ }\bibfield  {title}
  {\bibinfo {title} {Single-mode approximation for fractional chern insulators
  and the fractional quantum hall effect on the torus},\ }\href
  {https://link.aps.org/doi/10.1103/physrevb.90.045114} {\bibfield  {journal}
  {\bibinfo  {journal} {Phys. Rev. B}\ }\textbf {\bibinfo {volume} {90}}
  (\bibinfo {year} {2014})}\BibitemShut {NoStop}%
\bibitem [{\citenamefont {Hu}\ \emph {et~al.}(2024)\citenamefont {Hu},
  \citenamefont {Xiao},\ and\ \citenamefont {Ran}}]{PhysRevB.109.245125}%
  \BibitemOpen
  \bibfield  {author} {\bibinfo {author} {\bibfnamefont {X.}~\bibnamefont
  {Hu}}, \bibinfo {author} {\bibfnamefont {D.}~\bibnamefont {Xiao}},\ and\
  \bibinfo {author} {\bibfnamefont {Y.}~\bibnamefont {Ran}},\ }\bibfield
  {title} {\bibinfo {title} {Hyperdeterminants and composite fermion states in
  fractional chern insulators},\ }\href
  {https://doi.org/10.1103/PhysRevB.109.245125} {\bibfield  {journal} {\bibinfo
   {journal} {Phys. Rev. B}\ }\textbf {\bibinfo {volume} {109}},\ \bibinfo
  {pages} {245125} (\bibinfo {year} {2024})}\BibitemShut {NoStop}%
\bibitem [{\citenamefont {Girvin}\ \emph {et~al.}(1986)\citenamefont {Girvin},
  \citenamefont {MacDonald},\ and\ \citenamefont
  {Platzman}}]{PhysRevB.33.2481}%
  \BibitemOpen
  \bibfield  {author} {\bibinfo {author} {\bibfnamefont {S.~M.}\ \bibnamefont
  {Girvin}}, \bibinfo {author} {\bibfnamefont {A.~H.}\ \bibnamefont
  {MacDonald}},\ and\ \bibinfo {author} {\bibfnamefont {P.~M.}\ \bibnamefont
  {Platzman}},\ }\bibfield  {title} {\bibinfo {title} {Magneto-roton theory of
  collective excitations in the fractional quantum hall effect},\ }\href
  {https://doi.org/10.1103/PhysRevB.33.2481} {\bibfield  {journal} {\bibinfo
  {journal} {Phys. Rev. B}\ }\textbf {\bibinfo {volume} {33}},\ \bibinfo
  {pages} {2481} (\bibinfo {year} {1986})}\BibitemShut {NoStop}%
\bibitem [{\citenamefont {Kohn}(1961)}]{PhysRev.123.1242}%
  \BibitemOpen
  \bibfield  {author} {\bibinfo {author} {\bibfnamefont {W.}~\bibnamefont
  {Kohn}},\ }\bibfield  {title} {\bibinfo {title} {Cyclotron resonance and de
  haas-van alphen oscillations of an interacting electron gas},\ }\href
  {https://doi.org/10.1103/PhysRev.123.1242} {\bibfield  {journal} {\bibinfo
  {journal} {Phys. Rev.}\ }\textbf {\bibinfo {volume} {123}},\ \bibinfo {pages}
  {1242} (\bibinfo {year} {1961})}\BibitemShut {NoStop}%
\bibitem [{\citenamefont {Mukherjee}\ and\ \citenamefont
  {Mandal}(2015)}]{PhysRevLett.114.156802}%
  \BibitemOpen
  \bibfield  {author} {\bibinfo {author} {\bibfnamefont {S.}~\bibnamefont
  {Mukherjee}}\ and\ \bibinfo {author} {\bibfnamefont {S.~S.}\ \bibnamefont
  {Mandal}},\ }\bibfield  {title} {\bibinfo {title} {Anomalously low
  magnetoroton energies of the unconventional fractional quantum hall states of
  composite fermions},\ }\href {https://doi.org/10.1103/PhysRevLett.114.156802}
  {\bibfield  {journal} {\bibinfo  {journal} {Phys. Rev. Lett.}\ }\textbf
  {\bibinfo {volume} {114}},\ \bibinfo {pages} {156802} (\bibinfo {year}
  {2015})}\BibitemShut {NoStop}%
\bibitem [{\citenamefont {Wolf}\ \emph {et~al.}(2024)\citenamefont {Wolf},
  \citenamefont {Chao}, \citenamefont {MacDonald},\ and\ \citenamefont
  {Su}}]{wolf2024intrabandcollectiveexcitationsfractional}%
  \BibitemOpen
  \bibfield  {author} {\bibinfo {author} {\bibfnamefont {T.~M.~R.}\
  \bibnamefont {Wolf}}, \bibinfo {author} {\bibfnamefont {Y.-C.}\ \bibnamefont
  {Chao}}, \bibinfo {author} {\bibfnamefont {A.~H.}\ \bibnamefont
  {MacDonald}},\ and\ \bibinfo {author} {\bibfnamefont {J.~J.}\ \bibnamefont
  {Su}},\ }\href@noop {} {\bibinfo {title} {Intraband collective excitations in
  fractional chern insulators are dark}} (\bibinfo {year} {2024}),\ \Eprint
  {https://arxiv.org/abs/2406.10709} {arXiv:2406.10709} \BibitemShut {NoStop}%
\bibitem [{\citenamefont {Cavicchi}\ \emph {et~al.}(2024)\citenamefont
  {Cavicchi}, \citenamefont {Reijnders}, \citenamefont {Katsnelson},\ and\
  \citenamefont {Polini}}]{cavicchi2024opticalpropertiesplasmonsorbital}%
  \BibitemOpen
  \bibfield  {author} {\bibinfo {author} {\bibfnamefont {L.}~\bibnamefont
  {Cavicchi}}, \bibinfo {author} {\bibfnamefont {K.~J.~A.}\ \bibnamefont
  {Reijnders}}, \bibinfo {author} {\bibfnamefont {M.~I.}\ \bibnamefont
  {Katsnelson}},\ and\ \bibinfo {author} {\bibfnamefont {M.}~\bibnamefont
  {Polini}},\ }\href@noop {} {\bibinfo {title} {Optical properties, plasmons,
  and orbital skyrme textures in twisted tmds}} (\bibinfo {year} {2024}),\
  \Eprint {https://arxiv.org/abs/2410.18025} {arXiv:2410.18025} \BibitemShut
  {NoStop}%
\bibitem [{\citenamefont {Gromov}\ and\ \citenamefont
  {Son}(2017{\natexlab{a}})}]{Gromov:2017qeb}%
  \BibitemOpen
  \bibfield  {author} {\bibinfo {author} {\bibfnamefont {A.}~\bibnamefont
  {Gromov}}\ and\ \bibinfo {author} {\bibfnamefont {D.~T.}\ \bibnamefont
  {Son}},\ }\bibfield  {title} {\bibinfo {title} {{Bimetric Theory of
  Fractional Quantum Hall States}},\ }\href
  {https://doi.org/10.1103/PhysRevX.7.041032} {\bibfield  {journal} {\bibinfo
  {journal} {Phys. Rev. X}\ }\textbf {\bibinfo {volume} {7}},\ \bibinfo {pages}
  {041032} (\bibinfo {year} {2017}{\natexlab{a}})}\BibitemShut {NoStop}%
\bibitem [{\citenamefont {Haldane}\ \emph {et~al.}(2021)\citenamefont
  {Haldane}, \citenamefont {Rezayi},\ and\ \citenamefont
  {Yang}}]{Haldane_2021}%
  \BibitemOpen
  \bibfield  {author} {\bibinfo {author} {\bibfnamefont {F.~D.~M.}\
  \bibnamefont {Haldane}}, \bibinfo {author} {\bibfnamefont {E.~H.}\
  \bibnamefont {Rezayi}},\ and\ \bibinfo {author} {\bibfnamefont
  {K.}~\bibnamefont {Yang}},\ }\bibfield  {title} {\bibinfo {title} {Graviton
  chirality and topological order in the half-filled landau level},\ }\href
  {https://link.aps.org/doi/10.1103/physrevb.104.l121106} {\bibfield  {journal}
  {\bibinfo  {journal} {Phys. Rev. B}\ }\textbf {\bibinfo {volume} {104}}
  (\bibinfo {year} {2021})}\BibitemShut {NoStop}%
\bibitem [{\citenamefont {Son}(2015)}]{PhysRevX.5.031027}%
  \BibitemOpen
  \bibfield  {author} {\bibinfo {author} {\bibfnamefont {D.~T.}\ \bibnamefont
  {Son}},\ }\bibfield  {title} {\bibinfo {title} {Is the composite fermion a
  dirac particle?},\ }\href {https://doi.org/10.1103/PhysRevX.5.031027}
  {\bibfield  {journal} {\bibinfo  {journal} {Phys. Rev. X}\ }\textbf {\bibinfo
  {volume} {5}},\ \bibinfo {pages} {031027} (\bibinfo {year}
  {2015})}\BibitemShut {NoStop}%
\bibitem [{\citenamefont
  {Haldane}(2011)}]{haldane2011selfdualitylongwavelengthbehaviorlandaulevel}%
  \BibitemOpen
  \bibfield  {author} {\bibinfo {author} {\bibfnamefont {F.~D.~M.}\
  \bibnamefont {Haldane}},\ }\href@noop {} {\bibinfo {title} {Self-duality and
  long-wavelength behavior of the landau-level guiding-center structure
  function, and the shear modulus of fractional quantum hall fluids}} (\bibinfo
  {year} {2011}),\ \Eprint {https://arxiv.org/abs/1112.0990} {arXiv:1112.0990}
  \BibitemShut {NoStop}%
\bibitem [{\citenamefont {Abanov}\ and\ \citenamefont
  {Gromov}(2014)}]{PhysRevB.90.014435}%
  \BibitemOpen
  \bibfield  {author} {\bibinfo {author} {\bibfnamefont {A.~G.}\ \bibnamefont
  {Abanov}}\ and\ \bibinfo {author} {\bibfnamefont {A.}~\bibnamefont
  {Gromov}},\ }\bibfield  {title} {\bibinfo {title} {Electromagnetic and
  gravitational responses of two-dimensional noninteracting electrons in a
  background magnetic field},\ }\href
  {https://doi.org/10.1103/PhysRevB.90.014435} {\bibfield  {journal} {\bibinfo
  {journal} {Phys. Rev. B}\ }\textbf {\bibinfo {volume} {90}},\ \bibinfo
  {pages} {014435} (\bibinfo {year} {2014})}\BibitemShut {NoStop}%
\bibitem [{\citenamefont {Maciejko}\ \emph {et~al.}(2013)\citenamefont
  {Maciejko}, \citenamefont {Hsu}, \citenamefont {Kivelson}, \citenamefont
  {Park},\ and\ \citenamefont {Sondhi}}]{PhysRevB.88.125137}%
  \BibitemOpen
  \bibfield  {author} {\bibinfo {author} {\bibfnamefont {J.}~\bibnamefont
  {Maciejko}}, \bibinfo {author} {\bibfnamefont {B.}~\bibnamefont {Hsu}},
  \bibinfo {author} {\bibfnamefont {S.~A.}\ \bibnamefont {Kivelson}}, \bibinfo
  {author} {\bibfnamefont {Y.}~\bibnamefont {Park}},\ and\ \bibinfo {author}
  {\bibfnamefont {S.~L.}\ \bibnamefont {Sondhi}},\ }\bibfield  {title}
  {\bibinfo {title} {Field theory of the quantum hall nematic transition},\
  }\href {https://doi.org/10.1103/PhysRevB.88.125137} {\bibfield  {journal}
  {\bibinfo  {journal} {Phys. Rev. B}\ }\textbf {\bibinfo {volume} {88}},\
  \bibinfo {pages} {125137} (\bibinfo {year} {2013})}\BibitemShut {NoStop}%
\bibitem [{\citenamefont {Liou}\ \emph {et~al.}(2019)\citenamefont {Liou},
  \citenamefont {Haldane}, \citenamefont {Yang},\ and\ \citenamefont
  {Rezayi}}]{CGFQHLiou}%
  \BibitemOpen
  \bibfield  {author} {\bibinfo {author} {\bibfnamefont {S.-F.}\ \bibnamefont
  {Liou}}, \bibinfo {author} {\bibfnamefont {F.~D.~M.}\ \bibnamefont
  {Haldane}}, \bibinfo {author} {\bibfnamefont {K.}~\bibnamefont {Yang}},\ and\
  \bibinfo {author} {\bibfnamefont {E.~H.}\ \bibnamefont {Rezayi}},\ }\bibfield
   {title} {\bibinfo {title} {Chiral gravitons in fractional quantum hall
  liquids},\ }\href {https://doi.org/10.1103/PhysRevLett.123.146801} {\bibfield
   {journal} {\bibinfo  {journal} {Phys. Rev. Lett.}\ }\textbf {\bibinfo
  {volume} {123}},\ \bibinfo {pages} {146801} (\bibinfo {year}
  {2019})}\BibitemShut {NoStop}%
\bibitem [{\citenamefont {Nguyen}\ \emph {et~al.}(2022)\citenamefont {Nguyen},
  \citenamefont {Haldane}, \citenamefont {Rezayi}, \citenamefont {Son},\ and\
  \citenamefont {Yang}}]{MMSSNguyen}%
  \BibitemOpen
  \bibfield  {author} {\bibinfo {author} {\bibfnamefont {D.~X.}\ \bibnamefont
  {Nguyen}}, \bibinfo {author} {\bibfnamefont {F.~D.~M.}\ \bibnamefont
  {Haldane}}, \bibinfo {author} {\bibfnamefont {E.~H.}\ \bibnamefont {Rezayi}},
  \bibinfo {author} {\bibfnamefont {D.~T.}\ \bibnamefont {Son}},\ and\ \bibinfo
  {author} {\bibfnamefont {K.}~\bibnamefont {Yang}},\ }\bibfield  {title}
  {\bibinfo {title} {Multiple magnetorotons and spectral sum rules in
  fractional quantum hall systems},\ }\href
  {https://doi.org/10.1103/PhysRevLett.128.246402} {\bibfield  {journal}
  {\bibinfo  {journal} {Phys. Rev. Lett.}\ }\textbf {\bibinfo {volume} {128}},\
  \bibinfo {pages} {246402} (\bibinfo {year} {2022})}\BibitemShut {NoStop}%
\bibitem [{\citenamefont {Yuzhu}\ and\ \citenamefont
  {Bo}(2023)}]{yuzhu2023geometric}%
  \BibitemOpen
  \bibfield  {author} {\bibinfo {author} {\bibfnamefont {W.}~\bibnamefont
  {Yuzhu}}\ and\ \bibinfo {author} {\bibfnamefont {Y.}~\bibnamefont {Bo}},\
  }\bibfield  {title} {\bibinfo {title} {Geometric fluctuation of conformal
  hilbert spaces and multiple graviton modes in fractional quantum hall
  effect},\ }\href {https://doi.org/10.1038/s41467-023-38036-0} {\bibfield
  {journal} {\bibinfo  {journal} {Nat. Commun.}\ }\textbf {\bibinfo {volume}
  {14}},\ \bibinfo {pages} {2317} (\bibinfo {year} {2023})}\BibitemShut
  {NoStop}%
\bibitem [{\citenamefont {Yang}(2016)}]{geometric_yang}%
  \BibitemOpen
  \bibfield  {author} {\bibinfo {author} {\bibfnamefont {K.}~\bibnamefont
  {Yang}},\ }\bibfield  {title} {\bibinfo {title} {Acoustic wave absorption as
  a probe of dynamical geometric response of fractional quantum hall liquids},\
  }\href {https://doi.org/10.1103/PhysRevB.93.161302} {\bibfield  {journal}
  {\bibinfo  {journal} {Phys. Rev. B}\ }\textbf {\bibinfo {volume} {93}},\
  \bibinfo {pages} {161302} (\bibinfo {year} {2016})}\BibitemShut {NoStop}%
\bibitem [{\citenamefont {Liu}\ \emph {et~al.}(2018)\citenamefont {Liu},
  \citenamefont {Gromov},\ and\ \citenamefont {Papi\ifmmode~\acute{c}\else
  \'{c}\fi{}}}]{PhysRevB.98.155140}%
  \BibitemOpen
  \bibfield  {author} {\bibinfo {author} {\bibfnamefont {Z.}~\bibnamefont
  {Liu}}, \bibinfo {author} {\bibfnamefont {A.}~\bibnamefont {Gromov}},\ and\
  \bibinfo {author} {\bibfnamefont {Z.}~\bibnamefont
  {Papi\ifmmode~\acute{c}\else \'{c}\fi{}}},\ }\bibfield  {title} {\bibinfo
  {title} {Geometric quench and nonequilibrium dynamics of fractional quantum
  hall states},\ }\href {https://doi.org/10.1103/PhysRevB.98.155140} {\bibfield
   {journal} {\bibinfo  {journal} {Phys. Rev. B}\ }\textbf {\bibinfo {volume}
  {98}},\ \bibinfo {pages} {155140} (\bibinfo {year} {2018})}\BibitemShut
  {NoStop}%
\bibitem [{\citenamefont {Du}\ \emph {et~al.}(2022)\citenamefont {Du},
  \citenamefont {Mehta}, \citenamefont {Nguyen},\ and\ \citenamefont
  {Son}}]{Du_2022}%
  \BibitemOpen
  \bibfield  {author} {\bibinfo {author} {\bibfnamefont {Y.-H.}\ \bibnamefont
  {Du}}, \bibinfo {author} {\bibfnamefont {U.}~\bibnamefont {Mehta}}, \bibinfo
  {author} {\bibfnamefont {D.}~\bibnamefont {Nguyen}},\ and\ \bibinfo {author}
  {\bibfnamefont {D.~T.}\ \bibnamefont {Son}},\ }\bibfield  {title} {\bibinfo
  {title} {Volume-preserving diffeomorphism as nonabelian higher-rank gauge
  symmetry},\ }\href {http://dx.doi.org/10.21468/SciPostPhys.12.2.050}
  {\bibfield  {journal} {\bibinfo  {journal} {SciPost Physics}\ }\textbf
  {\bibinfo {volume} {12}} (\bibinfo {year} {2022})}\BibitemShut {NoStop}%
\bibitem [{\citenamefont {Pinczuk}\ \emph {et~al.}(1993)\citenamefont
  {Pinczuk}, \citenamefont {Dennis}, \citenamefont {Pfeiffer},\ and\
  \citenamefont {West}}]{PhysRevLett.70.3983}%
  \BibitemOpen
  \bibfield  {author} {\bibinfo {author} {\bibfnamefont {A.}~\bibnamefont
  {Pinczuk}}, \bibinfo {author} {\bibfnamefont {B.~S.}\ \bibnamefont {Dennis}},
  \bibinfo {author} {\bibfnamefont {L.~N.}\ \bibnamefont {Pfeiffer}},\ and\
  \bibinfo {author} {\bibfnamefont {K.}~\bibnamefont {West}},\ }\bibfield
  {title} {\bibinfo {title} {Observation of collective excitations in the
  fractional quantum hall effect},\ }\href
  {https://doi.org/10.1103/PhysRevLett.70.3983} {\bibfield  {journal} {\bibinfo
   {journal} {Phys. Rev. Lett.}\ }\textbf {\bibinfo {volume} {70}},\ \bibinfo
  {pages} {3983} (\bibinfo {year} {1993})}\BibitemShut {NoStop}%
\bibitem [{\citenamefont {Liang}\ \emph {et~al.}(2024)\citenamefont {Liang},
  \citenamefont {Liu}, \citenamefont {Yang}, \citenamefont {Huang},
  \citenamefont {Wurstbauer}, \citenamefont {Dean}, \citenamefont {West},
  \citenamefont {Pfeiffer}, \citenamefont {Du},\ and\ \citenamefont
  {Pinczuk}}]{liang2024evidence}%
  \BibitemOpen
  \bibfield  {author} {\bibinfo {author} {\bibfnamefont {J.}~\bibnamefont
  {Liang}}, \bibinfo {author} {\bibfnamefont {Z.}~\bibnamefont {Liu}}, \bibinfo
  {author} {\bibfnamefont {Z.}~\bibnamefont {Yang}}, \bibinfo {author}
  {\bibfnamefont {Y.}~\bibnamefont {Huang}}, \bibinfo {author} {\bibfnamefont
  {U.}~\bibnamefont {Wurstbauer}}, \bibinfo {author} {\bibfnamefont {C.~R.}\
  \bibnamefont {Dean}}, \bibinfo {author} {\bibfnamefont {K.~W.}\ \bibnamefont
  {West}}, \bibinfo {author} {\bibfnamefont {L.~N.}\ \bibnamefont {Pfeiffer}},
  \bibinfo {author} {\bibfnamefont {L.}~\bibnamefont {Du}},\ and\ \bibinfo
  {author} {\bibfnamefont {A.}~\bibnamefont {Pinczuk}},\ }\bibfield  {title}
  {\bibinfo {title} {Evidence for chiral graviton modes in fractional quantum
  hall liquids},\ }\href {https://doi.org/10.1038/s41586-024-07201-w}
  {\bibfield  {journal} {\bibinfo  {journal} {Nature}\ }\textbf {\bibinfo
  {volume} {628}},\ \bibinfo {pages} {78} (\bibinfo {year} {2024})}\BibitemShut
  {NoStop}%
\bibitem [{\citenamefont {Kang}\ \emph {et~al.}(2001)\citenamefont {Kang},
  \citenamefont {Pinczuk}, \citenamefont {Dennis}, \citenamefont {Pfeiffer},\
  and\ \citenamefont {West}}]{PhysRevLett.86.2637}%
  \BibitemOpen
  \bibfield  {author} {\bibinfo {author} {\bibfnamefont {M.}~\bibnamefont
  {Kang}}, \bibinfo {author} {\bibfnamefont {A.}~\bibnamefont {Pinczuk}},
  \bibinfo {author} {\bibfnamefont {B.~S.}\ \bibnamefont {Dennis}}, \bibinfo
  {author} {\bibfnamefont {L.~N.}\ \bibnamefont {Pfeiffer}},\ and\ \bibinfo
  {author} {\bibfnamefont {K.~W.}\ \bibnamefont {West}},\ }\bibfield  {title}
  {\bibinfo {title} {Observation of multiple magnetorotons in the fractional
  quantum hall effect},\ }\href {https://doi.org/10.1103/PhysRevLett.86.2637}
  {\bibfield  {journal} {\bibinfo  {journal} {Phys. Rev. Lett.}\ }\textbf
  {\bibinfo {volume} {86}},\ \bibinfo {pages} {2637} (\bibinfo {year}
  {2001})}\BibitemShut {NoStop}%
\bibitem [{\citenamefont {Wu}\ \emph {et~al.}(2019)\citenamefont {Wu},
  \citenamefont {Lovorn}, \citenamefont {Tutuc}, \citenamefont {Martin},\ and\
  \citenamefont {MacDonald}}]{wu2019topological}%
  \BibitemOpen
  \bibfield  {author} {\bibinfo {author} {\bibfnamefont {F.}~\bibnamefont
  {Wu}}, \bibinfo {author} {\bibfnamefont {T.}~\bibnamefont {Lovorn}}, \bibinfo
  {author} {\bibfnamefont {E.}~\bibnamefont {Tutuc}}, \bibinfo {author}
  {\bibfnamefont {I.}~\bibnamefont {Martin}},\ and\ \bibinfo {author}
  {\bibfnamefont {A.~H.}\ \bibnamefont {MacDonald}},\ }\bibfield  {title}
  {\bibinfo {title} {Topological insulators in twisted transition metal
  dichalcogenide homobilayers},\ }\href
  {https://doi.org/10.1103/PhysRevLett.122.086402} {\bibfield  {journal}
  {\bibinfo  {journal} {Phys. Rev. Lett.}\ }\textbf {\bibinfo {volume} {122}},\
  \bibinfo {pages} {086402} (\bibinfo {year} {2019})}\BibitemShut {NoStop}%
\bibitem [{\citenamefont {Xu}\ \emph {et~al.}(2024)\citenamefont {Xu},
  \citenamefont {Li}, \citenamefont {Xu}, \citenamefont {Bi},\ and\
  \citenamefont {Zhang}}]{xu2024maximally}%
  \BibitemOpen
  \bibfield  {author} {\bibinfo {author} {\bibfnamefont {C.}~\bibnamefont
  {Xu}}, \bibinfo {author} {\bibfnamefont {J.}~\bibnamefont {Li}}, \bibinfo
  {author} {\bibfnamefont {Y.}~\bibnamefont {Xu}}, \bibinfo {author}
  {\bibfnamefont {Z.}~\bibnamefont {Bi}},\ and\ \bibinfo {author}
  {\bibfnamefont {Y.}~\bibnamefont {Zhang}},\ }\bibfield  {title} {\bibinfo
  {title} {Maximally localized wannier functions, interaction models, and
  fractional quantum anomalous hall effect in twisted bilayer mote2},\ }\href
  {https://doi.org/10.1073/pnas.2316749121} {\bibfield  {journal} {\bibinfo
  {journal} {Proc. Natl. Acad. Sci.}\ }\textbf {\bibinfo {volume} {121}},\
  \bibinfo {pages} {e2316749121} (\bibinfo {year} {2024})}\BibitemShut
  {NoStop}%
\bibitem [{\citenamefont {Sharma}\ \emph {et~al.}(2024)\citenamefont {Sharma},
  \citenamefont {Peng},\ and\ \citenamefont {Sheng}}]{sharma2024topological}%
  \BibitemOpen
  \bibfield  {author} {\bibinfo {author} {\bibfnamefont {P.}~\bibnamefont
  {Sharma}}, \bibinfo {author} {\bibfnamefont {Y.}~\bibnamefont {Peng}},\ and\
  \bibinfo {author} {\bibfnamefont {D.~N.}\ \bibnamefont {Sheng}},\ }\href@noop
  {} {\bibinfo {title} {Topological quantum phase transitions driven by
  displacement fields in the twisted mote2 bilayers}} (\bibinfo {year}
  {2024}),\ \Eprint {https://arxiv.org/abs/2405.08181} {arXiv:2405.08181}
  \BibitemShut {NoStop}%
\bibitem [{\citenamefont {Reddy}\ \emph {et~al.}(2023)\citenamefont {Reddy},
  \citenamefont {Alsallom}, \citenamefont {Zhang}, \citenamefont {Devakul},\
  and\ \citenamefont {Fu}}]{reddy2023fractional}%
  \BibitemOpen
  \bibfield  {author} {\bibinfo {author} {\bibfnamefont {A.~P.}\ \bibnamefont
  {Reddy}}, \bibinfo {author} {\bibfnamefont {F.}~\bibnamefont {Alsallom}},
  \bibinfo {author} {\bibfnamefont {Y.}~\bibnamefont {Zhang}}, \bibinfo
  {author} {\bibfnamefont {T.}~\bibnamefont {Devakul}},\ and\ \bibinfo {author}
  {\bibfnamefont {L.}~\bibnamefont {Fu}},\ }\bibfield  {title} {\bibinfo
  {title} {Fractional quantum anomalous hall states in twisted bilayer
  ${\mathrm{mote}}_{2}$ and ${\mathrm{wse}}_{2}$},\ }\href
  {https://doi.org/10.1103/PhysRevB.108.085117} {\bibfield  {journal} {\bibinfo
   {journal} {Phys. Rev. B}\ }\textbf {\bibinfo {volume} {108}},\ \bibinfo
  {pages} {085117} (\bibinfo {year} {2023})}\BibitemShut {NoStop}%
\bibitem [{\citenamefont {Shen}\ \emph {et~al.}(2024)\citenamefont {Shen},
  \citenamefont {Wang}, \citenamefont {Guo}, \citenamefont {Xu}, \citenamefont
  {Duan},\ and\ \citenamefont
  {Xu}}]{shen2024stabilizingfractionalcherninsulators}%
  \BibitemOpen
  \bibfield  {author} {\bibinfo {author} {\bibfnamefont {X.}~\bibnamefont
  {Shen}}, \bibinfo {author} {\bibfnamefont {C.}~\bibnamefont {Wang}}, \bibinfo
  {author} {\bibfnamefont {R.}~\bibnamefont {Guo}}, \bibinfo {author}
  {\bibfnamefont {Z.}~\bibnamefont {Xu}}, \bibinfo {author} {\bibfnamefont
  {W.}~\bibnamefont {Duan}},\ and\ \bibinfo {author} {\bibfnamefont
  {Y.}~\bibnamefont {Xu}},\ }\href@noop {} {\bibinfo {title} {Stabilizing
  fractional chern insulators via exchange interaction in moir\'e systems}}
  (\bibinfo {year} {2024}),\ \Eprint {https://arxiv.org/abs/2405.12294}
  {arXiv:2405.12294} \BibitemShut {NoStop}%
\bibitem [{\citenamefont {Lu}\ \emph {et~al.}(2024{\natexlab{c}})\citenamefont
  {Lu}, \citenamefont {Wu}, \citenamefont {Chen},\ and\ \citenamefont
  {Meng}}]{lu2024vestigialgaplessbosondensity}%
  \BibitemOpen
  \bibfield  {author} {\bibinfo {author} {\bibfnamefont {H.}~\bibnamefont
  {Lu}}, \bibinfo {author} {\bibfnamefont {H.-Q.}\ \bibnamefont {Wu}}, \bibinfo
  {author} {\bibfnamefont {B.-B.}\ \bibnamefont {Chen}},\ and\ \bibinfo
  {author} {\bibfnamefont {Z.~Y.}\ \bibnamefont {Meng}},\ }\href@noop {}
  {\bibinfo {title} {Vestigial gapless boson density wave emerging between $\nu
  = 1/2$ fractional chern insulator and finite-momentum supersolid}} (\bibinfo
  {year} {2024}{\natexlab{c}}),\ \Eprint {https://arxiv.org/abs/2408.07111}
  {arXiv:2408.07111} \BibitemShut {NoStop}%
\bibitem [{\citenamefont {Yu}\ \emph {et~al.}(2024)\citenamefont {Yu},
  \citenamefont {Herzog-Arbeitman}, \citenamefont {Wang}, \citenamefont
  {Vafek}, \citenamefont {Bernevig},\ and\ \citenamefont
  {Regnault}}]{yu2024fractional}%
  \BibitemOpen
  \bibfield  {author} {\bibinfo {author} {\bibfnamefont {J.}~\bibnamefont
  {Yu}}, \bibinfo {author} {\bibfnamefont {J.}~\bibnamefont
  {Herzog-Arbeitman}}, \bibinfo {author} {\bibfnamefont {M.}~\bibnamefont
  {Wang}}, \bibinfo {author} {\bibfnamefont {O.}~\bibnamefont {Vafek}},
  \bibinfo {author} {\bibfnamefont {B.~A.}\ \bibnamefont {Bernevig}},\ and\
  \bibinfo {author} {\bibfnamefont {N.}~\bibnamefont {Regnault}},\ }\bibfield
  {title} {\bibinfo {title} {Fractional chern insulators versus nonmagnetic
  states in twisted bilayer ${\mathrm{mote}}_{2}$},\ }\href
  {https://doi.org/10.1103/PhysRevB.109.045147} {\bibfield  {journal} {\bibinfo
   {journal} {Phys. Rev. B}\ }\textbf {\bibinfo {volume} {109}},\ \bibinfo
  {pages} {045147} (\bibinfo {year} {2024})}\BibitemShut {NoStop}%
\bibitem [{\citenamefont {Lu}\ \emph {et~al.}(2024{\natexlab{d}})\citenamefont
  {Lu}, \citenamefont {Wu}, \citenamefont {Chen}, \citenamefont {Sun},\ and\
  \citenamefont {Meng}}]{lu2024interactiondrivenrotoncondensationc}%
  \BibitemOpen
  \bibfield  {author} {\bibinfo {author} {\bibfnamefont {H.}~\bibnamefont
  {Lu}}, \bibinfo {author} {\bibfnamefont {H.-Q.}\ \bibnamefont {Wu}}, \bibinfo
  {author} {\bibfnamefont {B.-B.}\ \bibnamefont {Chen}}, \bibinfo {author}
  {\bibfnamefont {K.}~\bibnamefont {Sun}},\ and\ \bibinfo {author}
  {\bibfnamefont {Z.~Y.}\ \bibnamefont {Meng}},\ }\href@noop {} {\bibinfo
  {title} {Interaction-driven roton condensation in c = 2/3 fractional quantum
  anomalous hall state}} (\bibinfo {year} {2024}{\natexlab{d}}),\ \Eprint
  {https://arxiv.org/abs/2403.03258} {arXiv:2403.03258} \BibitemShut {NoStop}%
\bibitem [{\citenamefont {Waters}\ \emph {et~al.}(2024)\citenamefont {Waters},
  \citenamefont {Okounkova}, \citenamefont {Su}, \citenamefont {Zhou},
  \citenamefont {Yao}, \citenamefont {Watanabe}, \citenamefont {Taniguchi},
  \citenamefont {Xu}, \citenamefont {Zhang}, \citenamefont {Folk} \emph
  {et~al.}}]{waters2024interplay}%
  \BibitemOpen
  \bibfield  {author} {\bibinfo {author} {\bibfnamefont {D.}~\bibnamefont
  {Waters}}, \bibinfo {author} {\bibfnamefont {A.}~\bibnamefont {Okounkova}},
  \bibinfo {author} {\bibfnamefont {R.}~\bibnamefont {Su}}, \bibinfo {author}
  {\bibfnamefont {B.}~\bibnamefont {Zhou}}, \bibinfo {author} {\bibfnamefont
  {J.}~\bibnamefont {Yao}}, \bibinfo {author} {\bibfnamefont {K.}~\bibnamefont
  {Watanabe}}, \bibinfo {author} {\bibfnamefont {T.}~\bibnamefont {Taniguchi}},
  \bibinfo {author} {\bibfnamefont {X.}~\bibnamefont {Xu}}, \bibinfo {author}
  {\bibfnamefont {Y.-H.}\ \bibnamefont {Zhang}}, \bibinfo {author}
  {\bibfnamefont {J.}~\bibnamefont {Folk}}, \emph {et~al.},\ }\bibfield
  {title} {\bibinfo {title} {Interplay of electronic crystals with integer and
  fractional chern insulators in moir$\backslash$'e pentalayer graphene},\
  }\href {https://arxiv.org/abs/2408.10133} {\bibfield  {journal} {\bibinfo
  {journal} {arXiv preprint arXiv:2408.10133}\ } (\bibinfo {year}
  {2024})}\BibitemShut {NoStop}%
\bibitem [{\citenamefont {Bonsall}\ and\ \citenamefont
  {Maradudin}(1977)}]{PhysRevB.15.1959}%
  \BibitemOpen
  \bibfield  {author} {\bibinfo {author} {\bibfnamefont {L.}~\bibnamefont
  {Bonsall}}\ and\ \bibinfo {author} {\bibfnamefont {A.~A.}\ \bibnamefont
  {Maradudin}},\ }\bibfield  {title} {\bibinfo {title} {Some static and
  dynamical properties of a two-dimensional wigner crystal},\ }\href
  {https://doi.org/10.1103/PhysRevB.15.1959} {\bibfield  {journal} {\bibinfo
  {journal} {Phys. Rev. B}\ }\textbf {\bibinfo {volume} {15}},\ \bibinfo
  {pages} {1959} (\bibinfo {year} {1977})}\BibitemShut {NoStop}%
\bibitem [{\citenamefont {Tan}\ and\ \citenamefont
  {Devakul}(2024)}]{tan2024parentberrycurvatureideal}%
  \BibitemOpen
  \bibfield  {author} {\bibinfo {author} {\bibfnamefont {T.}~\bibnamefont
  {Tan}}\ and\ \bibinfo {author} {\bibfnamefont {T.}~\bibnamefont {Devakul}},\
  }\href@noop {} {\bibinfo {title} {Parent berry curvature and the ideal
  anomalous hall crystal}} (\bibinfo {year} {2024}),\ \Eprint
  {https://arxiv.org/abs/2403.04196} {arXiv:2403.04196} \BibitemShut {NoStop}%
\bibitem [{\citenamefont {Dong}\ \emph
  {et~al.}(2024{\natexlab{a}})\citenamefont {Dong}, \citenamefont {Wang},
  \citenamefont {Wang}, \citenamefont {Soejima}, \citenamefont {Zaletel},
  \citenamefont {Vishwanath},\ and\ \citenamefont
  {Parker}}]{dong2024anomaloushallcrystalsrhombohedral}%
  \BibitemOpen
  \bibfield  {author} {\bibinfo {author} {\bibfnamefont {J.}~\bibnamefont
  {Dong}}, \bibinfo {author} {\bibfnamefont {T.}~\bibnamefont {Wang}}, \bibinfo
  {author} {\bibfnamefont {T.}~\bibnamefont {Wang}}, \bibinfo {author}
  {\bibfnamefont {T.}~\bibnamefont {Soejima}}, \bibinfo {author} {\bibfnamefont
  {M.~P.}\ \bibnamefont {Zaletel}}, \bibinfo {author} {\bibfnamefont
  {A.}~\bibnamefont {Vishwanath}},\ and\ \bibinfo {author} {\bibfnamefont
  {D.~E.}\ \bibnamefont {Parker}},\ }\href@noop {} {\bibinfo {title} {Anomalous
  hall crystals in rhombohedral multilayer graphene i: Interaction-driven chern
  bands and fractional quantum hall states at zero magnetic field}} (\bibinfo
  {year} {2024}{\natexlab{a}}),\ \Eprint {https://arxiv.org/abs/2311.05568}
  {arXiv:2311.05568} \BibitemShut {NoStop}%
\bibitem [{\citenamefont {Dong}\ \emph
  {et~al.}(2024{\natexlab{b}})\citenamefont {Dong}, \citenamefont {Patri},\
  and\ \citenamefont {Senthil}}]{PhysRevLett.133.206502}%
  \BibitemOpen
  \bibfield  {author} {\bibinfo {author} {\bibfnamefont {Z.}~\bibnamefont
  {Dong}}, \bibinfo {author} {\bibfnamefont {A.~S.}\ \bibnamefont {Patri}},\
  and\ \bibinfo {author} {\bibfnamefont {T.}~\bibnamefont {Senthil}},\
  }\bibfield  {title} {\bibinfo {title} {Theory of quantum anomalous hall
  phases in pentalayer rhombohedral graphene moir\'e structures},\ }\href
  {https://link.aps.org/doi/10.1103/PhysRevLett.133.206502} {\bibfield
  {journal} {\bibinfo  {journal} {Phys. Rev. Lett.}\ }\textbf {\bibinfo
  {volume} {133}},\ \bibinfo {pages} {206502} (\bibinfo {year}
  {2024}{\natexlab{b}})}\BibitemShut {NoStop}%
\bibitem [{\citenamefont {Lu}\ \emph {et~al.}(2024{\natexlab{e}})\citenamefont
  {Lu}, \citenamefont {Wu}, \citenamefont {Chen},\ and\ \citenamefont
  {Meng}}]{lu2024fractionalquantumanomaloushall}%
  \BibitemOpen
  \bibfield  {author} {\bibinfo {author} {\bibfnamefont {H.}~\bibnamefont
  {Lu}}, \bibinfo {author} {\bibfnamefont {H.-Q.}\ \bibnamefont {Wu}}, \bibinfo
  {author} {\bibfnamefont {B.-B.}\ \bibnamefont {Chen}},\ and\ \bibinfo
  {author} {\bibfnamefont {Z.~Y.}\ \bibnamefont {Meng}},\ }\href@noop {}
  {\bibinfo {title} {From a fractional quantum anomalous hall state to a
  smectic state with equal hall conductance}} (\bibinfo {year}
  {2024}{\natexlab{e}}),\ \Eprint {https://arxiv.org/abs/2404.06745}
  {arXiv:2404.06745} \BibitemShut {NoStop}%
\bibitem [{\citenamefont {Dong}\ \emph
  {et~al.}(2024{\natexlab{c}})\citenamefont {Dong}, \citenamefont {Patri},\
  and\ \citenamefont {Senthil}}]{PhysRevB.110.205130}%
  \BibitemOpen
  \bibfield  {author} {\bibinfo {author} {\bibfnamefont {Z.}~\bibnamefont
  {Dong}}, \bibinfo {author} {\bibfnamefont {A.~S.}\ \bibnamefont {Patri}},\
  and\ \bibinfo {author} {\bibfnamefont {T.}~\bibnamefont {Senthil}},\
  }\bibfield  {title} {\bibinfo {title} {Stability of anomalous hall crystals
  in multilayer rhombohedral graphene},\ }\href
  {https://doi.org/10.1103/PhysRevB.110.205130} {\bibfield  {journal} {\bibinfo
   {journal} {Phys. Rev. B}\ }\textbf {\bibinfo {volume} {110}},\ \bibinfo
  {pages} {205130} (\bibinfo {year} {2024}{\natexlab{c}})}\BibitemShut
  {NoStop}%
\bibitem [{Note1()}]{Note1}%
  \BibitemOpen
  \bibinfo {note} {One can also project the phase space into the momentum space
  in FQH and consider the LLL in the momentum space.}\BibitemShut {Stop}%
\bibitem [{Note2()}]{Note2}%
  \BibitemOpen
  \bibinfo {note} {In this work, the ideal FCI refers to the FCI hosted in a
  perfect flatband with topology and geometry identical with LLL.}\BibitemShut
  {Stop}%
\bibitem [{Note3()}]{Note3}%
  \BibitemOpen
  \bibinfo {note} {In practice, the truncation is set to be $3|\protect \bm
  {G}|$ to ensure the convergence of $\protect \bm {q}$ summation. In fact, due
  to the exponential decay of the form factor, the contribution outside the 1st
  BZ in summation of $\protect \bm {q}$ is strongly suppressed, and hence one
  can directly apply the $(q_x\pm iq_y)^2$ and set the truncation of the
  summation up to the 1st BZ.}\BibitemShut {Stop}%
\bibitem [{\citenamefont {Luttinger}(1960)}]{PhysRev.119.1153}%
  \BibitemOpen
  \bibfield  {author} {\bibinfo {author} {\bibfnamefont {J.~M.}\ \bibnamefont
  {Luttinger}},\ }\bibfield  {title} {\bibinfo {title} {Fermi surface and some
  simple equilibrium properties of a system of interacting fermions},\ }\href
  {https://doi.org/10.1103/PhysRev.119.1153} {\bibfield  {journal} {\bibinfo
  {journal} {Phys. Rev.}\ }\textbf {\bibinfo {volume} {119}},\ \bibinfo {pages}
  {1153} (\bibinfo {year} {1960})}\BibitemShut {NoStop}%
\bibitem [{\citenamefont {Delacr\'etaz}\ \emph {et~al.}(2022)\citenamefont
  {Delacr\'etaz}, \citenamefont {Du}, \citenamefont {Mehta},\ and\
  \citenamefont {Son}}]{PhysRevResearch.4.033131}%
  \BibitemOpen
  \bibfield  {author} {\bibinfo {author} {\bibfnamefont {L.~V.}\ \bibnamefont
  {Delacr\'etaz}}, \bibinfo {author} {\bibfnamefont {Y.-H.}\ \bibnamefont
  {Du}}, \bibinfo {author} {\bibfnamefont {U.}~\bibnamefont {Mehta}},\ and\
  \bibinfo {author} {\bibfnamefont {D.~T.}\ \bibnamefont {Son}},\ }\bibfield
  {title} {\bibinfo {title} {Nonlinear bosonization of fermi surfaces: The
  method of coadjoint orbits},\ }\href
  {https://doi.org/10.1103/PhysRevResearch.4.033131} {\bibfield  {journal}
  {\bibinfo  {journal} {Phys. Rev. Res.}\ }\textbf {\bibinfo {volume} {4}},\
  \bibinfo {pages} {033131} (\bibinfo {year} {2022})}\BibitemShut {NoStop}%
\bibitem [{Note4()}]{Note4}%
  \BibitemOpen
  \bibinfo {note} {We are thankful for Steven Simon for pointing out
  this.}\BibitemShut {Stop}%
\bibitem [{\citenamefont {Hirjibehedin}\ \emph {et~al.}(2005)\citenamefont
  {Hirjibehedin}, \citenamefont {Dujovne}, \citenamefont {Pinczuk},
  \citenamefont {Dennis}, \citenamefont {Pfeiffer},\ and\ \citenamefont
  {West}}]{Hirjibehedin_2005}%
  \BibitemOpen
  \bibfield  {author} {\bibinfo {author} {\bibfnamefont {C.~F.}\ \bibnamefont
  {Hirjibehedin}}, \bibinfo {author} {\bibfnamefont {I.}~\bibnamefont
  {Dujovne}}, \bibinfo {author} {\bibfnamefont {A.}~\bibnamefont {Pinczuk}},
  \bibinfo {author} {\bibfnamefont {B.~S.}\ \bibnamefont {Dennis}}, \bibinfo
  {author} {\bibfnamefont {L.~N.}\ \bibnamefont {Pfeiffer}},\ and\ \bibinfo
  {author} {\bibfnamefont {K.~W.}\ \bibnamefont {West}},\ }\bibfield  {title}
  {\bibinfo {title} {Splitting of long-wavelength modes of the fractional
  quantum hall liquid at},\ }\href
  {http://dx.doi.org/10.1103/PhysRevLett.95.066803} {\bibfield  {journal}
  {\bibinfo  {journal} {Phys. Rev. Lett.}\ }\textbf {\bibinfo {volume} {95}}
  (\bibinfo {year} {2005})}\BibitemShut {NoStop}%
\bibitem [{\citenamefont {Saigal}\ \emph {et~al.}(2024)\citenamefont {Saigal},
  \citenamefont {Klebl}, \citenamefont {Lambers}, \citenamefont {Bahmanyar},
  \citenamefont {Anti\ifmmode~\acute{c}\else \'{c}\fi{}}, \citenamefont
  {Kennes}, \citenamefont {Wehling},\ and\ \citenamefont
  {Wurstbauer}}]{saigal2024collective}%
  \BibitemOpen
  \bibfield  {author} {\bibinfo {author} {\bibfnamefont {N.}~\bibnamefont
  {Saigal}}, \bibinfo {author} {\bibfnamefont {L.}~\bibnamefont {Klebl}},
  \bibinfo {author} {\bibfnamefont {H.}~\bibnamefont {Lambers}}, \bibinfo
  {author} {\bibfnamefont {S.}~\bibnamefont {Bahmanyar}}, \bibinfo {author}
  {\bibfnamefont {V.}~\bibnamefont {Anti\ifmmode~\acute{c}\else \'{c}\fi{}}},
  \bibinfo {author} {\bibfnamefont {D.~M.}\ \bibnamefont {Kennes}}, \bibinfo
  {author} {\bibfnamefont {T.~O.}\ \bibnamefont {Wehling}},\ and\ \bibinfo
  {author} {\bibfnamefont {U.}~\bibnamefont {Wurstbauer}},\ }\bibfield  {title}
  {\bibinfo {title} {Collective charge excitations between moir\'e minibands in
  twisted ${\mathrm{wse}}_{2}$ bilayers probed with resonant inelastic light
  scattering},\ }\href {https://doi.org/10.1103/PhysRevLett.133.046902}
  {\bibfield  {journal} {\bibinfo  {journal} {Phys. Rev. Lett.}\ }\textbf
  {\bibinfo {volume} {133}},\ \bibinfo {pages} {046902} (\bibinfo {year}
  {2024})}\BibitemShut {NoStop}%
\bibitem [{\citenamefont {Wang}\ \emph
  {et~al.}(2024{\natexlab{c}})\citenamefont {Wang}, \citenamefont {Devakul},
  \citenamefont {Zaletel},\ and\ \citenamefont
  {Fu}}]{wang2024diversemagneticordersquantum}%
  \BibitemOpen
  \bibfield  {author} {\bibinfo {author} {\bibfnamefont {T.}~\bibnamefont
  {Wang}}, \bibinfo {author} {\bibfnamefont {T.}~\bibnamefont {Devakul}},
  \bibinfo {author} {\bibfnamefont {M.~P.}\ \bibnamefont {Zaletel}},\ and\
  \bibinfo {author} {\bibfnamefont {L.}~\bibnamefont {Fu}},\ }\href@noop {}
  {\bibinfo {title} {Diverse magnetic orders and quantum anomalous hall effect
  in twisted bilayer mote2 and wse2}} (\bibinfo {year} {2024}{\natexlab{c}}),\
  \Eprint {https://arxiv.org/abs/2306.02501} {arXiv:2306.02501} \BibitemShut
  {NoStop}%
\bibitem [{\citenamefont {Yu}\ \emph {et~al.}(2023)\citenamefont {Yu},
  \citenamefont {Herzog-Arbeitman}, \citenamefont {Wang}, \citenamefont
  {Vafek}, \citenamefont {Bernevig},\ and\ \citenamefont
  {Regnault}}]{yu2023fractionalcherninsulatorsvs}%
  \BibitemOpen
  \bibfield  {author} {\bibinfo {author} {\bibfnamefont {J.}~\bibnamefont
  {Yu}}, \bibinfo {author} {\bibfnamefont {J.}~\bibnamefont
  {Herzog-Arbeitman}}, \bibinfo {author} {\bibfnamefont {M.}~\bibnamefont
  {Wang}}, \bibinfo {author} {\bibfnamefont {O.}~\bibnamefont {Vafek}},
  \bibinfo {author} {\bibfnamefont {B.~A.}\ \bibnamefont {Bernevig}},\ and\
  \bibinfo {author} {\bibfnamefont {N.}~\bibnamefont {Regnault}},\ }\href@noop
  {} {\bibinfo {title} {Fractional chern insulators vs. non-magnetic states in
  twisted bilayer mote$_2$}} (\bibinfo {year} {2023}),\ \Eprint
  {https://arxiv.org/abs/2309.14429} {arXiv:2309.14429} \BibitemShut {NoStop}%
\bibitem [{\citenamefont {Bernevig}\ \emph {et~al.}(2021)\citenamefont
  {Bernevig}, \citenamefont {Song}, \citenamefont {Regnault},\ and\
  \citenamefont {Lian}}]{Bernevig_2021}%
  \BibitemOpen
  \bibfield  {author} {\bibinfo {author} {\bibfnamefont {B.~A.}\ \bibnamefont
  {Bernevig}}, \bibinfo {author} {\bibfnamefont {Z.-D.}\ \bibnamefont {Song}},
  \bibinfo {author} {\bibfnamefont {N.}~\bibnamefont {Regnault}},\ and\
  \bibinfo {author} {\bibfnamefont {B.}~\bibnamefont {Lian}},\ }\bibfield
  {title} {\bibinfo {title} {Twisted bilayer graphene. iii. interacting
  hamiltonian and exact symmetries},\ }\href
  {http://dx.doi.org/10.1103/PhysRevB.103.205413} {\bibfield  {journal}
  {\bibinfo  {journal} {Phys. Rev. B}\ }\textbf {\bibinfo {volume} {103}}
  (\bibinfo {year} {2021})}\BibitemShut {NoStop}%
\bibitem [{\citenamefont {Andrews}\ \emph {et~al.}(2024)\citenamefont
  {Andrews}, \citenamefont {Raja}, \citenamefont {Mishra}, \citenamefont
  {Zaletel},\ and\ \citenamefont {Roy}}]{PhysRevB.109.245111}%
  \BibitemOpen
  \bibfield  {author} {\bibinfo {author} {\bibfnamefont {B.}~\bibnamefont
  {Andrews}}, \bibinfo {author} {\bibfnamefont {M.}~\bibnamefont {Raja}},
  \bibinfo {author} {\bibfnamefont {N.}~\bibnamefont {Mishra}}, \bibinfo
  {author} {\bibfnamefont {M.~P.}\ \bibnamefont {Zaletel}},\ and\ \bibinfo
  {author} {\bibfnamefont {R.}~\bibnamefont {Roy}},\ }\bibfield  {title}
  {\bibinfo {title} {Stability of fractional chern insulators with a non-landau
  level continuum limit},\ }\href {https://doi.org/10.1103/PhysRevB.109.245111}
  {\bibfield  {journal} {\bibinfo  {journal} {Phys. Rev. B}\ }\textbf {\bibinfo
  {volume} {109}},\ \bibinfo {pages} {245111} (\bibinfo {year}
  {2024})}\BibitemShut {NoStop}%
\bibitem [{\citenamefont {Claassen}\ \emph {et~al.}(2015)\citenamefont
  {Claassen}, \citenamefont {Lee}, \citenamefont {Thomale}, \citenamefont
  {Qi},\ and\ \citenamefont {Devereaux}}]{Claassen_2015}%
  \BibitemOpen
  \bibfield  {author} {\bibinfo {author} {\bibfnamefont {M.}~\bibnamefont
  {Claassen}}, \bibinfo {author} {\bibfnamefont {C.~H.}\ \bibnamefont {Lee}},
  \bibinfo {author} {\bibfnamefont {R.}~\bibnamefont {Thomale}}, \bibinfo
  {author} {\bibfnamefont {X.-L.}\ \bibnamefont {Qi}},\ and\ \bibinfo {author}
  {\bibfnamefont {T.~P.}\ \bibnamefont {Devereaux}},\ }\bibfield  {title}
  {\bibinfo {title} {Position-momentum duality and fractional quantum hall
  effect in chern insulators},\ }\href
  {http://dx.doi.org/10.1103/PhysRevLett.114.236802} {\bibfield  {journal}
  {\bibinfo  {journal} {Phys. Rev. Lett.}\ }\textbf {\bibinfo {volume} {114}}
  (\bibinfo {year} {2015})}\BibitemShut {NoStop}%
\bibitem [{\citenamefont {Gromov}\ and\ \citenamefont
  {Son}(2017{\natexlab{b}})}]{PhysRevX.7.041032}%
  \BibitemOpen
  \bibfield  {author} {\bibinfo {author} {\bibfnamefont {A.}~\bibnamefont
  {Gromov}}\ and\ \bibinfo {author} {\bibfnamefont {D.~T.}\ \bibnamefont
  {Son}},\ }\bibfield  {title} {\bibinfo {title} {Bimetric theory of fractional
  quantum hall states},\ }\href {https://doi.org/10.1103/PhysRevX.7.041032}
  {\bibfield  {journal} {\bibinfo  {journal} {Phys. Rev. X}\ }\textbf {\bibinfo
  {volume} {7}},\ \bibinfo {pages} {041032} (\bibinfo {year}
  {2017}{\natexlab{b}})}\BibitemShut {NoStop}%
\bibitem [{\citenamefont {Goldman}\ and\ \citenamefont
  {Fradkin}(2018)}]{PhysRevB.98.165137}%
  \BibitemOpen
  \bibfield  {author} {\bibinfo {author} {\bibfnamefont {H.}~\bibnamefont
  {Goldman}}\ and\ \bibinfo {author} {\bibfnamefont {E.}~\bibnamefont
  {Fradkin}},\ }\bibfield  {title} {\bibinfo {title} {Dirac composite fermions
  and emergent reflection symmetry about even-denominator filling fractions},\
  }\href {https://doi.org/10.1103/PhysRevB.98.165137} {\bibfield  {journal}
  {\bibinfo  {journal} {Phys. Rev. B}\ }\textbf {\bibinfo {volume} {98}},\
  \bibinfo {pages} {165137} (\bibinfo {year} {2018})}\BibitemShut {NoStop}%
\bibitem [{\citenamefont {Balram}\ \emph {et~al.}(2022)\citenamefont {Balram},
  \citenamefont {Liu}, \citenamefont {Gromov},\ and\ \citenamefont
  {Papi\'c}}]{Balram:2021opn}%
  \BibitemOpen
  \bibfield  {author} {\bibinfo {author} {\bibfnamefont {A.~C.}\ \bibnamefont
  {Balram}}, \bibinfo {author} {\bibfnamefont {Z.}~\bibnamefont {Liu}},
  \bibinfo {author} {\bibfnamefont {A.}~\bibnamefont {Gromov}},\ and\ \bibinfo
  {author} {\bibfnamefont {Z.}~\bibnamefont {Papi\'c}},\ }\bibfield  {title}
  {\bibinfo {title} {{Very-High-Energy Collective States of Partons in
  Fractional Quantum Hall Liquids}},\ }\href
  {https://doi.org/10.1103/PhysRevX.12.021008} {\bibfield  {journal} {\bibinfo
  {journal} {Phys. Rev. X}\ }\textbf {\bibinfo {volume} {12}},\ \bibinfo
  {pages} {021008} (\bibinfo {year} {2022})}\BibitemShut {NoStop}%
\bibitem [{\citenamefont {Nguyen}\ and\ \citenamefont
  {Son}(2021)}]{PhysRevResearch.3.033217}%
  \BibitemOpen
  \bibfield  {author} {\bibinfo {author} {\bibfnamefont {D.~X.}\ \bibnamefont
  {Nguyen}}\ and\ \bibinfo {author} {\bibfnamefont {D.~T.}\ \bibnamefont
  {Son}},\ }\bibfield  {title} {\bibinfo {title} {Dirac composite fermion
  theory of general jain sequences},\ }\href
  {https://doi.org/10.1103/PhysRevResearch.3.033217} {\bibfield  {journal}
  {\bibinfo  {journal} {Phys. Rev. Res.}\ }\textbf {\bibinfo {volume} {3}},\
  \bibinfo {pages} {033217} (\bibinfo {year} {2021})}\BibitemShut {NoStop}%
\bibitem [{\citenamefont {Dong}\ \emph {et~al.}(2023)\citenamefont {Dong},
  \citenamefont {Wang}, \citenamefont {Ledwith}, \citenamefont {Vishwanath},\
  and\ \citenamefont {Parker}}]{Dong_2023}%
  \BibitemOpen
  \bibfield  {author} {\bibinfo {author} {\bibfnamefont {J.}~\bibnamefont
  {Dong}}, \bibinfo {author} {\bibfnamefont {J.}~\bibnamefont {Wang}}, \bibinfo
  {author} {\bibfnamefont {P.~J.}\ \bibnamefont {Ledwith}}, \bibinfo {author}
  {\bibfnamefont {A.}~\bibnamefont {Vishwanath}},\ and\ \bibinfo {author}
  {\bibfnamefont {D.~E.}\ \bibnamefont {Parker}},\ }\bibfield  {title}
  {\bibinfo {title} {Composite fermi liquid at zero magnetic field in twisted
  mote2},\ }\href {http://dx.doi.org/10.1103/PhysRevLett.131.136502} {\bibfield
   {journal} {\bibinfo  {journal} {Phys. Rev. Lett.}\ }\textbf {\bibinfo
  {volume} {131}} (\bibinfo {year} {2023})}\BibitemShut {NoStop}%
\bibitem [{\citenamefont {Murthy}\ and\ \citenamefont
  {Shankar}(2003)}]{murthy2003hamiltonian}%
  \BibitemOpen
  \bibfield  {author} {\bibinfo {author} {\bibfnamefont {G.}~\bibnamefont
  {Murthy}}\ and\ \bibinfo {author} {\bibfnamefont {R.}~\bibnamefont
  {Shankar}},\ }\bibfield  {title} {\bibinfo {title} {Hamiltonian theories of
  the fractional quantum hall effect},\ }\href
  {https://doi.org/10.1103/RevModPhys.75.1101} {\bibfield  {journal} {\bibinfo
  {journal} {Rev. Mod. Phys.}\ }\textbf {\bibinfo {volume} {75}},\ \bibinfo
  {pages} {1101} (\bibinfo {year} {2003})}\BibitemShut {NoStop}%
\end{thebibliography}%

\clearpage
\onecolumngrid
\vspace{1cm}
\begin{center}
{\bf\large Supplemental Materials: ``Magnetorotons in moir\'e fractional Chern insulators''}
\end{center}
\tableofcontents
\section{I. Continuum model and exact diagonalization}
The continuum Hamiltonian of the twisted $\mathrm{MoTe}_2$ is
\begin{equation}
\mathcal{H}_{\tau}(\boldsymbol{r})=\left(\begin{array}{cc}
-\frac{\hbar^2\left(\boldsymbol{k}-\tau \boldsymbol{\kappa}_{+}\right)^2}{2 m^*}+\Delta_{\mathfrak{b}}(\boldsymbol{r})+V_z/2 & \Delta_T(\boldsymbol{r}) \\
\Delta_T^{\dagger}(\boldsymbol{r}) & -\frac{\hbar^2\left(\boldsymbol{k}-\tau \boldsymbol{\kappa}_{-}\right)^2}{2 m^*}+\Delta_{\mathfrak{t}}(\boldsymbol{r})-V_z/2
\end{array}\right),
\end{equation}
where $\tau = \pm 1$ is the spin-valley indices, 
\begin{equation}
\begin{aligned}
\Delta_{\mathfrak{t}, \mathfrak{b}}(\boldsymbol{r}) & =2 V \sum_{j=1,3,5} \cos \left(\boldsymbol{G}_j \cdot \boldsymbol{r}+\ell \psi\right), \\
\Delta_T(\boldsymbol{r}) & =w\left(1+e^{-i \tau \boldsymbol{G}_2 \cdot \boldsymbol{r}}+e^{-i \tau \boldsymbol{G}_3 \cdot \boldsymbol{r}}\right).
\end{aligned}
\end{equation}
$\Delta_{\mathfrak{t},\mathfrak{b}}(\boldsymbol{r})$ is the intralayer moir\'e potential and $\Delta_{T}(\boldsymbol{r})$ is the interlayer tunneling terms. $\ell = \pm 1$ is the layer indices. $\boldsymbol{G}_j, j = 1,\cdots,6$ are the reciprocal lattice vectors.  In our numerical calculations, we take the parameters from fitting the first-principles calculation as in: $V=20.8~\text{meV}$, $\psi=107.7^\circ$ and $w=-23.8~\text{meV}$. 
\begin{figure}[h]
    \centering
    \includegraphics[width=0.5\linewidth]{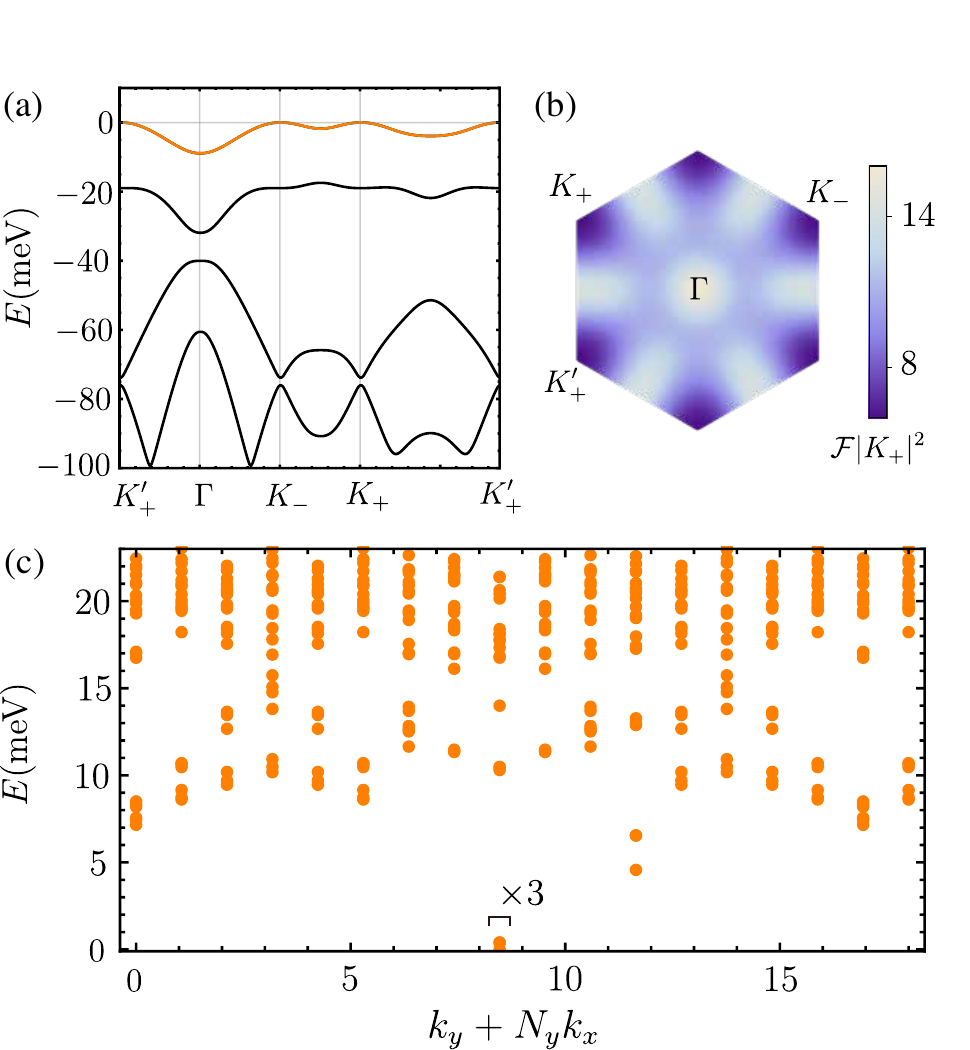}
    \caption{(a) Band structure of twisted $\mathrm{MoTe}_2$ at $\theta = 3.89^\circ$. The topmost valence band is highlighted in orange. (b) Berry curvature $\mathcal{F}$ of the topmost valence band. (c) Many-body spectrum from the $N = 18$ exact diagonalization in $\epsilon = 5,\theta = 3.89^\circ,d = 30\text{nm}$.}
    \label{fig:ED_cm}
\end{figure}

From the calculation of the continuum model, the topmost valence band possesses a nearly flat band structure with $\mathcal{C} = 1$. Under the condition of $\epsilon = 5,\theta = 3.89^\circ,d = 30\text{nm}$, the ground state exhibits three-fold ground states at $\Gamma$ point, serving as the indicator that the system is in the FCI phase. See Fig.~\ref{fig:ED_cm} for the band structure and ED spectrum.
\section{II. Numerical details of single mode approximation in FCIs}
\begin{figure}[h]
     \centering
     \includegraphics[width=0.35\linewidth]{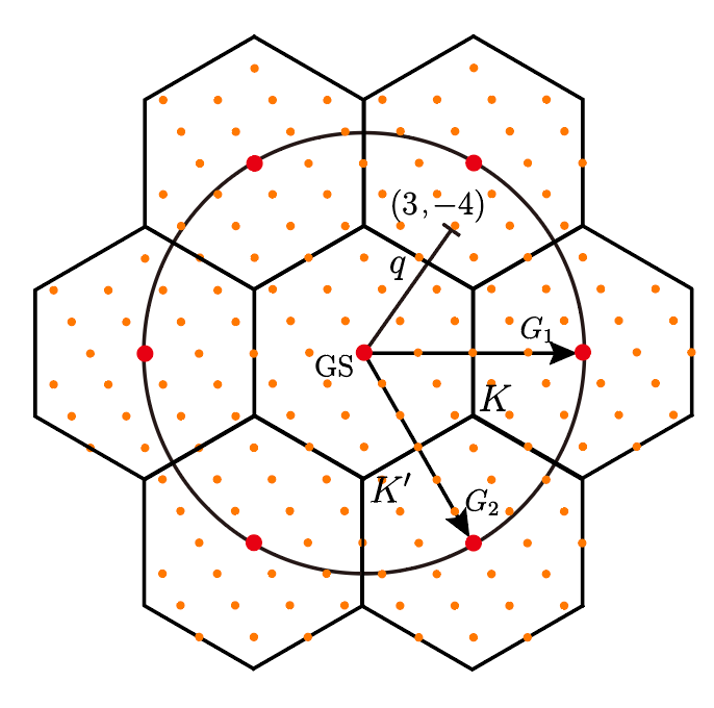}
     \caption{Schematic representation of the SMA construction in the twisted TMD. The central hexagon region is the 1st BZ and the adjacent hexagons are the 2nd BZs. The orange points are the discretization (e.g. $N_1 = 4,N_2 = 6$) in exact diagonalization calculations. The coordinate $(3,-4)$ represents point $\frac{3}{4}\boldsymbol{G}_1-\frac{4}{6}\boldsymbol{G}_2$. The red points are the $\Gamma$ points where the ground states reside. }
     \label{fig:discretization}
 \end{figure}

In this section, we discuss in numerical detail how to construct the GMP wavefunction from the SMA in FCIs. In general, we construct the GMP wavefunction according to 
\begin{equation}
    |\psi_{\boldsymbol{q}}\rangle = \hat{\rho}(\boldsymbol{q})|\psi^m_0\rangle, m = 1,2,3
\end{equation}
as shown in Fig.~\ref{fig:discretization}.

In the $-\frac{2}{3}$ filling, we have three-fold ground degeneracy, and the distribution of the ground states dependents on the discretization, or more specifically the generalized Pauli principle. This imposes a possible issue for the ansatz: when the ground state does not reside at the $\Gamma$ point, which leads to the non-zero overlap between the GMP wavefunction and the ground states, suggesting the ansatz is not effective at those momenta. To avoid this problem, we choose the $3\times 6$ points at most of the calculations, in which the three-fold ground states are all located in the $\Gamma$ points to ensure the GMP wavefunctions are orthogonal to the ground states at other momenta. We also calculated $N = 24,30$ in practice and did not find obvious differences between different clusters.

 Another issue is that the single mode approximation is not precise at large $\boldsymbol{q}$, hence we restrict ourselves inside the second BZ, say $|\boldsymbol{q}|<|\boldsymbol{G}|$.
 \section{III. Geometric Excitations in Thermodynamic Limit}

\begin{figure}[h]
    \centering
    \includegraphics[width=0.8\linewidth]{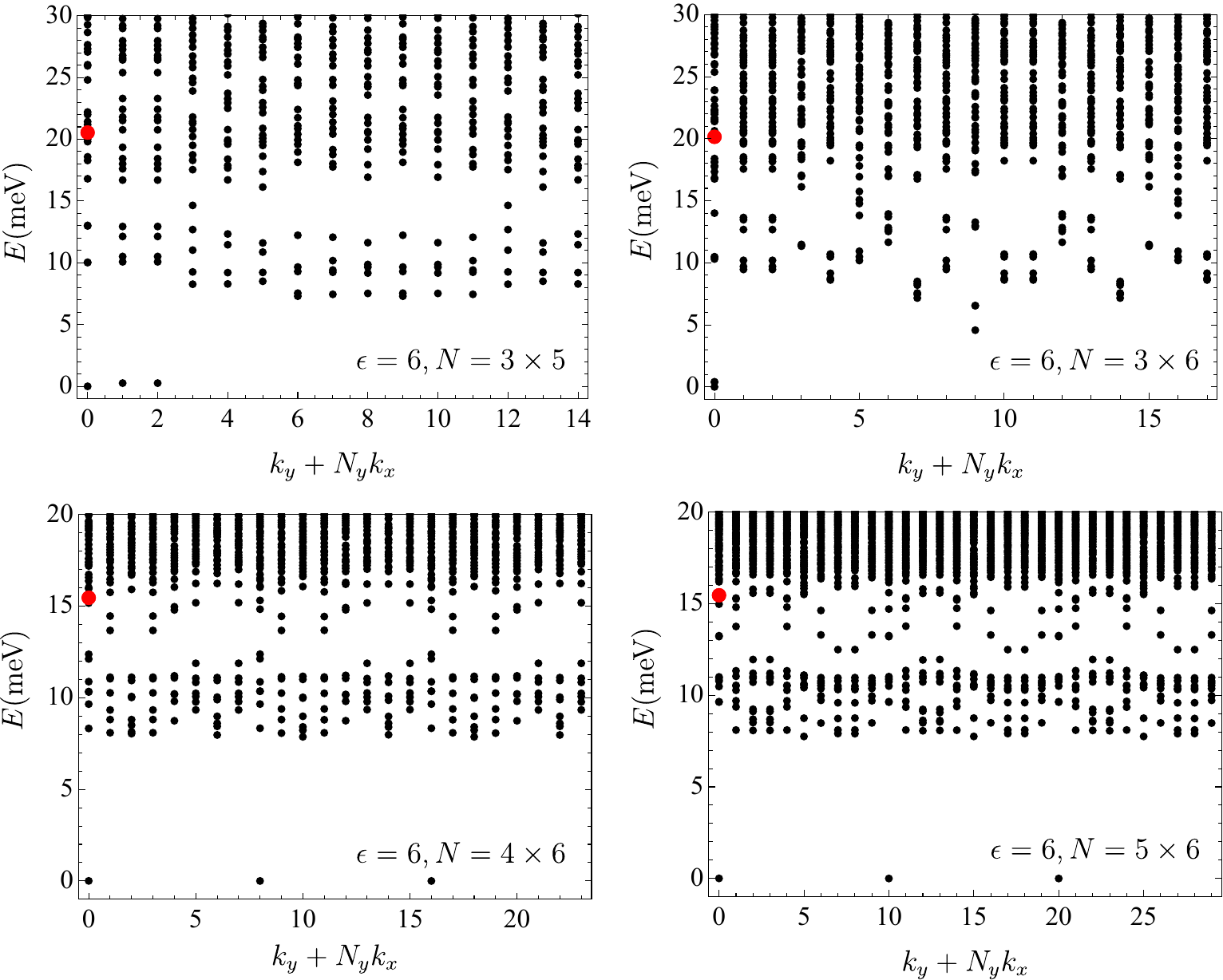}
    \caption{Position of geometric excitations in the many-body spectrum for $N = 3\times5, 3\times 6,4\times 6,5\times 6$. The energy of geometric excitations are highlighted in red.}
    \label{fig:spectrum_N}
\end{figure}

This section provides information on the energy of angular momentum-2 geometric excitations in the many-body spectrum for various system sizes. The parameters in Sec. I are used. As shown in Fig.~\ref{fig:spectrum_N}, we find that for small system sizes the energy of the geometric excitations (the maximum peak of the dynamical spectrum of the operator $\hat{\mathcal{O}}$), highlighted in red, lies inside the continuum excitations region. However, as the increasing of system size, the energy of geometric excitations gets close to the lower bound of the continuum excitations region. For system size $4\times 6$ and $5\times 6$, the geometric excitations are lying below the continuum excitation region of the many-body spectrum, which suggests in the system the geometric excitations serve as the low-lying excitations compared with the continuum of high energy excitations, and is possible to distinguish from the continuum. It provides evidence that under these conditions, the boundary location of the geometric excitations prevents the diverging of the density of states, henceforth the peaks of geometric excitations may survive in the thermodynamic limit. This suggests the possibility of experimental detection in the thermodynamic limit.  

\section{IV. Geometric Excitations in the band system}
In this section, we try to present the details of geometric excitations in the moir\'e flatband system and deduce the general form of the operator that is responsible for the system's geometric dynamics. We comes to the conclusion that the geometric excitations in FCI arise due to the incompressibility in the phase space. In particular, we discuss in the context of the $2D$ moir\'e band system mentioned above. 

\subsection{a. Phase space volume-preserving diffeomorphism}

% Starting from the lattice system with charge conservation and translation symmetry, we have a well-defined filling $\nu$ in $k$ space, 
% \begin{equation}
%     \frac{\mathcal{V}_F}{(2\pi)^d} = \nu
% \end{equation}
% $\mathcal{V}_F$ is the volume circled by the Fermi surface. In the absence of interaction, by requiring charge conservation, the volume bounded by the fermi surface remains fixed, and in general, the geometric excitations naturally arise from the incompressibility of the phase space volume. Let us distort the Fermi surface without changing its volume, which formally means a volume-preserving diffeomorphism (VPD) is applied to the Fermi surface. 
We first give the formal definition of volume-preserving diffeomorphism. Starting from a diffeomorphism in the $d$ dimensional space, we have
\begin{equation}
\label{eqn:VPD}
    f:\textbf{x}_\alpha\to \textbf{x}_\alpha^\prime = \textbf{x}_\alpha+\epsilon v_\alpha(\textbf{x}),\alpha = 1,\cdots,d,
\end{equation}
$\textbf{x}$ is the $d$ dimensional coordinates. In the following of considering the transformation of other quantities under the diffeomorphism, we will focus on the \textit{active} diffeomorphism instead of simply doing the coordinate transformation, which is called the \textit{passive} diffeomorphism. The active diffeomorphism means after dragging the coordinates along the manifolds, we perform the coordinate transformation back to the original coordinates and compare the functions at the same point. The active diffeomorphism then naturally relates different geometric quantities in the same coordinates. We consider the Riemannian flat space. Under active diffeomorphism, we write down the transformation of a flat metric
\begin{equation}
\label{VPD:aandg}
\begin{aligned}
    \delta^\prime_{\alpha\beta}(\textbf{x}) & = \delta_{\alpha\beta} + \epsilon \partial_\alpha v^\beta(\textbf{x})+\epsilon \partial_\beta v^\alpha(\textbf{x}) , \alpha,\beta = 1,\cdots, d.
    \end{aligned}
\end{equation}
The local volume form in the space $\text{Vol}_\delta = \sqrt{\det \delta}\mathrm{d}x_1\wedge \mathrm{d}x_2\wedge\cdots \wedge\mathrm{d}x_d$ now changes under the active diffeomorphism
\begin{equation}
    \mathrm{Vol}_\delta = \sqrt{\det \delta}\mathrm{d}x_1\wedge \mathrm{d}x_2\wedge\cdots \wedge\mathrm{d}x_d\to \mathrm{Vol}_{\delta^\prime} = \sqrt{\det \delta^\prime(\textbf{x})}\mathrm{d}x_1\wedge \mathrm{d}x_2\wedge\cdots \wedge\mathrm{d}x_d = \sqrt{\det \delta}(1-\epsilon\partial_\mu v^\mu(\textbf{x}))\mathrm{d}x_1\wedge \mathrm{d}x_2\wedge\cdots \wedge\mathrm{d}x_d,
\end{equation}
By requiring the invariant space volume $\text{Vol}_\delta = \text{Vol}_{\delta^\prime}$, we have $\partial_\mu v^\mu(\textbf{x}) = 0$. A general solution would be $v^\mu = \epsilon^{\mu \nu} \partial_\nu h(\textbf{x})$, where $h$ is a differentiated function of $\textbf{x}$. The volume perserving diffeomorphism $f$ (VPD) forms a group $\text{SDiff}(\mathbb{R}^d)$. Let us now consider the 4D phase space with $\textbf{x} = (x_1,x_2,k_1,k_2)$. In the phase space, $x_\alpha,k_\alpha, \alpha = 1,2$ are conjugated variables to each other
$ [x_\alpha,p_\beta] = \delta_{\alpha\beta} $. The canonical transformation $F$ that preserves the commutator is 
\begin{equation}
    F_\alpha: x_{\alpha}^\prime \to x_\alpha+\epsilon\nabla_{k_\alpha}h(\textbf{x}),\quad     k_\alpha^\prime \to k_\alpha-\epsilon\nabla_{x_\alpha}h(\textbf{x}),\alpha = 1,2.
\end{equation}
According to the VPD defined above, the canonical transformation $F$ preserves the phase space volume element $\mathrm{d}x_\alpha\wedge \mathrm{d}k_\alpha$ invariant, and indicates the incompressibility in phase space. 
% \begin{equation}
% [\hat{\boldsymbol{k}}^\prime,\hat{\boldsymbol{x}}^\prime] = 1.
% \end{equation}
% For a curved space, the constraint of VPD would be $\nabla_\mu v^\mu(k) = 0$, where $\nabla$ is the covariant derivative that involves the Riemannian connection. And thus an alternative expression for VPD would be $g_{\mu\nu}\delta_f g^{\mu\nu} = 0$. 

The algebra of the canonical transformation is generated by $\hat{\mathcal{G}}_{h} = \varepsilon^{ij}\nabla_{i}h(\textbf{x})\cdot \nabla_j,i,j = \boldsymbol{x,p}$, and 
the generator satisfies 
\begin{equation}
    [\hat{\mathcal{G}}_{h_1}, \hat{\mathcal{G}}_{h2}] = \hat{\mathcal{G}}_{[h_{1},h_{2}]},
\end{equation}
where $[h_{1},h_2] = \nabla_{\boldsymbol{x}}h_1(\textbf{x})\cdot \nabla_{\boldsymbol{p}}h_2(\textbf{x})-\nabla_{\boldsymbol{p}}h_1(\textbf{x})\cdot \nabla_{\boldsymbol{x}}h_2(\textbf{x})$. It is more convenient to phrase the algebra on a suitable basis. This can be achieved by choosing the monomial basis $h(\textbf{x}) = -\sum_{\ell,m} h^{j,k}_{n,m}x_1^{j+1} x_2^{k+1} k_1^{n+1} k_2^{m+1}$, and the generator takes the form $\hat{\mathcal{G}}_h = \sum_{j,k,m,n}h_{n,m}^{j,k}\hat{\mathcal{L}}_{n,m}^{j,k}$, where the $\hat{\mathcal{L}}_{n,m}^{j,k}$ obey the algebra
\begin{equation}
\begin{aligned}
\left[\hat{\mathcal{L}}_{n_1 ,m_1}^{j_1, k_1}, \hat{\mathcal{L}}_{n_2, m_2}^{j_2, k_2}\right]= & \left(\left(j_2+1\right)\left(n_1+1\right)-\left(j_1+1\right)\left(n_2+1\right)\right) \hat{\mathcal{L}}_{n_1+n_2, m_1+m_2+1}^{j_1+j_2, k_1+k_2+1} \\
& +\left(\left(k_2+1\right)\left(m_1+1\right)-\left(k_1+1\right)\left(m_2+1\right)\right) \hat{\mathcal{L}}_{n_1+n_2+1, m_1+m_2}^{j_1+j_2+1, k_1+k_2} .
\end{aligned}
\end{equation}
In general, the geometric excitations generated by $\hat{\mathcal{L}}_{n,m}^{j,k}|\text{0}\rangle$ can be considered as the distortion mode of the space-momentum mixed space. It is direct to check the algebra contains 2 sets of commuting subalgebra $\text{sdiff}(\mathbb{R}^2)$, or 2D classical $w_\infty$ algebra formed by $\hat{\mathcal{\mathcal{L}}}_{-1,m}^{-1,k}$ and $\hat{\mathcal{\mathcal{L}}}_{n,-1}^{j,-1}$, within each sector the higher spin mode can be defined as the ``phase space higher spin mode''. 
\begin{equation}
    \left[\hat{\mathcal{L}}_{m}^k, \hat{\mathcal{L}}_{n}^j\right]=((m+1)(j+1)-(n+1)(k+1)) \hat{\mathcal{L}}_{m+n}^{k+j},\quad \hat{\mathcal{L}}_{m}^k = \hat{\mathcal{L}}_{m,-1}^{k,-1}.
\end{equation}
The band projection in LLL relates $x_1\to k_2, x_1\to k_1$, thus unifying 2 sets of 2D $w_\infty$ algebra and projecting the 2D phase space to the momentum or real subspace. The corresponding phase space geometric excitations then become the real space or momentum space geometric excitations.

In general, the interaction does not commute with the generators $\mathcal{L}$, henceforth the VPD modes are likely to consume energy and acquire an energy gap. We will now consider the change of action or Hamiltonian under VPD by taking account into the interaction.  For simplicity, in the following discussion, we assume the interaction $V(\boldsymbol{q})$ is isotropic, which amounts to neglecting the anisotropy in the dielectric constant and projected to the single valence band. The projected interaction term dominating the kinetic term is 

% In general, the anisotropic Galilean metric and the traceless anti-symmmetric tensor is described by introducing the complex vector $\boldsymbol{\omega}$ and $\eta^{\mu \nu}=\bar{\omega}^\mu \omega^\nu+\bar{\omega}^\nu \omega^\mu, i\varepsilon^{\mu\nu} = \bar{\omega}^\mu \omega^\nu-\bar{\omega}^\nu \omega^\mu$ and for isotropic case $\bar{\omega}=(1, i)^{T} / \sqrt{2}$, $\eta = \mathbb{I},\varepsilon = i\sigma_y$. 

\begin{equation}\label{eqn:interaction}
\begin{aligned}
    V & = \frac{1}{2 A} \sum_{\boldsymbol{q}} V(\boldsymbol{q}) \hat{\bar{\rho}}(\boldsymbol{q}) \hat{\bar{\rho}}(-\boldsymbol{q})= \frac{1}{2 A} \sum_{\boldsymbol{k}, \boldsymbol{k}^{\prime}, \boldsymbol{q}}  V(\boldsymbol{q})\Lambda_{\boldsymbol{k}, \boldsymbol{q}}\Lambda_{ \boldsymbol{k}^{\prime}, -\boldsymbol{q}}\hat{c}_{\boldsymbol{k}+\boldsymbol{q}}^{\dagger} \hat{c}_{\boldsymbol{k}^{\prime}-\boldsymbol{q}}^{\dagger} \hat{c}_{\boldsymbol{k}^{\prime}} \hat{c}_{\boldsymbol{k}},
    \end{aligned}
\end{equation}
$\Lambda_{\boldsymbol{k},\boldsymbol{q}}$ is the form factor $\left\langle u_{ \boldsymbol{k}+\boldsymbol{q}}| u_{\boldsymbol{k}}\right\rangle$, which is not gauge invariant and need gauge fixing \cite{Bernevig_2021,PhysRevB.109.245111}. Let's now assume the form factor is fast decay versus $\boldsymbol{q}$, so that we can set the truncation of $\boldsymbol{q}$ small enough to validate the small $\boldsymbol{q}$ expansion. By directly expanding, we have
\begin{equation}
\begin{aligned}
\label{eqn:expansion_ff}
    \Lambda_{\boldsymbol{k},\boldsymbol{q}}&= 1-i\langle u_{\boldsymbol{k}}|\partial_\mu u_{\boldsymbol{k}}\rangle\mathrm{d}q^{\mu}+\frac{1}{2}\langle u_{\boldsymbol{k}}|\partial_{\mu}\partial_{\nu}u_{\boldsymbol{k}}\rangle \mathrm{d}q^\mu\mathrm{d}q^\nu+\mathcal{O}(q^3) \\
    & = 1-iA_\mu(\boldsymbol{k})\mathrm{d}q^\mu+\frac{1}{2}A_{\mu}(\boldsymbol{k})A_{\nu}(\boldsymbol{k})\mathrm{d}q^{\mu}\mathrm{d}q^\nu-\frac{1}{2}g_{\mu\nu}(\boldsymbol{k})\mathrm{d}q^\mu \mathrm{d}q^\nu+\frac{1}{4}(\partial_\nu A_\mu(\boldsymbol{k})+\partial_\mu A_\nu(\boldsymbol{k}))\mathrm{d}q^\mu\mathrm{d}q^\nu+\mathcal{O}(q^3).
    \end{aligned}
\end{equation}
where $A_{\mu}(\boldsymbol{k})$ is the Berry connection and $g_{\mu\nu}(\boldsymbol{k})$ is the Fubini-Study metric. Formally, in an isolated general Bloch band, the geometry of the Bloch state is characterized by the map $P:\text{BZ}^2\to \mathbb{C}P^{n-1}$, where $\text{BZ}^2$ is the two-dimensional Brillouin zone and $\mathbb{C}P^{n-1}$ is the $n-1$ dimensional complex projective space characterized by $n-1$ parameters. It is then natural to consider the Fubini-Study metric on the Brillouin zone as the pullback of the standard $\mathbb{C}P^{n-1}$ metric and Berry curvature $\Omega(\boldsymbol{k})$ as the pullback of the symplectic 2-form in $\mathbb{C}P^{n-1}$. Under the distortion $f$, the gauge field and the quantum metric in momentum space are then transformed as:

\begin{equation}
\label{VPD:aandg}
    \delta_f A_\alpha(\boldsymbol{k}) = \epsilon (v^\mu \partial_\mu A_\alpha(\boldsymbol{k})+A^\mu\partial_\mu v_\alpha(\boldsymbol{k}) ),
    \end{equation}
    
    \begin{equation}
    \label{eqn:change_metric}
    \delta_f g_{\alpha\beta}(\boldsymbol{k}) =  \epsilon\left(v^\sigma \partial_\sigma g_{\alpha \beta}(\boldsymbol{k})+g_{\mu \beta} \partial_\alpha v^\mu(\boldsymbol{k})+g_{\mu \alpha} \partial_\beta v^\mu(\boldsymbol{k})\right) .
    \end{equation}
    
% The expansion in general indicates the form factor contains an Aharonov-Bohm phase factor $\exp(i\int_{\boldsymbol{k}}^{\boldsymbol{k}+\boldsymbol{q}}A(\boldsymbol{p})\mathrm{d}\boldsymbol{p})$ and a real gauge invariant factor $\exp(\frac{1}{2}\int^{\boldsymbol{k}+\boldsymbol{q}}_{\boldsymbol{k}}g(\boldsymbol{p})\mathrm{d}^2\boldsymbol{p})$ that measures the distance between $\boldsymbol{k}$ and $\boldsymbol{k}+\boldsymbol{q}$ at wavefunction space. 

The change of the Hamiltonian under the VPD $f$, which is the operator $\hat{\mathcal{O}} = \delta_f H(A,g)$ in the main context, measures the energy and the dynamical property of geometric excitations. 
\begin{equation}
    \hat{\mathcal{O}}(\boldsymbol{k}) = \frac{\delta V(A,g)}{\delta A}\delta_f A(\boldsymbol{k})+\frac{\delta V(A,g)}{\delta g}\delta_f g(\boldsymbol{k}).
\end{equation}
We will focus on several simple cases to derive the explicit form of the operator.
% \begin{equation}
%     \hat{\mathcal{O}}^\ell_{\text{GE}} = \frac{\delta H(g, A)}{\delta f}.
% \end{equation}

% As we will show, the explicit form of the operator can be derived. 

% Using $\Lambda_{\boldsymbol{k},\boldsymbol{q}}\approx 1-\frac{1}{2}g(\boldsymbol{k})\boldsymbol{q}\boldsymbol{q}$, where $g_{\mu \nu}(\boldsymbol{k})=\frac{1}{2}\left[\left\langle\partial_\mu u_{\boldsymbol{k}} | \partial_\nu u_{\boldsymbol{k}}\right\rangle-\left\langle\partial_\mu u_{\boldsymbol{k}} |u_{\boldsymbol{k}}\right\rangle\left\langle u_{\boldsymbol{k}} | \partial_\nu u_{\boldsymbol{k}}\right\rangle+(\mu \leftrightarrow \nu)\right]$ is the Fubini-Study metric of the Chern band and $\boldsymbol{q}\boldsymbol{q}$ is the symmetric rank-2 tensor with component $q^\mu q^\nu$ \cite{PhysRevResearch.5.L012015,PhysRevB.90.165139,cheng2013quantumgeometrictensorfubinistudy}.

\subsection{b. Lowest Landau level in momentum space}
As a warmup, we first consider the lowest Landau level with the magnetic field $\nabla\times A = -B\cdot\hat{z}$. In this case, the phase and distance factors can be analytically derived by introducing the guiding center $\boldsymbol{R}$ and Landau orbit $\bar{\boldsymbol{R}}$
\begin{equation}
    \boldsymbol{R}=\boldsymbol{r}+\ell_B^2 \boldsymbol{\pi} \times \hat{z} = \boldsymbol{r}-\bar{\boldsymbol{R}}
\end{equation}
$\boldsymbol{\pi} = \boldsymbol{p}+e\boldsymbol{A}$ is the momentum that commutes with $\boldsymbol{R}$. 
The Hamiltonian can be written into 
\begin{equation}
    H = \frac{\boldsymbol{\pi}^2}{2m}=\omega_c(a^\dagger a+\frac{1}{2}), \quad a=\frac{1}{\ell_B \sqrt{2}}\left(\bar{R}_y-i \bar{R}_x\right)
\end{equation}
The magnetic translation operator formed by the guiding center $t(\boldsymbol{r}) = \exp(i\boldsymbol{r}\cdot \frac{1}{\ell_B^2} (\hat{z} \times \boldsymbol{R}))$ satisfies the magnetic translational algebra
\begin{equation}
    t(\boldsymbol{r}_1)t(\boldsymbol{r}_2) = \exp(i(\boldsymbol{r}_1\times \boldsymbol{r}_2)\cdot \hat{z}/\ell_B^2)t(\boldsymbol{r}_2)t(\boldsymbol{r}_1)
\end{equation}
the magnetic algebra effectively defines a reduced magnetic lattice with unit cell spanned by  $\boldsymbol{a}_1,\boldsymbol{a}_2$ that encloses $2\pi\ell_B^2$ area penetrated by a flux quanta. Notice the commutator of Landau orbits and guiding center satisfy
\begin{equation}
\label{eqn:commu_gc}
    \left[R_i, R_j\right]=-i \epsilon_{i j} \ell_B^2 ,\quad \left[\bar{R}_i, \bar{R}_j\right]=i \epsilon_{i j} \ell_B^2 ,\quad [R_i, \tilde{R}_j]=0
\end{equation}

Since the Hamiltonian only depends on Landau orbits, the eigenvalues can be classified into the Landau levels $E_n = \omega_c(n+\frac{1}{2})$, since the Landau orbits are commute with guiding centers, we can further label the states inside a Landau level by calculating the eigenvalues of the magnetic translational operators, hence in general a state can be expressed into $|n,\boldsymbol{k}\rangle$, where $n$ is the $n\text{th}$ Landau level and $\boldsymbol{k}$ is the momentum in the reduced magnetic BZ. Any state with momentum $\boldsymbol{k}$ inside the reduced magnetic BZ 
can be generated by applying a boost on a zero-momentum state
\begin{equation}
    |n,\boldsymbol{k}\rangle = \exp(i\boldsymbol{k}\cdot\boldsymbol{R} )|n,\boldsymbol{0}\rangle
\end{equation}
The formfactor of the LLL then goes to
\begin{equation}
\begin{aligned}
     \Lambda_{\boldsymbol{k},\boldsymbol{q}} & = \langle u_{\boldsymbol{k}+\boldsymbol{q}}|u_{\boldsymbol{k}}\rangle = \langle 0,\boldsymbol{k}+\boldsymbol{q}|\exp(i\boldsymbol{q}\cdot \boldsymbol{r})|0,\boldsymbol{k}\rangle = \langle 0,\boldsymbol{k}+\boldsymbol{q}|\exp(i\boldsymbol{q}\cdot (\boldsymbol{R}+\bar{\boldsymbol{R}}))|0,\boldsymbol{k}\rangle\\
    & = \langle 0 |\exp(i\bar{\boldsymbol{R}})|0\rangle\langle\boldsymbol{k}+\boldsymbol{q}|\exp(i\boldsymbol{q}\cdot \boldsymbol{R})|\boldsymbol{k}\rangle = \exp\left[\frac{\ell_B^2}{4}(2i\boldsymbol{k}\times \boldsymbol{q}-|\boldsymbol{q}|^2)\right]\\
    \end{aligned}
\end{equation}
where we have use BHC formula and Eq.(\ref{eqn:commu_gc}) when calculating $\langle\boldsymbol{k}+\boldsymbol{q}|\exp(i\boldsymbol{q}\cdot \boldsymbol{R})|\boldsymbol{k}\rangle$
\begin{equation}
\begin{aligned}
\langle\boldsymbol{k}+\boldsymbol{q}|\exp(i\boldsymbol{q}\cdot \boldsymbol{R})|\boldsymbol{k}\rangle & = \langle\boldsymbol{0}|\exp(-i(\boldsymbol{q}+\boldsymbol{k})\cdot \boldsymbol{R})\exp(i\boldsymbol{q}\cdot \boldsymbol{R})\exp(i\boldsymbol{k}\cdot \boldsymbol{R})|\boldsymbol{0}\rangle\\
& = \langle\boldsymbol{0}|\exp(-i(\boldsymbol{q}+\boldsymbol{k})\cdot \boldsymbol{R})\exp(i(\boldsymbol{q}+\boldsymbol{k})\cdot \boldsymbol{R})\exp(i\frac{1}{2}\ell_B^2\epsilon_{ij}q_i k_j)|\boldsymbol{0}\rangle\\
& = \exp\left(i\frac{\ell_B^2}{2}\boldsymbol{k}\times \boldsymbol{q}\right)
\end{aligned}
\end{equation}
As we have demonstrated above, the formfactor consists of a phase factor under the gauge $A = (-k_2,k_1)\frac{B}{2}$ and a distance factor with quantum metric $g = \frac{1}{2B}\mathbb{I}$. According to the condiction of VPD, we choose $h(\textbf{x}) = k_xk_y$ and the corresponding VPD $f$ is generated to be $f: k_1\to k_1+\epsilon k_1, k_2\to k_2-\epsilon k_2$ as the infinitesimal version of VPD $(k_1,k_2)\to (a k_1,k_2/a)$. Using Eq.(\ref{eqn:change_metric}), the quantum metric transforms into $g\to  g^\prime = \frac{1}{2B}(\mathbb{I}+\epsilon\sigma_z), A\to A^\prime = A$ under VPD. The invariant of Berry connection ensures the AB phase factor $\exp(i\int A\mathrm{d}q) = \exp(i\frac{1}{2}\boldsymbol{k}\times \boldsymbol{q})$ is invariant under VPD.
% Let us now consider the time-dependent anisotropy deformation for the system and one of the canonical forms of the deformation is $\boldsymbol{\omega} = (1,i)^T/\sqrt{2}\to \boldsymbol{\omega}^\prime = ((1+\delta a)^{1/2},i/(1+\delta a)^{1/2})^T/\sqrt{2}$, here $\delta a\to 0$ is the small time-dependent value to characterize the effect of anisotropy and by expanding $\eta$ to the linear order of $\delta a$, we have $\eta^\prime \approx \mathbb{I}+\delta a \sigma_z$ and $\varepsilon^\prime = \varepsilon$. The invariant anti-symmetric tensor ensures the phase factor from the line integral of the Berry connection is invariant under the transformation. 

Under the deformation, the form factor transforms as 
\begin{equation}
    \Lambda_{\boldsymbol{k},\boldsymbol{q}}\to \Lambda^\prime_{\boldsymbol{k},\boldsymbol{q}} = \exp\left[\frac{\ell_B^2}{4}(2i\boldsymbol{k}\times \boldsymbol{q}-((1+\epsilon)q_1^2+q_2^2(1-\epsilon))\right]\approx \Lambda_{\boldsymbol{k},\boldsymbol{q}}-\epsilon \frac{\ell_B^2}{4}(q_1^2-q_2^2)\Lambda_{\boldsymbol{k},\boldsymbol{q}}
\end{equation}
inserting into Eq.(\ref{eqn:interaction}), we have the form of operator that coupled to geometric excitations
\begin{equation}
    \hat{\mathcal{O}}=\delta_fV(A,g) \approx \frac{1}{2A}\sum_{\boldsymbol{q}}\frac{\ell_B^2}{2}(q_2^2-q_1^2)V(\boldsymbol{q})\hat{\bar{\rho}}(\boldsymbol{q})\hat{\bar{\rho}}(-\boldsymbol{q})
\end{equation}
the operator can be recast into a more general form
\begin{equation}
    \hat{\mathcal{O}} = \frac{1}{2A}\sum_{\boldsymbol{q}}V(\boldsymbol{q})(\sigma_z\boldsymbol{q})\boldsymbol{q}^T\hat{\bar{\rho}}(\boldsymbol{q})\hat{\bar{\rho}}(-\boldsymbol{q})
\end{equation}
The result is applicable to a general gauge that interpolates between the Landau gauge and the symmetric gauge.

\subsection{c. Topological trivial band}
We now consider the Chern band with $\mathcal{C} = 0$ which mimics the topological trivial flatband that supports CDW in the main context. To facilitate the gauge fixing, we set $\Omega_{\boldsymbol{k}} = 0$. The triviality of $\Omega_{\boldsymbol{k}}$ allows us to erase the gauge field using the gauge transformation $A_{\boldsymbol{k}}^\prime\to A_{\boldsymbol{k}}-\nabla \phi_{\boldsymbol{k}} = 0$. Since $\langle u_{\boldsymbol{k}}|\partial_\mu u_{\boldsymbol{k}}\rangle = 0$, we have $\langle u_{\boldsymbol{k}}|\partial_\mu\partial_\nu u_{\boldsymbol{k}}\rangle+\langle \partial_\mu u_{\boldsymbol{k}}|\partial_\nu u_{\boldsymbol{k}}\rangle = 0$, and henceforth $\text{Im}\langle u_{\boldsymbol{k}}|\partial_\mu\partial_\nu u_{\boldsymbol{k}}\rangle = -\text{Im}\langle \partial_\mu u_{\boldsymbol{k}}|\partial_\nu u_{\boldsymbol{k}}\rangle$. By using the fact that $\langle \partial_\mu u_{\boldsymbol{k}}|\partial_\nu u_{\boldsymbol{k}}\rangle$ is Hermitian, the imaginary part of it is anti-symmetric. Finally, we conclude that 
$\text{Im}\langle  u_{\boldsymbol{k}}|\partial_\mu \partial_\nu u_{\boldsymbol{k}}\rangle$ is also anti-symmetric and do not contribute to the expansion in Eq.(\ref{eqn:expansion_ff}). 
The long-wavelength expansion of the formfactor now reduces to 
\begin{equation}
    \Lambda_{\boldsymbol{k},\boldsymbol{q}} \approx 1+\frac{1}{2}\text{Re}\langle u_{\boldsymbol{k}}|\partial_\mu\partial_\nu u_{\boldsymbol{k}}\rangle\mathrm{d}q^\mu \mathrm{d}q^\nu + \cdots = 1-\frac{1}{2}g_{\mu\nu}(\boldsymbol{k})\mathrm{d}q^\mu \mathrm{d}q^\nu+\cdots ,
\end{equation}
% Again the definition of VPD tells us $\nabla_\mu v^\mu(\boldsymbol{k}) = 0$, to solve the constraint analytically, we will impose a stronger condition on the quantum metric by assuming it varies slowly with $\boldsymbol{k}$, so that we can ignore the derivative of metric and assume a vanishing quantum Christoffel connection. 
% After setting $\nabla = \partial$, we solve the VPD constraints and find $v^\mu(\boldsymbol{k}) = \varepsilon^{\mu\nu}\partial_\nu \lambda(\boldsymbol{k})$. where $\varepsilon$ is the Levi-Civita symbol and $\lambda(\boldsymbol{k})$ is a differentiate scalar function. 

We pick up a canonical transformation generated by $h(\textbf{x}) = x_1k_1-x_2k_2$ in the phase space to generate the phase space angular momentum-2 mode. 
\begin{equation}
    k_1^\prime \to k_1+\epsilon\nabla_{x_1}(x_1k_1-x_2k_2),     k_2^\prime \to k_2+\epsilon\nabla_{x_2}(x_1k_1-x_2k_2)  ,
\end{equation}
% The deformed metric under the corresponding VPD is now
% \begin{equation}
%     \delta_f g_{\mu\nu}(\boldsymbol{k}) = \epsilon( \varepsilon^{\rho \sigma}\left(\partial_\rho g_{\mu\nu }+g_{\rho \nu} \partial_\mu+g_{\mu \rho} \partial_\nu\right) \partial_\sigma \lambda)(\boldsymbol{k})
% \end{equation}
% here we pick a canonical form of $\lambda(\boldsymbol{k}) = k_xk_y$ analogous to the LLL case to preserve the lowest order of momentum. 
Explicitly in coordinates, the change of quantum metric is
\begin{equation}
\begin{aligned}
    \delta_f g_{11}(\boldsymbol{k}) = \epsilon (k_1\partial_1g_{11}-k_2\partial_2 g_{11}+2g_{11})(\boldsymbol{k}),\quad  \delta_f g_{22}(\boldsymbol{k}) = \epsilon(k_1\partial_1 g_{22}-k_2\partial_2 g_{22}-2g_{22})(\boldsymbol{k}) ,
    \end{aligned}
    \end{equation}
\begin{equation}
\begin{aligned}
    \delta_f g_{12}(\boldsymbol{k}) = \delta_f g_{21}(\boldsymbol{k}) = \epsilon(k_1\partial_1g_{12}-k_2 \partial_2 g_{12})(\boldsymbol{k}).
    \end{aligned}
\end{equation}

% Again by applying the time-dependent anisotropy deformation for the system $\boldsymbol{\omega} = (1,i)^T/\sqrt{2}\to \boldsymbol{\omega}^\prime = ((1+\delta a)^{1/2},i/(1+\delta a)^{1/2})^T/\sqrt{2}$, the form factor transforms as 
the formfactor under VPD is then
\begin{equation}
\label{ff_anisotropy}
\Lambda_{\boldsymbol{k},\boldsymbol{q}}\to \Lambda_{\boldsymbol{k},\boldsymbol{q}}^\prime \approx 1-\frac{1}{2}g^\prime(\boldsymbol{k}) \boldsymbol{q}\boldsymbol{q}\approx 1-\frac{1}{2}g(\boldsymbol{k})\boldsymbol{qq}-\epsilon\left(g(\boldsymbol{k})(\sigma_z\boldsymbol{q})( \boldsymbol{q})+\frac{1}{2}(\sigma_z\boldsymbol{k}\cdot \nabla )g(\boldsymbol{k})\boldsymbol{q}\boldsymbol{q}\right).
\end{equation}
the above expression in general involves metric derivatives, which can not be ignored when the band is strongly dispersive, which might indicate the excitations become unstable away from the flatband limit. This also suggests that when the band gets dispersive, the fluctuation of geometry cannot be ignored and the higher spin mode can be interpreted to live in a ``curved space''. We drop it under our assumption and leave it for future study. We find that Eq.(\ref{ff_anisotropy}) actually indicates the coupling between the quantum metric $g$ and the flat external auxiliary metric (which has been considered as a band mass tensor in \cite{geometric_yang}) $\eta = \mathbb{I}$
\begin{equation}
\Lambda_{\boldsymbol{k},\boldsymbol{q}}^\prime\approx 1-\frac{1}{2}g(\boldsymbol{k})(\eta \boldsymbol{q})(\boldsymbol{q})
\end{equation}
by deforming the auxiliary flat metric and fixing the quantum metric, we are able to derive the same operator. The deformation of $\eta$ instead of $g$ actually amounts to the coordinate transformation and is a \textit{passive} perspective to the VPD. This indicates the general form of coupling between quantum metric and external metric $g\eta$. 
A similar form of coupling has also been revealed in \cite{Claassen_2015} by introducing a quadratic confining potential and bimetric gravity theory \cite{PhysRevX.7.041032} which appears in the potential term that leads to the spontaneous symmetry breaking in the isotropic phase. 

% We now deform the external metric $\eta_{\mu\nu}$ by a tunable parameter $a = 1+\delta a, \delta a\to 0$ which measures the anisotropy of the $k-$space. One of the canonical forms of the choice is 
% \begin{equation}
%     \eta = \left(\begin{array}{ll}
% 1 & 0 \\
% 0 & 1
% \end{array}\right)\to \left(\begin{array}{ll}
% a & 0 \\
% 0 & \frac{1}{a}
% \end{array}\right)\approx \left(\begin{array}{ll}
% 1 & 0 \\
% 0 & 1
% \end{array}\right)+\delta a\left(\begin{array}{ll}
% 1 & 0 \\
% 0 & -1
% \end{array}\right)
% \end{equation}
The operator can be easily found under VPD
\begin{equation}
    \hat{\mathcal{O}} = \delta_fV(A,g) =\frac{1}{2A}\sum_{\boldsymbol{q},\boldsymbol{k},\boldsymbol{k}^\prime}V(\boldsymbol{q})\left[(\sigma_z\boldsymbol{q})\frac{g(\boldsymbol{k})+g(\boldsymbol{k}^\prime)}{2}\boldsymbol{q}^T\right]\Lambda_{\boldsymbol{k},\boldsymbol{q}}\Lambda_{\boldsymbol{k}^\prime,-\boldsymbol{q}}\hat{c}_{\boldsymbol{k}+\boldsymbol{q}}^{\dagger} \hat{c}_{\boldsymbol{k}^{\prime}-\boldsymbol{q}}^{\dagger} \hat{c}_{\boldsymbol{k}^{\prime}} \hat{c}_{\boldsymbol{k}}
\end{equation}
which is the operator responsible for the geometric dynamics of a topological trivial band. Here $\boldsymbol{q} = (q_x,q_y)$.

\section{IV. Trial excitonic wavefunction in moir\'e FCI}
This section postulates the trial many-body wavefunctions of intraband neutral excitations in twisted $\mathrm{MoTe}_2$. The following discussion is based on the projection on the topmost valence band. 
Several obstacles exist to constructing the wavefunctions and the corresponding field theory. First of all, the electron is now living in the zero field conditions, or equivalently the flux-attaching process can not be directly applied. Secondly, in the case that the electron is living in the moir\'e  lattice, it is essential for the wavefunction to accommodate the moir\'e potential and should be compatible with the crystalline symmetry. 

Our phenomenological approach is motivated by the parton construction \cite{PhysRevB.98.165137,Balram:2021opn,PhysRevResearch.3.033217}, where the physical electron $c$ can be decomposed into a bosonic parton $\chi$ with charge $q_1$ and a fermionic parton $\phi$ \cite{Dong_2023} with charge $q_2 = 1-q_1$

\begin{equation}
    c = \chi\cdot\phi
\end{equation}

The $U(1)$ symmetry of the composite form requires the introduction of the auxiliary gauge field $a$. This gluing gauge field assigns the $+1$ charge to the bosonic partons and $-1$ charge to the fermionic partons. The theory in terms of new degrees of freedom is then decomposed into 
\begin{equation}
\mathcal{L}  = \mathcal{L}_{\text{bosonic}}(\chi,q_1A+a)+\mathcal{L}_{\text{fermionic}}(\phi,q_2A-a)   
\end{equation}
Reorganizing the gauge field that interacts with bosonic and fermionic partons to be
\begin{equation}
    \alpha_{\chi}  = q_1A+a,\quad \alpha_{\phi}  = q_2A-a
\end{equation}
$\phi$ parton now is feeling the flux from $\alpha_\phi$, $d\alpha_\phi = d(q_2A-a)= -da$ in the zero external flux case, this means that the existence of the auxiliary gauge field compensates for the zero-flux of the external field. We first condense the bosonic parton to the superfluid phase that higgs $\alpha_\chi$ and then focus on the field theory of the fermionic parton, which is exactly the field theory that resembles the FQH case. 

Coincidentally, such a decomposed picture can be visualized in the factorized form of the single particle wavefunction in the ideal $\mathcal{C} = 1$ Chern band in the twisted $\mathrm{MoTe_2}$
\begin{equation}
\psi_{\boldsymbol{k} \ell}(\boldsymbol{r})=\phi_{\boldsymbol{k}}(\boldsymbol{r}) \zeta_\ell(\boldsymbol{r}) = f(z) e^{-K(\boldsymbol{r})} \zeta_\ell(\boldsymbol{r})
\end{equation}
$\ell = \pm 1$ is the layer indices. $\zeta(r)$ is $k$-independent and encodes the information of the orbital and layer basis.  $\phi_k(r)$ describes a fermion residing in the inhomogeneous periodic magnetic field. The single particle wavefunctions indicate that the many-body wave function has the form of 
\begin{equation}
    \Psi\left(\left\{\boldsymbol{r}_i\right\}\right)=\Psi_\phi \Psi_\chi=\Phi\left(\boldsymbol{r}_i\right) \prod_i \zeta_{\ell_i}\left(\boldsymbol{r}_i\right)
\end{equation}
where $\Psi_\phi$ is the wavefunctions of the flux-feeling particles. The many-body excitonic function at $\frac{1}{3}$ filling ${\Psi^{\text{exc}}_{\nu = \frac{1}{3}}}$, in the same manner, is established to be
\begin{equation}
    {\Psi^{\text{exc}}_{\nu = \frac{1}{3}}}\left(\left\{\boldsymbol{r}_i\right\}\right) = \Phi_{\nu = \frac{1}{3}}^{\text{exc}}\left(\boldsymbol{r}_i\right)\prod_i \zeta_{\ell_i}\left(\boldsymbol{r}_i\right).
\end{equation}
where $\Phi_{\nu = \frac{1}{3}}^{\text{exc}}$ is the many-body excitonic wavefunction of the fermionic parton.

Here the excited wavefunction of $\phi$-parton $\Phi_{\phi,\nu=1/3}^{\text{ex}}(\{z_i\})$ can be extracted from our SMA ansatz $|\psi_{\text{ex},\bm q}\rangle=\pmb{\bm\bar\rho}_e(\bm q)|\psi_0\rangle$, with the help of the connection between the LLL single-body electronic density operator and the CF density operators. Basically, we can follow the projective construction of composite-fermion states to enlarge the electronic Hilbert space by tensor product with the vortex Hilbert space \cite{murthy2003hamiltonian,PhysRevB.109.245125}. And the physical electronic states must satisfy the constraint
\begin{equation}\label{eq:projective_construction}
    |\psi_e\rangle\otimes|\psi_v\rangle=\pmb{\bm{\mathcal P}}_v|\psi_{\text{CF}}\rangle,
\end{equation}
where $\pmb{\bm{\mathcal P}}_v\equiv|\psi_v\rangle\langle\psi_v|$ is the many-body projector of vortex states (as some gauge degree of freedom), which can also be appreciated as LLL projection $\mathcal P_{\text{LLL}}$ when feeding into Jain's prescription. Generally, however, the electronic many-body wavefunction satisfying Eq.\eqref{eq:projective_construction} is shown to take the form of \emph{hyperdeterminants} of some tensors \cite{PhysRevB.109.245125}. For the $\nu=1/3$ states here, the vortex states should be bosonic with filling $\nu=1/2$, and thus it is natural to take a choice of $\nu=1/2$ bosonic Laughlin states, and the electronic wavefunction reads
\begin{equation}\label{eq:GS_wavefunction}
    \Phi_{\nu=1/3}(\{z_{e,i}\})=\int\mathcal D\omega_{v,j}\mathcal D\zeta_{\text{CF},k}\,\psi_{\nu=1/2,v}^{\text{Laughlin}}(\{\omega_{v,j}\})^*\psi_{\text{CF}}(\{\zeta_{\text{CF},k}\})\langle z_{e,i},\omega_{v,j}|\zeta_{\text{CF},k}\rangle
\end{equation}
for the coherent basis of electrons and vortex $\{z_{e,i}\}$ and $\{\omega_{v,j}\}$, and the position basis (delta function) of CFs $\{\zeta_{\text{CF},k}\}$. Here $i,j,k$ are particle index and $\int\mathcal D z_{\alpha}\equiv\int\prod_i\dfrac{\mathrm{d}z_{\alpha,i}}{2\pi\ell_\alpha^2}$ represents the coherent state integration. The fusion coefficients $\langle z_{e,i},\omega_{v,j}|\zeta_{\text{CF},k}\rangle$ can be evaluated with the help of magnetic translation algebra \cite{PhysRevB.109.245125}
\begin{equation*}
    \langle z_{e,i},\omega_{v,j}|\zeta_{\text{CF},k}\rangle=\sqrt{\dfrac{1+c}{1-c}} e^{-\frac{1}{4\ell_v^2}\bar\omega\omega+\frac{1}{2\ell_v^2}\bar\omega(-\frac{z}{c}+\frac{1+1}{c}\zeta)} e^{-\frac{1}{4\ell_{\text{CF}}^2}\frac{1-c}{1+c}\bar\zeta\zeta+\frac{1}{2\ell_{\text{CF}^2}}\frac{\bar\zeta z}{1-c}} e^{-\frac{\bar z z}{4\ell_e^2}}.
\end{equation*}

Following Murthy-Shankar's composite-fermion substitution \cite{murthy2003hamiltonian}
\begin{equation}\label{eq:CF_substitution}
    \bm{\mathcal R}_e=\bm{\mathcal R}+c\bm\eta,\quad \bm{\mathcal R_v}=\bm{\mathcal R}-\bm\eta/c,
\end{equation}
the action of the projected electronic density oparator $\pmb{\bm{\bar\rho}}_e(\bm q)$ on the ground state then can be lifted to the action within the CF Hilbert space (up to some phases). After fixing the gauge to be consisting with the expectation in the continuum limit, where bare guiding-center coordinates and cyclotron coordinates are all well-defined as Eq.\eqref{eq:CF_substitution}, we can lift the action of LLL electronic density operator to CF Hilbert space as
\begin{equation}
    \rho_e(\bm q)=\rho_{\bm{\mathcal R}}(\bm q)\rho_{\bm\eta}(-c\bm q),
\end{equation}
where $\rho_{\bm{\mathcal R}}(\bm q)=e^{i\bm q\cdot\bm{\mathcal R}}$ and $\rho_{\bm\eta}(\bm q)=e^{-i\bm q\cdot\bm\eta}$ are the density operator of the CF guiding-center/cyclotron degrees of freedom.
% as $|\zeta_{\text{CF},k}\rangle\rightarrow|\widetilde{\zeta}_{\text{CF},\bm q,k}\rangle\equiv\rho_{\bm{\mathcal R}}(\bm q)\rho_{\bm\eta}(-c\bm q)|\zeta_{\text{CF},k}\rangle$.
% , which can be computed by projecting the Dirac delta function at position $\zeta_{\text{CF},k}$ to CF LLs. % Precisely, the single-body density operator can be recognized with the displacement operator for the coherent-state basis $\rho_\alpha(\bm q_\alpha)=D_\alpha(\bm q_\alpha)$. 
Consequently, the excited electronic wavefunction following the SMA approximation can still be obtained following the projective constrution Eq.\eqref{eq:projective_construction}, so will be in a similar form as Eq.\eqref{eq:GS_wavefunction}, but with the new CF wavefunction, i.e., the excited CF wavefunction deviated from the original one due to the lifted action
\begin{equation}\label{eq:excited_CF_wavefunction}
    \psi_{\text{CF},\bm q}(\{\zeta_{\text{CF},k}\})\equiv\langle\zeta_{\text{CF},k}|\rho_{\bm{\mathcal R}}(\bm q)\rho_{\bm\eta}(-c\bm q)|\psi_{\text{CF}}\rangle.
\end{equation}
To extract the physical essense of this excited CF wavefunction, one just need to remind that $\bm{\mathcal R}$ controls the motion within the CF Landau level, while $\bm\eta$ controls the motion within the $\bm\eta$-plane constructed in Ref. \cite{PhysRevB.109.245125}, after the exact mapping from the Landau level indices to a fictitious plane with a different sample size controlled by the fillings hosting a LLL. As a result, any shifts of coherent states within the $\bm\eta$-plane due to $\rho_{\bm\eta}(\bm q)$ will corresponds to a CF LL index change, so Eq.\eqref{eq:excited_CF_wavefunction} does represent some superposition of the CF particle-hole excitations, matching the CF understanding of magnetoroton exicitations. The discrepancy of SMA to the exact computatoin of magnetoroton excitations will be on the envelope function of CF particle-hole pairs (which is ansatzed here).

\setcounter{secnumdepth}{3}
\setcounter{equation}{0}
\setcounter{figure}{0}
\renewcommand{\theequation}{S\arabic{equation}}
\renewcommand{\thefigure}{S\arabic{figure}}
\renewcommand\figurename{Supplementary Figure}
\renewcommand\tablename{Supplementary Table}
\newcommand\Scite[1]{[S\citealp{#1}]}
\makeatletter \renewcommand\@biblabel[1]{[S#1]} \makeatother

\end{document}